\RequirePackage{silence}
\documentclass[fleqn,usenatbib,useAMS]{mnras}

\usepackage{ae,aecompl}

\usepackage{graphicx}	
\usepackage{amsmath}	
\DeclareMathOperator{\sech}{sech}
\usepackage{amssymb}	
\usepackage{multicol}        
\usepackage{multirow}
\usepackage{bm}		
\usepackage{booktabs}
\usepackage{listings}
\usepackage{color}
\usepackage{threeparttable}
\usepackage{hyperref}
\usepackage{tikz}
\usetikzlibrary{calc}
\usetikzlibrary{decorations.markings}
\usetikzlibrary{shadings, calc, decorations.markings}

\DeclareRobustCommand*{\matr}[1]{\mathbfss{#1}}

\title[Barred spiral galaxies in modified gravity]{Barred spiral galaxies in modified gravity theories}

\author[M. Roshan et al.]{
Mahmood Roshan,$^{1,2}$\thanks{E-mail: \href{mailto:mroshan@um.ac.ir}{mroshan@um.ac.ir}}
Indranil Banik,$^{3}$\thanks{Alexander von Humboldt Fellow}
Neda Ghafourian,$^{1}$
Ingo Thies,$^{3}$
\newauthor
Benoit Famaey$^4$,
Elena Asencio$^3$,
and Pavel Kroupa$^{3,5}$
\\
$^{1}$Department of Physics, Faculty of Science, Ferdowsi University of Mashhad, P.O. Box 1436, Mashhad, Iran\\
$^{2}$School of Astronomy, Institute for Research in Fundamental Sciences (IPM), 19395-5531, Tehran, Iran\\
$^{3}$Helmholtz-Institut f\"ur Strahlen- und Kernphysik, Universit\"at Bonn, Nussallee 14-16, 53115 Bonn, Germany\\
$^{4}$Universit\'e de Strasbourg, CNRS, Observatoire astronomique de Strasbourg, UMR 7550, F-67000 Strasbourg, France\\
$^{5}$Astronomical Institute, Faculty of Mathematics and Physics, Charles University in Prague, V Hole\v{s}ovi\v{c}k\'ach 2, CZ-180 00 Praha 8,\\Czech Republic}

\pubyear{2021} 
\begin{document}
\label{firstpage}
\pagerange{\pageref{firstpage}--\pageref{lastpage}}
\maketitle

\begin{abstract}

When bars form within galaxy formation simulations in the standard cosmological context, dynamical friction with dark matter (DM) causes them to rotate rather slowly. However, almost all observed galactic bars are fast in terms of the ratio between corotation radius and bar length. Here, we explicitly display an $8\sigma$ tension between the observed distribution of this ratio and that in the EAGLE simulation at redshift 0. We also compare the evolution of Newtonian galactic discs embedded in DM haloes to their evolution in three extended gravity theories: Milgromian Dynamics (MOND), a model of non-local gravity, and a scalar-tensor-vector gravity theory (MOG). Although our models start with the same initial baryonic distribution and rotation curve, the long-term evolution is different. The bar instability happens more violently in MOND compared to the other models. There are some common features between the extended gravity models, in particular the negligible role played by dynamical friction $-$ which plays a key role in the DM model. Partly for this reason, all extended gravity models predict weaker bars and faster bar pattern speeds compared to the DM case. Although the absence of strong bars in our idealized, isolated extended gravity simulations is in tension with observations, they reproduce the strong observational preference for `fast' bar pattern speeds, which we could not do with DM. We confirm previous findings that apparently `ultrafast' bars can be due to bar-spiral arm alignment leading to an overestimated bar length, especially in extended gravity scenarios where the bar is already fast.

\end{abstract}



\section{Introduction}
\label{Introduction}

The missing gravity problem on galaxy and larger scales is one of the long-standing challenges in theoretical physics. After a few early hints, it was put forward almost a century ago in the Coma galaxy cluster \citep{Zwicky_1933, Zwicky_1937} and in the Local Group \citep{Kahn_Woltjer_1959}. From the 1970s onward, it has been taken as a serious problem appearing on galactic and cosmological scales \citep{Rubin_Ford_1970, Rogstad_1972, Roberts_1975, Bosma_1981} $-$ for an early review, see \citet{Faber_1979}. A related issue is that self-gravitating Newtonian discs are unstable \citep{Miller_1968, Hockney_1969, Hohl_1971}. After half a century, the solution is still not known.

The standard hypothesis is haloes of cold dark matter (CDM) particles surrounding each galaxy \citep{Ostriker_1973}. Their microphysical properties are ever more severely constrained by null detections in sensitive searches \citep[e.g.][]{Hoof_2020}. The CDM hypothesis is a main ingredient of $\Lambda$CDM, the current standard cosmological model \citep{White_1978, Efstathiou_1990, Ostriker_Steinhardt_1995}. Without the contribution of CDM, cosmic structures cannot be formed in the context of Einstein's general relativity (GR). Even with the assumption of CDM and a cosmological constant $\Lambda$, the $\Lambda$CDM paradigm still faces cosmological tensions, for instance with foreground lensing of the cosmic microwave background \citep[CMB,][]{Valentino_2019} or the present expansion rate of the Universe {\citep[e.g.][and references therein]{Valentino_2021}}.

On the smaller scale of individual galaxies, high-resolution hydrodynamical simulations of structure formation reveal several additional challenges \citep[e.g.][]{Kroupa_2010, Weinberg_2015, Bullock_2017}. Depending on their central baryonic surface density, observed spiral galaxies display a wide diversity of rotation curve shapes at a fixed mass scale, which would imply a large variety of central dark matter (DM) profiles ranging from cusps to cores \citep{Oman_2015}. It is very difficult to explain this diversity through stochastic feedback processes  \citep[][]{Ghari_2019} while maintaining other observed regularities such as the radial acceleration relation (RAR), a very tight relation between the gravity inferred from galaxy rotation curves and that expected from the baryons alone \citep{McGaugh_Lelli_2016, Lelli_2017}. The dynamics of elliptical galaxies appear to delineate the same RAR as spirals \citep{Chae_2020_elliptical, Shelest_2020}.

One of the most persistent small-scale challenges to $\Lambda$CDM is the plane of satellite galaxies around the Milky Way \citep[MW,][]{Kroupa_2005}. In the last decade, the situation has dramatically worsened with the discovery of similar planes around M31 \citep{Ibata_2013} and Centaurus \citep{Muller_2018, Muller_2021}, with hints of a plane around M83 \citep{Muller_2018_M83}. Gaia data on members of the MW satellite plane confirm its existence at extremely high significance \citep{Pawlowski_2020}, while the only two M31 satellite plane members with known proper motions also suggest the dynamical coherence of this structure \citep{Sohn_2020}. A comprehensive review of possible solutions in $\Lambda$CDM failed to find any that were viable \citep{Pawlowski_2014}. Such thin planes of satellite galaxies can actually be most naturally explained as tidal dwarfs formed out of the debris expelled by a previous galactic interaction \citep{Pawlowski_2012}. Such tidal dwarf galaxies are expected to be free of CDM, as shown using idealized simulations \citep{Barnes_1992, Wetzstein_2007} and in the IllustrisTNG \citep{Pillepich_2018} cosmological hydrodynamical simulation \citep{Haslbauer_2019}. Without the presence of CDM, some other explanation must be found for the high internal velocity dispersions of the MW and M31 satellites \citep{McGaugh_2010, McGaugh_2013}, and extended or modified gravity can be a solution.\footnote{Throughout this paper, we use modified gravity or extended gravity interchangeably when discussing alternatives to GR. However, extended gravity may have a firmer terminological basis because nature does not modify a law, but rather follows one particular law. More importantly, the original formulation of gravitation by Newton and Einstein was based only on the empirical data then available, i.e. for Solar System objects. With the availability of dynamical data on the scale of galaxies in the late 1970s and early 1980s, it became established that this description of gravitation fails, unless additional DM is hypothesized to exist. A plausible solution to this missing gravity problem is that the original classical formulation of gravitation needs to be extended, perhaps to include quantum corrections \citep{Milgrom_1999}.}

Another less commonly discussed problem is the fact that observed spiral galaxies harbour non-axisymmetric features like central bars and grand-design spirals, which are often difficult to fully reproduce with galaxy formation simulations in a cosmological context. Such simulations \citep[e.g.][]{Hopkins_2018} sometimes almost completely lack bars in non-quenched galaxies at redshift $z = 0$, probably related to their sub-grid feedback recipes being too efficient. Such efficient feedback is however needed in the simulations to get roughly the right baryon fraction in galaxies. However, other cosmological simulation projects like IllustrisTNG and Auriga do reproduce some properties of barred galaxies like the bar fraction, bar sizes, and luminosities \citep{Blazquez_2020, Fragkoudi_2020, Rosas_Guevara_2020}.

When bars do form in such simulations, dynamical friction with CDM particles causes them to rotate rather slowly \citep{Tremaine_1984}, while almost all observed bars are fast \citep{Debattista_2000, Aguerri_2015, Algorry_2017, Guo_2019}. This problem was recently revisited by \citet{Peschken_2019}, who applied the so-called \citet{Tremaine_1984} method to find the pattern speeds of bars in the Illustris cosmological simulation \citep{Vogelbserger_2014}. Figure~8 of \citet{Peschken_2019} shows that most of the bars are slow. Also, flocculent spirals are much more common in those simulations than regular grand-design spiral arms, even though the latter are observed in most disc galaxies \citep{Hart_2017}.

All this means that it is extremely valuable to seriously consider the other major approach to the missing gravity problem $-$ modifying gravity or inertia in the weak-field regime, hence modifying Newtonian dynamics. In this approach, there is in principle no need for DM particles, at least on galaxy scales. Introducing modifications to standard gravity has a long history dating back to Einstein when he tried to find a Palatini approach to GR. Attempts to modify GR were continued by others like Eddington, Schr\"odinger, and Cartan. Their motivations were generally theoretical in nature. Most extended theories of gravity that are currently explored are instead motivated by observations, usually in order to resolve cosmological issues like the nature of dark energy, inflation, and other problems \citep[for comprehensive reviews, see][]{Clifton_2012, Baker_2019}.

Modifications to the gravitational law specifically to address the missing gravity in galaxies started with \citet{Finzi_1963} and were continued by e.g. \citet{Tohline_1983}. These works did not gain serious attention from the astrophysics community. For the latter work, this is partly because in the same year, \citet{Milgrom_1983} presented another more successful approach in which the modification arises specifically at low acceleration. After almost 40 years, it is now well-known that this Milgromian dynamics (MOND) approach is remarkably successful at explaining spiral galaxy rotation curves and many other relevant observations \citep{Famaey_2012}. Indeed, galactic-scale dynamical discrepancies arise below a certain acceleration rather than beyond a fixed distance (e.g. {their} figure~10). The most interesting point about MOND is the introduction of a single new constant of nature with dimensions of acceleration. We will discuss the main features of MOND in Section \ref{MOND}. Its unique relativistic completion (sometimes dubbed FUNDAMOND) is still unknown, though the recently developed theory of \citet{Skordis_2019} is promising because the model predicts that tensor mode gravitational waves propagate at the speed of light $c$, even in the presence of structure. Moreover, it can in principle provide a term decoupled from the baryon-photon plasma in the early Universe to explain the CMB angular power spectrum and obtain a standard late-time matter power spectrum \citep{Skordis_2020}. Further exploration of this and other relativistic MOND theories would help to check their viability in a cosmological context, which is the single most important next step for MOND in general.

In this regard, the recent study of \citet*{Haslbauer_2020} describes a viable MOND cosmology where CDM is replaced by the same total mass in light sterile neutrinos, as originally proposed by \citet{Angus_2009}. This leads to the same behaviour as $\Lambda$CDM with regards to the CMB anisotropies, primordial light element abundances, and overall expansion history. The use of MOND for structure formation leads to significant differences, in particular by allowing us to reside in a very large and deep void with enhanced apparent Hubble constant. Such a void is actually observed \citep{Keenan_2013} and contradicts $\Lambda$CDM at $6.04\sigma$, which rises to $7.09\sigma$ in combination with the Hubble tension \citep{Haslbauer_2020}. However, those authors showed that these and other important local Universe observables can be explained with only $2.53\sigma$ tension in MOND. The use of hot dark matter (HDM) in this so-called $\nu$HDM model also allows the dynamics of galaxy clusters to be explained in a MOND context \citep{Angus_2010} without much affecting galaxies \citep{Angus_2010_galaxies}. $\nu$HDM also better explains the formation of galaxy clusters like the massive high-redshift interacting pair known as El Gordo \citep{Menanteau_2012}, whose properties arise naturally at about the right frequency in cosmological $\nu$HDM simulations \citep{Katz_2013}. El Gordo would be an extremely unlikely $6.16\sigma$ outlier in $\Lambda$CDM cosmology \citep*{Asencio_2021}, indicating that it underpredicts both overdensities and underdensities on large scales little affected by baryonic physics in galaxies.

Moffat's scalar-tensor-vector theory of gravity (known as MOG in the literature) is another modification to GR \citep{Moffat_2006}. This theory uses two scalar fields and a vector field to address the missing gravity problem. MOG is a covariant generalization of GR whose consequences can in principle be investigated in cosmology, which is possible in MOND only if some particular relativistic extension is adopted. Another interesting extended gravity theory is non-local gravity \citep[NLG,][]{Hehl_2009}. NLG uses the metric tensor without any other gravitational field. It provides some modifications to GR originating from the non-local features of gravity. Interestingly, these non-local corrections can mimic the behaviour of CDM, at least on galactic scales. In all cases, the parameters of these theories must be tuned so that the galaxy rotation curves $-$ which are well predicted by MOND $-$ become close enough to the MOND phenomenology, and thus reproduce the observed RAR. The detailed secular evolution of spiral galaxies might however be different in these various frameworks, especially the development of bar instabilities.

Therefore, our purpose in this paper is to compare the traditional CDM model with MOND, NLG, and MOG in simulated barred spiral galaxies, as pioneered by \citet{Tiret_2007} in MOND. These theories are among the main approaches presented in the literature to solve the missing gravity problem in galaxies. Where a galaxy's missing gravity comes from likely has serious implications for its secular evolution.

The outline of this paper is as follows: In Section \ref{nlgmog}, we briefly introduce MOND, NLG, and MOG. In Section \ref{methods}, we describe our numerical codes and procedures for constructing the initial conditions. The results of our simulations are presented in Section \ref{results}, where we also use them to compare the above-mentioned theories. Although we are mainly interested in comparing theories with each other in a non-cosmological context, we also consider the latest observational constraints, especially for the more mature $\Lambda$CDM simulations. In particular, Section \ref{R_statistics} quantifies a significant tension between the statistical properties of bars in the EAGLE simulations and in observed galaxies. We discuss our results in Section \ref{Discussion} and summarize them in Section \ref{conclusion}.

\section{Alternative theories of gravity without particle dark matter}\label{nlgmog}

In this paper, we compare the evolution of isolated spiral galaxies in the context of three well-known extended gravity theories, and compare them with the standard CDM halo models. In the following, we briefly introduce the extended gravity theories.

\subsection{Milgromian Dynamics (MOND)}
\label{MOND}

MOND \citep{Milgrom_1983} is the main alternative to galactic DM. It postulates that the gravitational field strength $g$ at distance $r$ from an isolated point mass $M$ transitions from the Newtonian ${GM/r^2}$ law at short range to:
\begin{eqnarray}
	g ~=~ \frac{\sqrt{GMa_{_0}}}{r} ~~~\text{for } ~ r \gg \sqrt{\frac{GM}{a_{_0}}} \, .
	\label{Deep_MOND_limit}
\end{eqnarray}
MOND introduces $a_{_0}$ as a fundamental acceleration scale of nature below which the deviation from Newtonian dynamics becomes significant. Empirically, $a_{_0} \approx 1.2 \times {10}^{-10}$ m/s$^2$ to match galaxy rotation curves \citep{Begeman_1991, Gentile_2011}. With this value of $a_{_0}$, MOND continues to fit galaxy rotation curves very well using only their directly observed baryonic matter \citep[e.g.][]{Li_2018, Kroupa_2018, Sanders_2019}. In particular, observations confirm the \emph{a priori} MOND prediction of very large departures from Newtonian dynamics in low surface brightness galaxies \citep[LSBs, e.g.][]{Blok_1997, McGaugh_1998}. More generally, there is a very tight empirical relation between the gravity inferred from rotation curves and that expected from the baryons alone in Newtonian dynamics \citep{McGaugh_Lelli_2016, Lelli_2017}. This RAR confirms the central prediction of \citet{Milgrom_1983}. One important consequence is that the asymptotic rotational velocity $v_f$ far from an isolated galaxy is related to its total baryonic mass according to
\begin{eqnarray}
	v_f ~=~ \sqrt[4]{GMa_{_0}} \, .
	\label{BTFR}
\end{eqnarray}
This relation is known as the baryonic Tully-Fisher relation (BTFR), which extends the work of \citet{Tully_Fisher_1977} and has been reviewed elsewhere \citep[e.g.][]{McGaugh_2020}. More complicated geometries should be handled using Equation \ref{QUMOND_governing_equation}.

In this contribution, we focus on the most common extended gravity interpretation of MOND. It can also be interpreted as extended inertia \citep{Milgrom_1994}, but the appropriate field equations are not clear. This is partly because they must be strongly non-local to be consistent with observations. One of the most pressing issues in this regard is to understand the barycentric behaviour of a composite body with high internal accelerations. This issue is completely resolved in an extended gravity interpretation of MOND $-$ gravity follows the standard inverse square law near the Sun and the Galactic centre, but a different law applies in the low-acceleration regions in between \citep{Bekenstein_1984}.

\subsection{Nonlocal gravity (NLG)}
\label{NLG}

In a series of papers, Mashhoon and collaborators have investigated fundamental issues related to non-locality in special relativity \citep[][and references therein]{Mashhoon_2017}. The idea is that the locality hypothesis is a \textit{useful approximation} \citep{Einstein_1950}. In principle, it could be violated for highly accelerated observers. This directly means that non-local effects should appear in GR as corrections to the field equations. A novel approach to implement non-local features into GR has been introduced by \citet{Hehl_2009}. Their approach exploits the similarity between Maxwell's equations and that of the teleparallel equivalent theory of GR \citep{Hehl_2009b}. In this way, the non-local corrections to GR can be constructed similarly to those in electrodynamics. Eventually, by postulating a specific form for the non-locality tensor $\matr{N}_{\mu\nu}$, a non-local version of GR in the teleparallel formalism has been introduced.\footnote{A teleparallel relativistic version of MOND has also been proposed \citep{Dambrosio_2020}.} In this theory, the gravitational behaviour of the system depends on its past. For recent developments in the theoretical aspects, we refer the reader to \citet{Puetzfeld_2019} and \citet{Puetzfeld_2020}.

In the Newtonian limit, the non-local terms show up as an extra effective `phantom' density in the right hand side of Poisson's equation. This effective density is given by:
\begin{eqnarray}
	\rho_p(\bm{r}) ~=~ \int q \left( \left| \bm{r} - \bm{r}' \right| \right) \rho_b \left( \bm{r}' \right) \, d^3\bm{r}' \, ,
	\label{dmnlg}
\end{eqnarray}
where $\rho_b$ is the density of the baryonic matter, and $q$ is a kernel which should be found from observations. $\rho_p$ mimics the conventional DM density at galactic scales. Using rotation curve fits, a suitable form for the kernel can be written as \citep{Rahvar_2014}:
\begin{eqnarray}
	q \left( \left| \bm{r} - \bm{r}' \right| \right) ~=~ \left( \frac{1}{4 \mathrm{\pi} \lambda_0} \right) \frac{1 + \mu_0 \left| \bm{r} - \bm{r}' \right|}{\left| \bm{r} - \bm{r}' \right|^2} \, \mathrm{e}^{-\mu_0 \left| \bm{r} - \bm{r}' \right|} \, ,
\end{eqnarray}
where $\lambda_0$ and $\mu_0$ are two free parameters with dimensions of length and inverse length, respectively. Using this kernel and the modified version of the Poisson equation, one may find the mutual gravitational force between two point masses located at $\bm{r}_1$ and $\bm{r}_2$:
\begin{eqnarray}
	\bm{g}_{\text{NLG}} ~=~ \left(1 + \alpha-\alpha \left[1 + \frac{\mu_0}{2} \left| \bm{r}_2-\bm{r}_1 \right| \right] \, \mathrm{e}^{-\mu_0 \left| \bm{r}_2 - \bm{r}_1 \right|} \right) \bm{g}_{\text{N}} \, ,
	\label{g_NLG}
\end{eqnarray}
where $\bm{g}_{\text{N}}$ is the Newtonian force, and $\alpha = 2/\left( \mu_0 \lambda_0 \right)$. These parameters should be obtained from observations like rotation curve data \citep[e.g.][]{Rahvar_2014}. In our simulations, we use the following parameters obtained by fitting to our DM numerical model: $\alpha = 10.0$ and $\mu_0 = 0.0525$~kpc$^{-1}$.

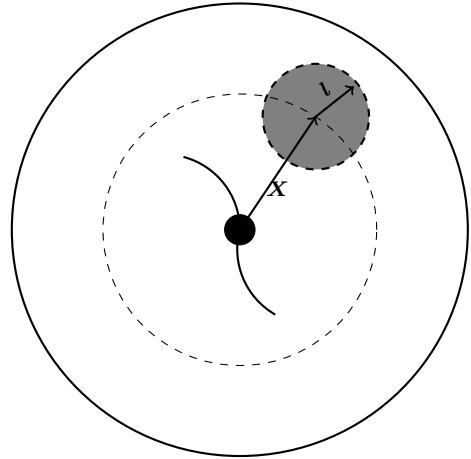
\begin{figure}
    \centering
    \begin{tikzpicture}
        \draw[thick] (2,2) circle (3cm);
        \filldraw[fill=gray,thick,dashed] (3.,3.5) circle (0.7cm);
        \filldraw[fill=black] (2,2) circle (0.2cm);
        \draw[dashed] (2,2) circle (1.8cm);
        \draw[->][thick](2,2) -- (3,3.5)node [midway,below] {$\bm X$};
        \draw[->][thick](3,3.5) -- (3.5,3.9) node[midway,above,sloped] {$\bm l$};
        \draw[thick] (2,2) arc (0:75:1cm);
        \draw[thick,rotate around={165:(2,2)}] (2,2) arc (0:75:1cm);
    \end{tikzpicture}
    \caption{Illustration of the characteristic length $l = \mu_0^{-1}$ (the radius of the grey circle) on the surface of a galactic disc at location $\bm{X}$ beyond which NLG effects first become dominant.}
    \label{fl}
\end{figure}

It is clear that there is a characteristic length $l = \mu_0^{-1}$ in NLG. This length is assumed to be fundamentally related to the non-local features of gravity. However, one should note that the existence of this length does not mean that there is a fixed distance in galaxies beyond which we expect NLG effects to dominate over conventional gravity. Specifically, significant NLG effects will appear only at positions $\bm{X}$ in the system where the correction to the gravitational force due to NLG becomes comparable to or larger than the Newtonian gravitational force. The NLG force is mainly obtained from the baryonic mass inside radius $l = \mu_0^{-1}$ around the point $\bm{X}$, as schematically shown in Figure~\ref{fl}. Expressed more precisely, the gravitational field in NLG can be written as:
\begin{eqnarray}
	\bm{g}_{\text{NLG}}(\bm{X}) ~=~ \bm{g}_{\text{N}}(\bm{X}) + \bm{\Delta}(\bm{X};\alpha,\mu_0) \,
\end{eqnarray}
where $\bm{\Delta}(\bm{X};\alpha,\mu_0)$ is the correction term due to $\rho_p$. To determine the length $X$ beyond which NLG effects appear, one should find the distance where the correction force first becomes comparable to the Newtonian force. In other words, we should solve the following equation:
\begin{eqnarray}
	\left|\bm{\Delta} \left( \bm{X};\alpha,\mu_0 \right) \right| ~\ga~ \left| \bm{g}_{\text{N}} \left( \bm{X} \right) \right| \, .
\end{eqnarray}
Depending on the value of $\alpha$ and the baryonic matter distribution, the resulting length $X$ could be completely different from $\mu_0^{-1}$. Let us assume that the baryonic matter density has characteristic length $r_b$ and characteristic mass $M_b$. Then, $X$ would in principle be a function of $\alpha$, $\mu_0$, $r_b$, and $M_b$, namely:
\begin{eqnarray}
	X ~=~ X(\alpha,\mu_0,r_b,M_b) \, .
\end{eqnarray}
At the phenomenological level, this combination of parameters should mostly reproduce the MOND phenomenology at equilibrium, although simulated galaxies could evolve differently from MOND. The current version of NLG respects the weak equivalence principle in the Newtonian limit. Consequently, the kernel needs to be a universal function and the free parameters should be the same for all galaxies \citep{Mashhoon_2017}. However, a general version of the theory might not respect the weak equivalence principle. In this case, the theory does not prevent the parameters $\alpha$ and $\mu_0$ from being mass-dependent.

The energy-momentum tensor $\matr{T}_{\mu\nu}$ is not conserved in NLG, i.e. $\nabla_{\mu} \matr{T}^{\mu\nu} = {\mathcal{I}}^{\nu}$, where ${\mathcal{I}}^{\nu}$ is a tensor containing the non-local features of gravity. The main kernel of NLG also appears in this tensor. In the current version of the theory, ${\mathcal{I}}^{\nu}$ is postulated to be a universal function without any direct dependence on the physical properties of the underlying self-gravitating system. It is possible to construct a model of NLG by postulating a mass-dependent kernel. In this case, Einstein's principle of equivalence will be violated because the right-hand side of the geodesic equation would depend on the internal structure of the free-falling body \citep{Roshan_2013}.

\subsection{MOG}
\label{MOG}

MOG is a scalar-tensor-vector theory of gravity \citep{Moffat_2006}. The scalar fields in this model are similar in spirit to the Brans-Dicke theory. Thus, one may say that the gravitational constant is time-dependent and appears as a scalar field in the field equations. There is also another scalar field that appears as a dynamical mass for the vector field in MOG. The existence of these three extra fields enables MOG to behave differently than GR on galactic and extragalactic scales.

It has been shown that the theory has a true sequence of cosmological epochs. In other words, the cosmos starts from a radiation-dominated universe and then enters a matter-dominated phase, before finally evolving towards an accelerated de Sitter universe \citep{Jamali_2017}. One main issue which should still be addressed by MOG is cosmic structure formation. The evolution of cosmic perturbations has been investigated in \citet{Jamali_2018}. It turns out that MOG is consistent with the redshift distortion data.

The CMB power spectrum in MOG has been investigated in \citet{Moffat_2011}. Their work shows that MOG leads to a serious enhancement of the baryon acoustic oscillations, which in general is a very important constraint on any cosmological model \citep{Pardo_2020}. A similar problem was claimed to occur in the relativistic version of MOND known as TeVeS \citep{Dodelson_2011}, but its inherent non-linearity means that modes of different wavelengths would mix to a substantial extent, likely smoothing these oscillations in a way that is difficult to determine without numerical simulations \citep[section 5.2 of][]{Mcgaugh_2015}. In the MOND context, it is also possible to extend the dynamics of the $k$-essence scalar field of TeVeS \citep{Bekenstein_2004} to play the role of DM in the CMB and in the linear regime of structure formation \citep{Skordis_2020}, or for a MOND-HDM hybrid model to explain the CMB similarly to $\Lambda$CDM \citep{Angus_2009, Haslbauer_2020}. The results presented in \citet{Moffat_2011} are also based on various assumptions and analytic descriptions. Therefore, it is necessary to modify standard codes like CAMB \citep{Lewis_2002} to find the CMB power spectrum in MOG. Due to the existence of three extra fields, this is not a simple task. However, it will be necessary to test the cosmological viability of such a theory.

On galactic scales, MOG introduces some modifications to Newtonian gravity. In the weak field limit, the gravitational force between two point masses located at $\bm{r}_1$ and $\bm{r}_2$ takes the following form \citep{Roshan_2014}:
\begin{eqnarray}
	\bm{g}_{\text{MOG}} ~=~ \left(1 + \alpha-\alpha\left[1 + \mu_0 |\bm{r}_2 - \bm{r}_1|\right] \, e^{-\mu_0 |\bm{r}_2 - \bm{r}_1|}\right)\bm{g}_{\text{N}} \, ,
	\label{g_MOG}
\end{eqnarray}
where $\alpha$ and $\mu_0$ are two free parameters that can be fixed using rotation curve data \citep{Moffat_2013}. These free parameters are related to the background values of the scalar fields in the theory. However, in our simulations, these parameters are fixed at $\alpha = 4.7$ and $\mu_0= 0.2125$ kpc$^{-1}$. In Section \ref{Initial_conditions}, we describe how these free parameters in MOG and NLG have been fixed. The consequences of this extended gravitational law on the evolution of spiral galaxies have been investigated in \citet{Ghafourian_2017} and \citet{Roshan_2018}. Notice that despite the similarity between the NLG and MOG point mass force laws (Equations \ref{g_NLG} and \ref{g_MOG}), there is a slight difference because one factor of $\mu_0/2$ in NLG is replaced by $\mu_0$ in MOG in a manner distinct from a redefinition.

On smaller scales like the Solar System, the MOG corrections are very small. This means that the theory remains valid in the Solar System while leading to significant deviations from Newtonian gravity on galactic scales. However, the velocity dispersion profile of the ultra-diffuse galaxy Dragonfly 44 contradicts MOG at ${5.49\sigma}$ confidence \citep{Haghi_2019}. Thus, the MOG theory can be consistent with observations only if $\alpha$ and $\mu_0$ vary in some environment-dependent way, greatly diminishing the possibility of making a priori predictions. This problem is most likely caused by the fact that galactic-scale dynamical discrepancies arise below a certain acceleration rather than beyond a fixed distance. Thus, consistency with observations requires the MOG length scale ${\mu_0}^{-1} \propto \sqrt{M}$ \citep[equation 10 of][]{Green_2019}, making the theory behave rather similarly to MOND (Equation \ref{Deep_MOND_limit}). This also causes severe theoretical issues since $M$ is not a well-defined covariant quantity, meaning the theory cannot be fundamental. We nevertheless explore MOG as representative of a wider class of theories with an extra Yukawa-like force.

\subsection{Phantom dark matter in extended gravity models}
\label{Phantom_dark_matter}

The gravitational forces in NLG and MOG are similar (Equations \ref{g_NLG} and \ref{g_MOG}), so one might naively expect similar behaviour regarding the disc evolution. This is not true in detail because although these models lead to similar rotation curves at $z = 0$, their effective `phantom' DM density can be substantially different. We quantify this by rewriting the Poisson equation in NLG and MOG as:
\begin{eqnarray}
	\nabla^2\Phi ~\equiv~ 4 \mathrm{\pi} G \left( \rho_b + \rho_p \right) \, ,
\end{eqnarray}
where all the corrections to Newtonian gravity are collected in the extra term $\rho_p$. Inspired by \citet{Milgrom_1986} in the MOND context, we call this term the \textit{`phantom' DM density} $\rho_p$, since combining it with the baryon density $\rho_b$ gives the density distribution whose Newtonian gravity equals that of the baryons alone in some other theory. Using Equation \ref{dmnlg}, $\rho_p$ in NLG is:
\begin{eqnarray}
	\rho_p\left( \bm{r} \right) ~=~ \frac{\alpha\mu_0}{4 \mathrm{\pi}}\int \frac{1 + \mu_0 \left| \bm{r} - \bm{r}' \right|}{\left| \bm{r} - \bm{r}' \right|^2} \, \mathrm{e}^{-\mu_0 \left| \bm{r} - \bm{r}' \right|} \rho_b \left( \bm{r}' \right) \, d^3\bm{r}' \, .
	\label{eff1}
\end{eqnarray}
The corresponding result for MOG is \citep{Roshan_2014}:
\begin{equation}
	\rho_p \left( \bm{r} \right) ~=~ \frac{\alpha {\mu_0}^2}{8\mathrm{\pi}}\int \frac{ \mathrm{e}^{-\mu_0 \left| \bm{r} - \bm{r}' \right|} }{\left| \bm{r} - \bm{r}' \right|} \,\rho_b(\bm{r}') \, d^3\bm{r}' \, .
	\label{eff2}
\end{equation}
In MOND, it is necessary to first determine $\bm{g}_\text{N}$ and then solve Equation \ref{QUMOND_governing_equation}. Since the distribution of $\rho_p$ is not an important part of our analysis, for simplicity we discuss below its distribution around an infinitely thin exponential disc in isolation with the same central surface density as used elsewhere in this contribution. The $\bm{g}_\text{N}$ of this configuration was derived in \citet{Freeman_1970}.

\begin{figure} 
	\includegraphics[width = 8.2cm]{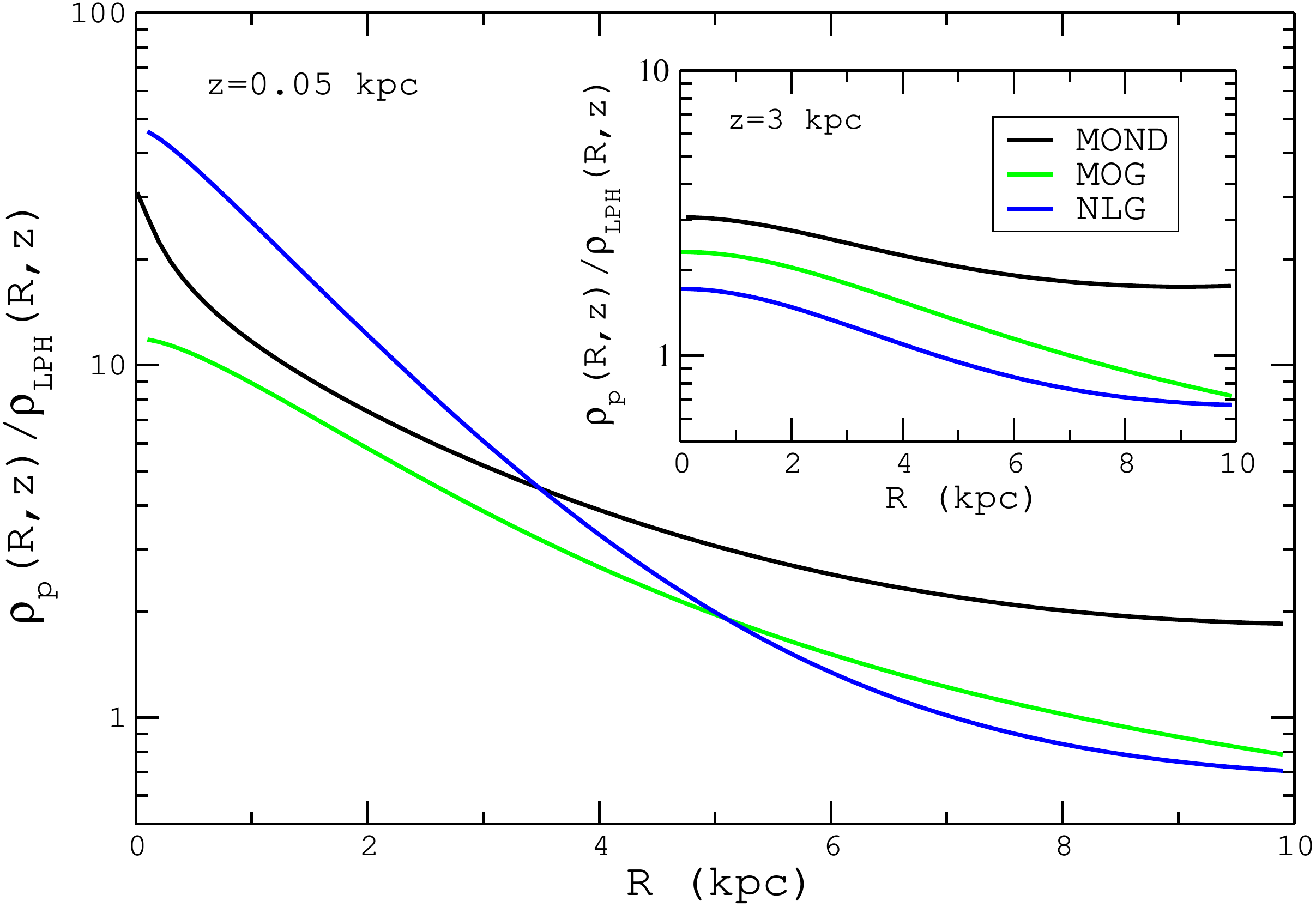}
	\caption{Initial effective `phantom' DM density relative to the Plummer DM density in terms of $R$ as measured at $z = 0.05$~kpc (main figure) and $z = 3$~kpc (inset), shown for an infinitely thin exponential disc whose scale length is 1~kpc. Densities are shown on a logarithmic scale. The distribution of $\rho_p$ for thin exponential MOND discs was visualized in e.g. \citet[][]{Lughausen_2013, Lughausen_2015}. We avoid the disc mid-plane because of the singular mass distribution there, which causes the MOND $\rho_p$ to diverge (see text).}
	\label{efdens}
\end{figure}

$\rho_p$ is not spherically symmetric for a thin baryonic disc, but the physical DM in our live Plummer halo (LPH) model is spherical (Section \ref{Initial_conditions}). Since the evolution of the bar instability should be closely related to the effective density ($\rho_b + \rho_p$) near the disc mid-plane, it is instructive to explore their properties here. Figure~\ref{efdens} shows $\rho_p$ relative to the Plummer DM density in our LPH model. The main panel illustrates $\rho_p$ at $z = 0.05$~kpc, while the inset shows $\rho_p$ at $z = 3$~kpc. We have excluded $z = 0$ because the phantom density in MOND is singular at $z = 0$. We see that at $z = 0.05$~kpc, the effective phantom DM density in the central regions is much higher in NLG than in MOG, with MOND giving an intermediate result. In all cases, $\rho_p$ is much higher than the physical DM in LPH. At larger $z$, the effective phantom density in MOG becomes higher than in NLG, with MOND giving a much higher result than either theory. All these theories have a larger $\rho_p$ than the LPH case out to at least several disc scale lengths, which for an exponential disc covers the vast majority of its baryonic mass. This means that the NLG, MOG, and especially MOND discs are effectively more massive at $z = 0.05$~kpc, making them more prone to the bar instability. From this perspective, one may expect faster bar growth in NLG and to a lesser extent in MOG compared to the LPH case, with MOND bars expected to grow fastest of all for reasons discussed below. This might explain differences between the early time evolution of the bar strength in these models (Section~\ref{Bar_instability}).

\begin{figure}
	\includegraphics[width = 8.2cm]{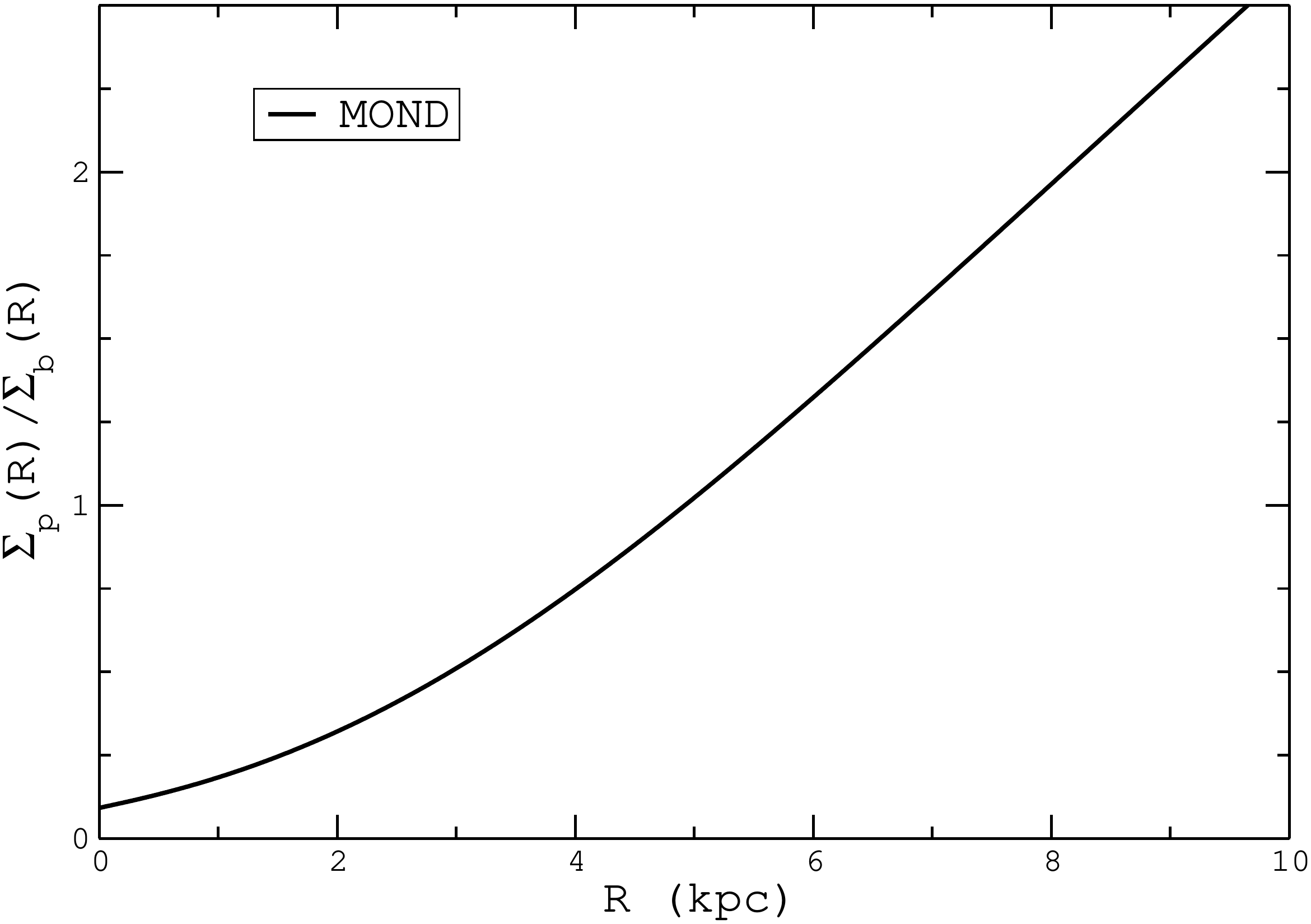}
	\caption{The surface density $\Sigma_p$ of the phantom disc in MOND, shown relative to the baryonic surface density $\Sigma_b$. Notice that $\Sigma_p \ll \Sigma_b$ near the centre, a consequence of the high central gravity relative to $a_{_0}$ (Section \ref{Initial_conditions}). However, the phantom disc is dominant further out. There is no phantom disc in any of the other considered theories (see text).}
	\label{Sigma_p_MOND}
\end{figure}

The DM halo in our LPH model does not have a singular disc component, even though the baryons are distributed in an infinitely thin disc. Similarly, MOG and NLG also have a finite $\rho_p$ at $z = 0$ because these both involve a distance-dependent modification to gravity. Since the gravity at some location $A$ just outside the plane of a thin disc is dominated by baryons very close to $A$, there is no modification to the effective surface density of the disc as perceived by a Newtonian observer. However, the modification to gravity is acceleration-dependent in MOND (Equation \ref{Interpolating_function_MOND}), so in this case there is a phantom DM disc with surface density $\Sigma_p > 0$. This is readily calculated from the local value of $\nu$, which as discussed in Section \ref{DICE} must include an allowance for the vertical gravity just outside the disc. We use Figure \ref{Sigma_p_MOND} to show the ratio between $\Sigma_p$ and the baryonic surface density $\Sigma_b$. The latter dominates at the very centre due to the rather strong gravity there (Section \ref{Initial_conditions}). However, the phantom disc rapidly becomes dominant further out. We expect it to have a destabilizing effect on the disc, as captured approximately by the factor of $\nu$ in the Toomre stability condition (Equation \ref{Toomre_condition_MOND}). Since the rotation curve and thus radial epicyclic frequency are rather similar in all our considered models, the MOND phantom disc could have important implications for the overall stability. Its importance would be much greater still for a galaxy with lower $\Sigma_b$ \citep{Milgrom_1989}.

\section{Numerical methods}
\label{methods}

Two different codes have been used in this paper: one for the MOND simulations, and the other for DM, MOG, and NLG simulations. In the following, we briefly introduce them.

\subsection{MOND simulations with Phantom of RAMSES}
\label{por}

Our MOND simulations use its rather computer-friendly quasi-linear formulation known as QUMOND \citep{QUMOND}:
\begin{eqnarray}
	\nabla \cdot \bm{g} ~=~ \nabla \cdot \left( \nu \bm{g}_{\text{N}} \right) \, .
	\label{QUMOND_governing_equation}
\end{eqnarray}
The function $\nu$ is the MOND boost to the Newtonian gravity $\bm{g}_{\text{N}}$ for a spherically symmetric problem, where $\bm{g} \equiv \nu \bm{g}_{\text{N}}$. In our notation, $p \equiv \left| \bm{p} \right|$ for any vector $\bm{p}$. We choose the `simple' form of the interpolating function \citep{Famaey_Binney_2005} to transition between the Newtonian and deep-MOND regimes:
\begin{eqnarray}
	\nu ~=~ \frac{1}{2} \, + \, \sqrt{\frac{1}{4} + \frac{a_{_0}}{g_{\text{N}}}} \, .
	\label{Interpolating_function_MOND}
\end{eqnarray}
This provides a good fit to a variety of data on galactic and extragalactic dynamics \citep{Gentile_2011, Banik_2018_Centauri}. It is rather similar to the function used by \citet{McGaugh_Lelli_2016} to fit the Spitzer Photometry and Accurate Rotation Curve dataset \citep[SPARC,][]{SPARC}. In the QUMOND approach, $\nu$ depends only on $g_{\text{N}}$ and is thus readily computable once standard techniques are used to obtain $\bm{g}_{\text{N}}$. The approach is quasi-linear because it requires only a linear grid relaxation stage to solve the standard Poisson equation, minimising the computational cost and the modifications required to existing Newtonian codes.

Our simulations implement Equation \ref{QUMOND_governing_equation} using the Phantom of RAMSES $N$-body and hydrodynamics solver \citep[\textsc{por},][]{Lughausen_2015}. \textsc{por} adapts the potential solver of the grid-based code \textsc{ramses}, which uses adaptive mesh refinement to improve efficiency \citep{Teyssier_2002}. \textsc{por} has previously been used to investigate polar ring galaxies \citep{Lughausen_2013}, shell galaxies \citep{Bilek_2015}, and the tidal streams of Sagittarius \citep{Thomas_2017} and Palomar 5 \citep{Thomas_2018}. Recently, it was used in hydrodynamical simulations of collapsing gas clouds to naturally yield exponential disc galaxies \citep{Wittenburg_2020}, and to obtain a quite realistic morphology for M33 after secularly evolving it for 10~Gyr \citep{Banik_2020_M33}. Galaxy interactions have also been simulated in \textsc{por} with hydrodynamics \citep{Renaud_2016} and without it \citep{Bilek_2018}.

For this project, we conduct pure $N$-body simulations by disabling the hydrodynamics mode (the flag $hydro$ is set to false). Self-gravity is enabled by setting $gravity\_type = 0$ and activating $poisson$ so the Poisson solver is utilised. Since the potential is solved on a grid but is generated by a finite number of particles, we enable the flag $pic$ to activate the particle-in-cell solver. The simulation is advanced in a cubic Cartesian grid with side length of 256 kpc. This is much larger than the simulated galaxy, making it accurate to assume both the deep-MOND limit and spherical symmetry of the potential on our computational boundary. To provide adequate spatial resolution, we use $7-12$ levels of refinement, i.e. the highest resolution is $256/2^{12} = 0.0625$ kpc. We set the $m\_refine$ parameter to 20 for all levels, forcing \textsc{por} to further refine a cell if it has $>20$ particles. This is the only refinement condition because $mass\_sph$ is set to 0. Since we do not set the $nsubcycle$ parameter, the default value of 2 is used, causing the timestep to be halved with each level of refinement. Further details of the \textsc{ramses} package can be found in \citet{Teyssier_2002}, along with default values of parameters that we do not set.

We convert the output files into human-readable text files using an algorithm that we have made publicly available.\footnote{\href{https://github.com/GFThomas/MOND/tree/master/extract\_por}{github.com/GFThomas/MOND/tree/master/extract\_por}} \textsc{por} is the most widely used publicly available $N$-body solver for MOND. \citet{Banik_2020_M33} provides further details regarding \textsc{por} and its application to thin disc galaxies, including links to download the algorithms used to prepare the initial conditions for both hydrodynamical and stellar-only thin disc simulations (Section \ref{DICE}). A user guide for \textsc{por} simulations has recently been published, including details on how to initialize isolated or interacting disc galaxy simulations consisting of only stars or also including hydrodynamics \citep{Nagesh_2021}. The guide also describes the extraction of both particle and gas data into human-readable form.

Recently, the \textsc{raymond} algorithm has also been released \citep{Candlish_2015}.\footnote{\href{https://ifa.uv.cl/sites/graeme/codes.html}{ifa.uv.cl/sites/graeme/codes.html}} It can solve both QUMOND and the original aquadratic Lagrangian formulation of MOND \citep[AQUAL,][]{Bekenstein_1984}. It is anticipated that discs which are stable in QUMOND would also be stable in AQUAL as the stability conditions are numerically quite similar \citep{Banik_2018_Toomre}.

\subsection{{\sc galaxy} code for simulations other than MOND}

The \textsc{galaxy} code is a standard and well-developed $N$-body code for galactic simulations in the conventional CDM picture \citep{Sellwood_2014}. It is capable of constructing the equilibrium initial conditions for a variety of different halo, bulge, and disc combinations. It uses standard algorithms developed over almost four decades. The time evolution is obtained by calculating the gravitational field at each time step. This can be done with different methods already implemented in the code; however, the main one is the grid method.

It is not easy to modify this code to include extended gravity effects. All scripts for setting initial conditions and the evolution are based on Newtonian gravity. Fortunately, different grid coordinate systems use different approaches to compute the time evolution. It is straightforward to implement MOG and NLG effects in galactic models which use the cylindrical polar three-dimensional mesh (P3D). This grid is constructed by $N_r$ coaxial cylinders with logarithmically spaced radii, $N_\phi$ equally spaced azimuthal planes, and $N_z$ planes spaced equally in the vertical $z$ direction. The intersection of these planes gives $N_r \times N_\phi \times N_z$ mesh points. The Fourier method is used to determine $\bm{g}$ in the azimuthal and vertical directions. The Plummer softening kernel $P(\xi)$ is used to prevent singularities, where $\xi \equiv \left| \bm{r}-\bm{r}' \right|$. This kernel recovers the Newtonian inverse distance potential at distances much larger than the softening length, when $P(\xi) \propto -1/\xi$. However, $P(\xi)$ tapers smoothly to zero at short distances. Since we are simulating a collisionless system, the exact form of the kernel at short distances does not matter.

As discussed in Sections \ref{NLG} and \ref{MOG}, the effects of NLG and MOG appear at large separations. Therefore, we only need to modify the Plummer softening kernel at large distances by replacing $-1/\xi$ with the potential of a point mass obtained in each theory. In particular, one needs to apply this change to the following subroutines in the \textsc{galaxy} code: \texttt{sftpot}, \texttt{radfor}, \texttt{azifor}, \texttt{vrtfor}.

For the DM galactic models, we use a hybrid mesh $-$ a spherical three-dimensional (S3D) system is used for the DM halo, and a P3D mesh for the baryonic exponential disc \citep[for more details, see][]{Roshan_2018, Roshan_2019}.

\subsection{Initial conditions}
\label{Initial_conditions}

\begin{table*}
	\caption{The properties of our galactic models. Column 1: Acronym used to identify the simulation. Column 2: $r_d$ is the radial scale length of the exponential disc. Column 3: Vertical scale height $z_0$ in terms of $r_d$. Column 4: Initial disc mass $m_d$ in units of $8.57 \times 10^9 M_\odot$. Column 5: Type of DM halo (if any). Column 6: The halo mass scaled by $m_d$. Column 7: Radial scale length of the halo in units of $r_d$. The outer radius of the Plummer halo is $24 \, r_d$ and the corresponding value for the rigid Hernquist model is $15.42 \, r_d$. Column 8: the basic time step in units of $\tau = 5.095$~Myr. Column 9: number of rings, spokes, and planes in the cylindrical polar grid. Column 10: the number of shells in the spherical grid. Column 11: the gravity softening length in units of $r_d$. The MOND models are advanced using the Phantom of RAMSES solver \citep{Lughausen_2015}, which adapts the potential solver of RAMSES \citep{Teyssier_2002} to implement QUMOND (Section \ref{por}) on a Cartesian grid with adaptively refined mesh.} 
	\label{table:table1}
	\begin{tabular}{ccccccccccc}
		\hline
		Run & $r_d$ (kpc) & $z_0/r_d$ & $m_d$ & Halo & $m_h/m_d$ & $r_h/r_d$ & $\delta t$   & Cylindrical polar grid  & Spherical grid & Softening length \\
		\hline
		MOND & 1 & 0.15 & 1 & None & $\ldots$ & $\ldots$ & $\ldots$ & \multicolumn{3}{c}{Cell size ranges from $\left( 0.0625 - 2 \right)$~kpc in powers of 2} \\ 
		LPH & 1 & 0.15 & 1 & Plummer & 8 & 12 & 0.01 & $193 \times 224 \times 45$ & $1001$ & $0.16 \, r_d$ \\
		RHH & 1 & 0.15 & 1 & Hernquist & 13.53 & 11.87 & 0.01 & $193 \times 224\times 45$ & $1001$ & $0.16 \, r_d$ \\
		NLG & 1 & 0.15 & 1 & None & $\ldots$ & $\ldots$ & 0.01 & $193 \times 224 \times 45$ & $1001$ & $0.16 \,r_d$ \\
		MOG & 1 & 0.15 & 1 & None & $\ldots$ & $\ldots$ & 0.01 & $193 \times 224 \times 45$ & $1001$ & $0.16 \,r_d$ \\
		\hline
	\end{tabular}
	\label{Galaxy_parameters}
\end{table*}

Every model we run contains an exponential disc with mass $m_d$ and scale length $r_d$. The surface density
\begin{eqnarray}
	\Sigma(R) ~=~ \frac{m_d}{2 \mathrm{\pi} r_d^2}\, \exp \left( -\frac{R}{r_d} \right) \, .
	\label{expnew}
\end{eqnarray}
The disc has a $\sech^2 \left( z/ \left( 2 z_0 \right) \right)$ density profile in the vertical direction $z$ with scale height $z_0$. Table \ref{table:table1} summarizes this and other properties of the models like their mass, scale length, and grid properties. For a single-component exponential disc with known aspect ratio, the only dimensionless parameter in MOND is the surface density. We use a model where the vertical Newtonian gravity at the disc centre is $\bm{g}_{_{N,z}} = 2 \mathrm{\pi} G \Sigma_0 = 10 \, a_{_0}$, where $\Sigma_0$ is the central surface density of the disc. {The MW parameters in table~1 of \citet{Banik_2017} give $\bm{g}_{_{N,z}} = 15 \, a_{_0}$ at the disc centre. For M31, the flatline rotation curve level of $v_f = 225$~km/s \citep{Carignan_2006} implies a MOND mass of ${v_f}^4/\left( G a_{_0} \right) = 1.6 \times 10^{11} M_\odot$ (Equation \ref{BTFR}), which for a disc scale length of 5.3~kpc \citep{Courteau_2011} implies the central $\bm{g}_{_{N,z}} = 7 \, a_{_0}$. Therefore, our adopted value of $10 \, a_{_0}$ corresponds to a galaxy whose surface density is intermediate between the major Local Group galaxies.} However, these galaxies are larger than in our model, increasing their dynamical time $-$ they would evolve slower than our simulated Milgromian disc. It is possible to scale the results of our MOND model to other disc parameters provided the central surface density is fixed (Section \ref{Scaling_results}).

For the DM case, we have two models $-$ one with a live Plummer halo (LPH), and the other with a rigid Hernquist halo (RHH). The latter is unphysical in the $\Lambda$CDM context, but interestingly is rather similar to the expected behaviour for superfluid DM since dynamical friction would be negligible in this case \citep{Berezhiani_2016, Berezhiani_2019}. The Plummer halo density profile is:
\begin{eqnarray}
	\rho_{LPH} \left( r \right) ~=~ \frac{3 m_h}{4 \mathrm{\pi} {r_h}^3}\left[ 1 + \left(\frac{r}{r_h}\right)^2 \right]^{-5/2} \, .
\end{eqnarray}
The analogous result for the Hernquist profile is:
\begin{eqnarray}
	\rho_{RHH} \left( r \right) ~=~ \frac{m_h}{2 \mathrm{\pi} r_h^3}\left(\frac{r_h}{r}\right)\left[ 1 + \left(\frac{r}{r_h}\right) \right]^{-3} \, .
\end{eqnarray}

To perform a meaningful comparison between galactic simulations in different theories, it is necessary to start with the same initial conditions. Specifically, the distribution and velocity profile of the baryonic matter should be the same in all models, which therefore need to have the same rotation curve. This is the main reason for using the Plummer and Hernquist models. Our experience shows that fitting extended gravity with Plummer and Hernquist haloes is much simpler than other known haloes. The Plummer halo has also been used in previous MOND simulations \citep[e.g.][]{Tiret_2007}.

\begin{figure} 
	\centerline{\includegraphics[width = 8cm]{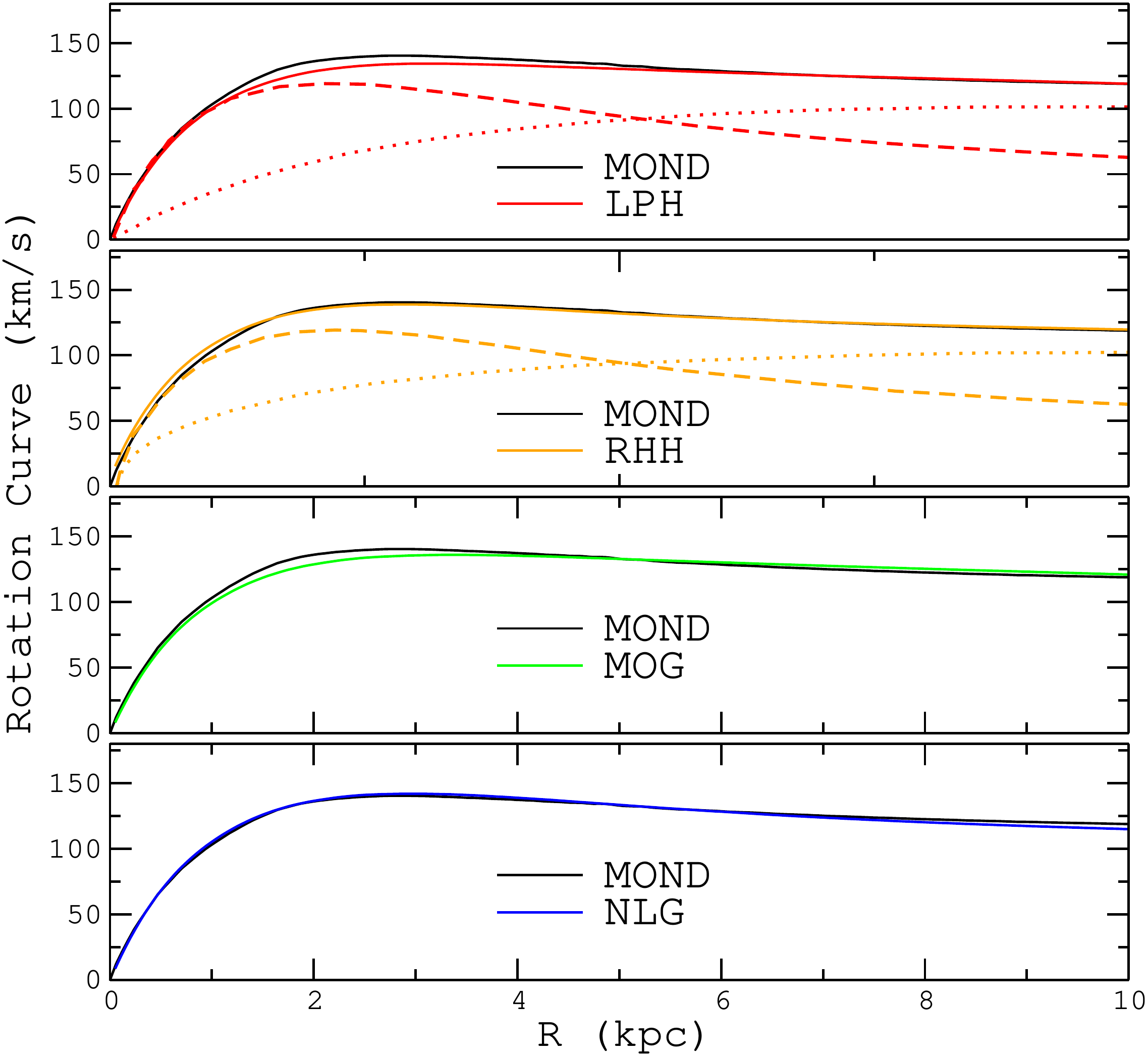}}
	\caption{Initial rotational velocities for (top to bottom): LPH (red), RHH (orange), MOG (green), and NLG (blue). All panels also show the Milgromian model (black). In the DM models, the dashed and dotted curves indicate the contributions of the disc and halo, respectively. The scale length is 1~kpc (Table \ref{Galaxy_parameters}).}
	\label{vhsb}
\end{figure}

The rotation curves of our models are shown in Figure~\ref{vhsb}. We take the MOND model as our main model, and try to fit the NLG and MOG models by choosing appropriate free parameters {$\alpha$ and $\mu_0$}. In the DM case, we change the halo properties to find a proper fit. As is clear from Figure~\ref{vhsb}, the rigid halo gives a suitable fit for the MOND curve. However, this is not the case for the live halo. One should note that in LPH, we adiabatically compress the Plummer halo to find a suitable equilibrium model. To do so, we use the original procedure already implemented in the \textsc{galaxy} code \citep{Sellwood_2005}. Consequently, the halo's properties are in principle different from the rigid case. At intermediate radii, the live halo model is thus unable to exactly reproduce the MOND model. Of course, one cannot expect exactly the same initial conditions for different theories as it is not mathematically possible. Though the halo is important at large radii, the baryonic disc provides the dominant contribution to the total rotation curve in the central regions, indicating that we are dealing with maximal discs in a standard context. {This can be quantified with the ratio between the total rotational velocity $v_c$ measured at $2.2\, r_d$ and the Newtonian circular velocity $v_{_N}$ due to the disc alone. For our models, this ratio is $v_{_N}/v_c \approx 0.91$.}

\begin{figure} 
	\centerline{\includegraphics[width = 8cm]{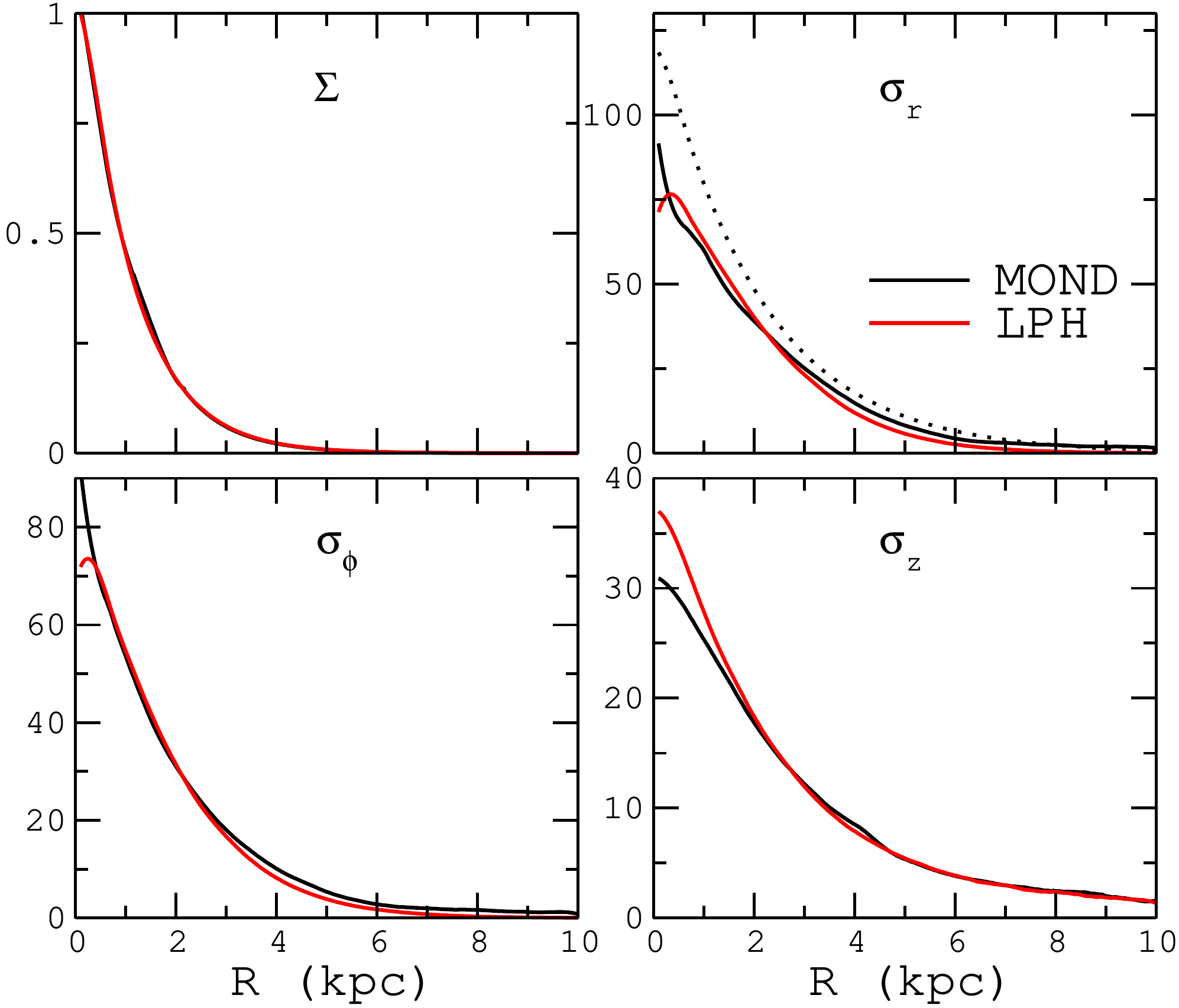}}
	\caption{\emph{Top left}: The initial surface density in units of the central surface density of the Milgromian model, shown for LPH and MOND. \emph{Other panels}: the initial velocity dispersion profiles of these models, expressed in km/s. The dotted curve in the panel for $\sigma_r$ shows an observational estimate (Equation \ref{sigma_r_obs}).}
	\label{ini_disp}
\end{figure}

It is also necessary to ensure that internal properties like the velocity dispersions are similar for different models. We use Figure~\ref{ini_disp} to show the initial surface density $\Sigma$ and the radial, azimuthal, and vertical velocity dispersions ($\sigma_r$, $\sigma_{\phi}$, and $\sigma_z$, respectively). The internal properties are very similar for all models constructed by the \textsc{galaxy} code. Thus, we plot only LPH and MOND. As expected, the surface density of both models is the same exponential disc law. It is clear that the velocity dispersions of both models are appropriately consistent except at very small and large radii, where the Milgromian model predicts modestly higher $\sigma_r$ and $\sigma_{\phi}$. This is because the gravity in the disc mid-plane becomes quite weak in these regions, enhancing the $\nu$ factor in MOND (Equation \ref{Interpolating_function_MOND}). However, the gravity at the disc surface is relatively strong at the disc centre, suppressing $\nu$ and therewith $\sigma_z$. At intermediate radii, the lower gravity implies a phantom dark disc (Section \ref{Phantom_dark_matter}), which very slightly increases $\sigma_z$. At even larger radii, the very low surface density of an exponential disc means the vertical restoring force is mostly a geometric one that can be understood by considering the potential as spherically symmetric. Since the rotation curve is similar in all our models by construction, we expect the initial $\sigma_z$ to be very similar at large radii in all cases.

The above discussion shows that our models can be compared with each other. For a comparison with real galaxies, it is also important for the velocity dispersion profile to broadly agree with observations. We therefore compare our assumed $\sigma_r \left( R \right)$ with the observationally inferred profile $\sigma_r^* \left( R \right)$ as deduced from observations \citep{Leroy_2008}. According to their equation B3,
\begin{eqnarray}
	\sigma_r^* \left( R \right) ~\approx~ 0.62 \sqrt{\frac{m_d \, G}{r_d}} \, \exp \left( -R/r_d \right) \, .
	\label{sigma_r_obs}
\end{eqnarray}
This relation is shown as a thin dotted line in the upper right panel of Figure \ref{ini_disp}. Considering that it overestimates $\sigma_r$ \citep{Mogotsi_2019}, it is clear that our adopted $\sigma_r \left( R \right)$ is reasonably consistent with observations.

\subsubsection{Initializing a Milgromian disc}
\label{DICE}

We set up a Milgromian disc using a code we make publicly available.\footnote{\href{https://github.com/GFThomas/MOND/tree/master/init\_conditions/disc}{github.com/GFThomas/MOND/tree/master/init\_conditions/disc}} The method was previously used to simulate M33 \citep{Banik_2020_M33}, but we briefly describe it here. We use an adapted version of the Newtonian code Disk Initial Conditions Environment \citep[\textsc{dice},][]{Perret_2014}. \textsc{dice} offers the advantage that the Jeans equations are not solved using the potential, which is difficult to define for an isolated system in MOND. \textsc{dice} uses only the Newtonian gravity $\bm{g}_{\text{N}}$, which it calculates using the principle of superposition accelerated by a fast Fourier transform. We approximately MONDify this using the algebraic MOND approximation, which states that the true gravity
\begin{eqnarray}
	\bm{g} ~\approx~ \nu \bm{g}_{\text{N}} \, .
	\label{ALM}
\end{eqnarray}
This approximation is exactly correct in spherical symmetry and works rather well in axisymmetric problems \citep{Angus_2012, Jones_2018}. It is expected to work particularly well just outside the disc \citep{Banik_2018_Toomre}. However, it becomes inaccurate within the disc due to the steep vertical gradient in $\nu$ caused by that in $\bm{g}_{_{N,z}}$. Naively applying Equation \ref{ALM} would imply a rapid change in $\bm{g}_{_r}$ with $z$, something that is physically unrealistic as it would cause $\nabla \times \bm{g} \neq 0$, allowing energy to be gained around a closed loop.\footnote{This problem does not arise if rigorously implementing QUMOND by applying Equation \ref{QUMOND_governing_equation} rather than the approximate Equation \ref{ALM}.} To avoid this, we fix the value of $\bm{g}_{_{N,z}}$ entering the calculation of $\nu$ (Equation \ref{Interpolating_function_MOND}) to $2 \tanh \left( 2 \right) \mathrm{\pi} G \Sigma$ if $\left| z \right|$ is small enough that the fraction of the local column density $\Sigma \left( r \right)$ at even smaller $\left| z \right|$ falls below $\tanh \left( 2 \right)$. This is based on the assumption that $\bm{g}_{_{N,z}} = 2 \mathrm{\pi} G \Sigma$ at the disc `surface', which is valid for a thin disc. Once $\nu$ is calculated in this revised way, we set $\bm{g} = \nu \bm{g}_{\text{N}}$.

\subsubsection{Local stability}

Although we are mainly interested in the global stability and evolution of galactic discs, it is useful to mention some points about their local stability in extended gravity theories. To suppress local fragmentation of our galaxy models, we set the Toomre parameter $Q > 1$ \citep{Toomre_1964}, with $Q$ defined as:
\begin{eqnarray}
	Q ~\equiv~ \frac{\sigma_r \kappa}{3.36 \, G\Sigma} \, .
	\label{Toomre_condition}
\end{eqnarray}
The rotation curve must be known to find $\kappa$, the radial epicyclic frequency.

The local stability of discs in MOG has been investigated in \citet{Roshan_2015}. They derived the dispersion relation for local perturbations and presented a generalized version of the Toomre criterion. Fortunately, the correction terms induced by MOG are negligible in spiral galaxies. This is reasonable because MOG effects appear at long distances $-$ we do not expect them in small scale local perturbations. Consequently, the standard Toomre criterion ($Q>1$) works very well for MOG galactic models. Due to the similar weak field limits of MOG and NLG, one may expect the same criterion for NLG models.

In the case of MOND, the appropriate generalization was derived in \citet{Banik_2018_Toomre}. Briefly, it states that Equation \ref{Toomre_condition} should be modified to:
\begin{eqnarray}
	\label{Toomre_condition_MOND}
	Q &\equiv& \frac{\sigma_r \kappa}{3.36 \, G \nu \left( 1 + \frac{K_0}{2}\right) \Sigma} \, , \text{ where} \\
	\nu &=& \nu \left( \sqrt{{\bm{g}_{N,r}}^2 + {\bm{g}_{N,z}}^2 } \right) \, , \text{ and} \\
	K_0 &\equiv& \frac{\partial \ln \nu}{\partial \ln g_N} \ .
\end{eqnarray}
Notice that the MOND boost factor $\nu$ depends on the total Newtonian gravity $g_N$ just outside the disc plane, i.e. both the radial Newtonian gravity $\bm{g}_{N,r}$ and the vertical component $\bm{g}_{N,z} \equiv 2 \mathrm{\pi} G \Sigma$ must be added in quadrature to yield $g_N$. In the deep-MOND limit, $K_0 \equiv -\frac{1}{2}$, while in the Newtonian limit $K_0 \equiv 0$. Apart from an order unity correction due to $K_0$, disc stability in QUMOND works similarly to Newtonian gravity with the local value of $G$ enhanced by the factor $\nu$. Note that $\nu$ can be arbitrarily large in MOND, especially in LSBs.

\begin{figure} 
	\centerline{\includegraphics[width = 8cm]{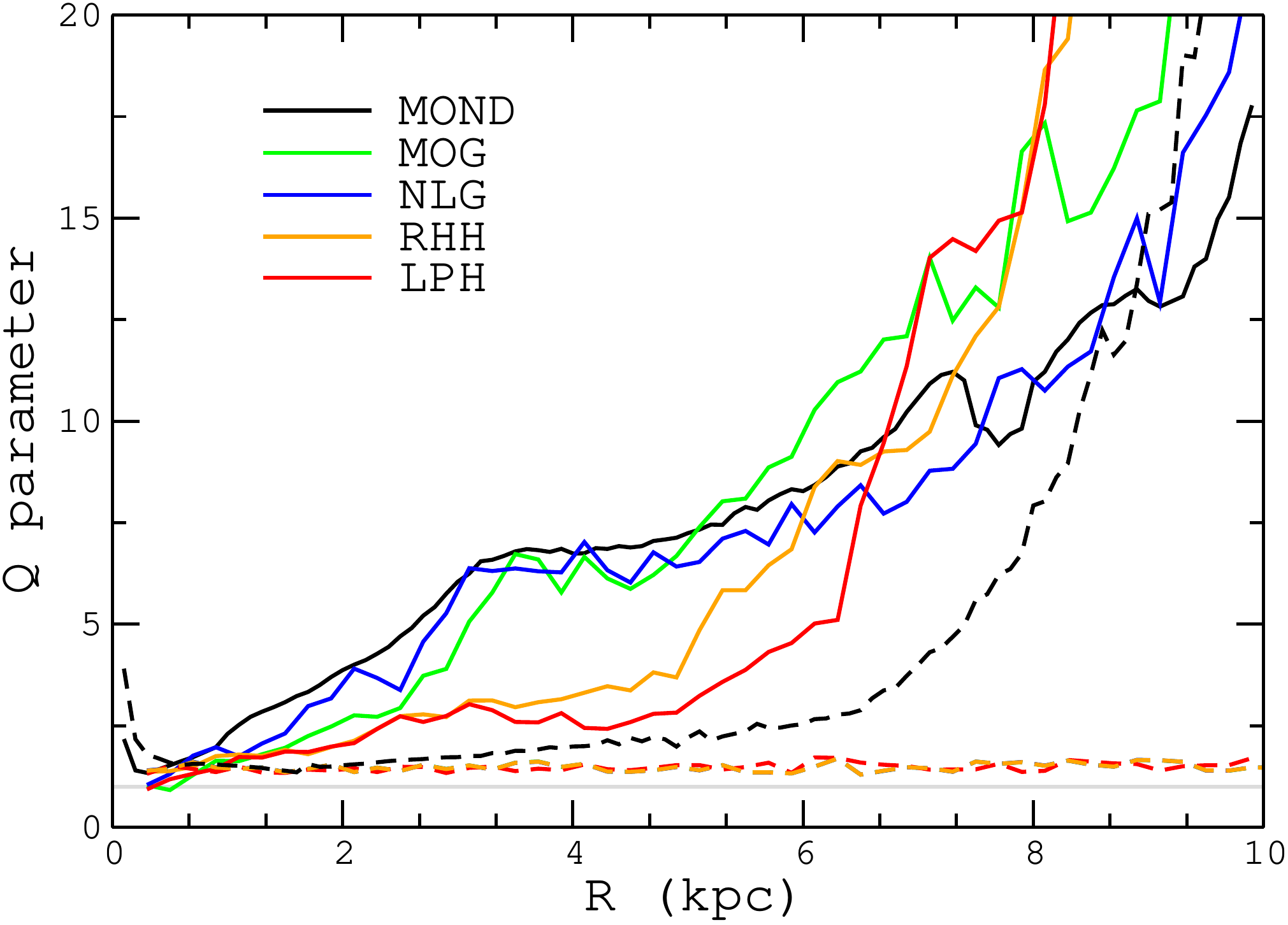}}
	\caption{Toomre $Q$ parameter (Equation \ref{Toomre_condition} or \ref{Toomre_condition_MOND}) at $t = 0$ (dashed curves) and $t = 4$~Gyr (solid curves). The solid grey line shows $Q = 1$.}
	\label{toom1}
\end{figure}

The local stability criterion is satisfied in all our models, as evident from the $Q$ parameter when $t = 0$ and $t = 4$~Gyr (Figure~\ref{toom1}). Since MOND predicts higher $\sigma_r$ when $t = 0$, we see that $Q \gg 1$ in the outer regions. At the end of the simulation, the solid curves show that the bar's activity has substantially increased $Q$, so all the discs remain locally stable throughout our simulations. In what follows, we therefore deal with their global stability.

\section{Results}
\label{results} 

We now discuss the time evolution of our models. As already mentioned, the bar instability and the bar pattern speed are two of the most important quantities directly related to the missing gravity problem. The buckling instability is another important feature in spiral galaxies. We compare these phenomena in our different models.

\subsection{Face-on and edge-on views}

\begin{figure*}
	\includegraphics[width = 13cm]{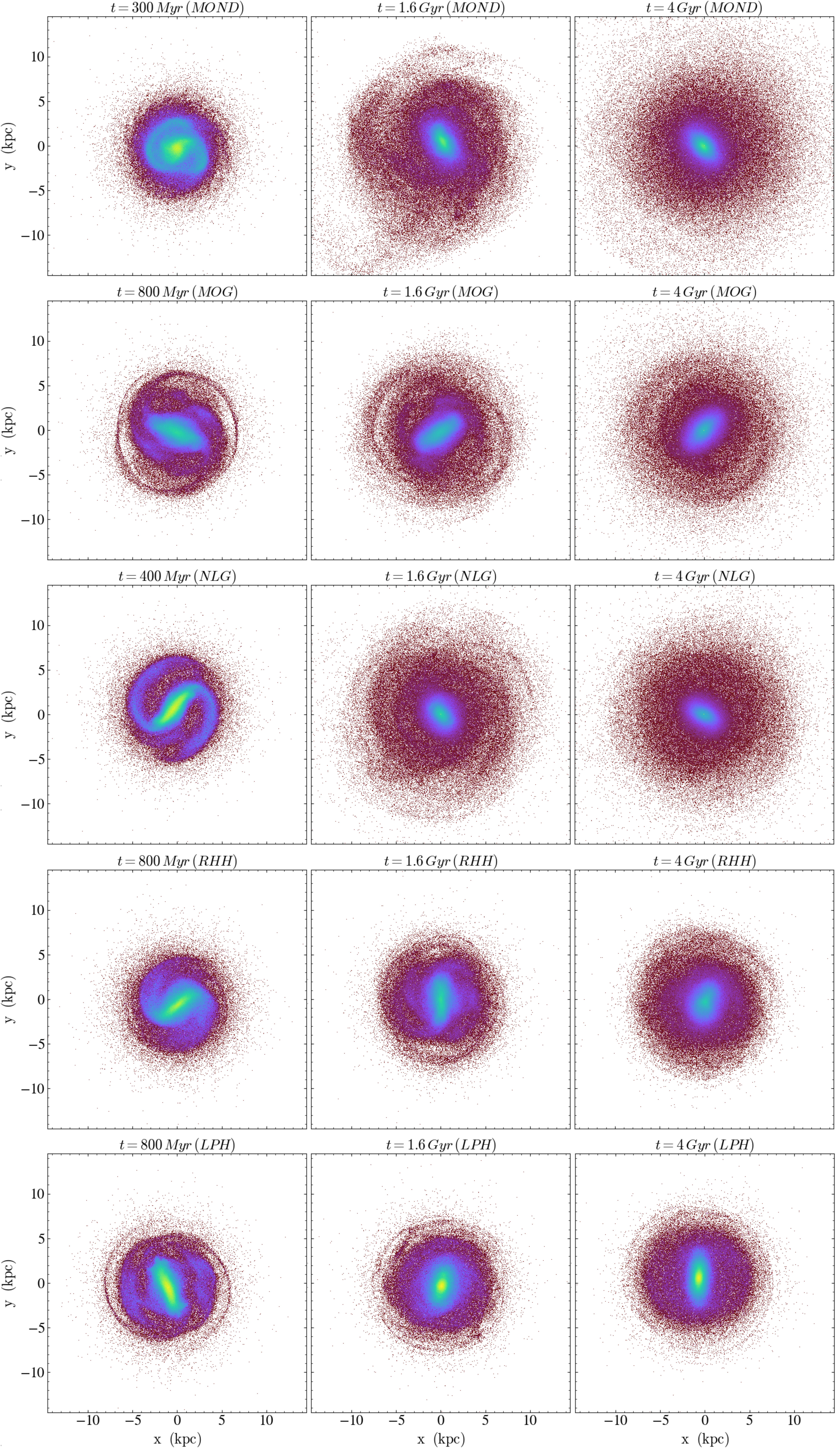}
	\caption{Evolution of the disc with $10^6$ particles projected on the $xy$ plane for each model (from top to bottom: MOND, MOG, NLG, RHH, and LPH). Radial expansion is apparent in all the extended gravity models, but not in the DM models. These plots are constructed using \textsc{yt} \citep{Turk_2011}.}
	\label{pos_others}
\end{figure*}

\begin{figure*}
	\includegraphics[width = 17.5cm]{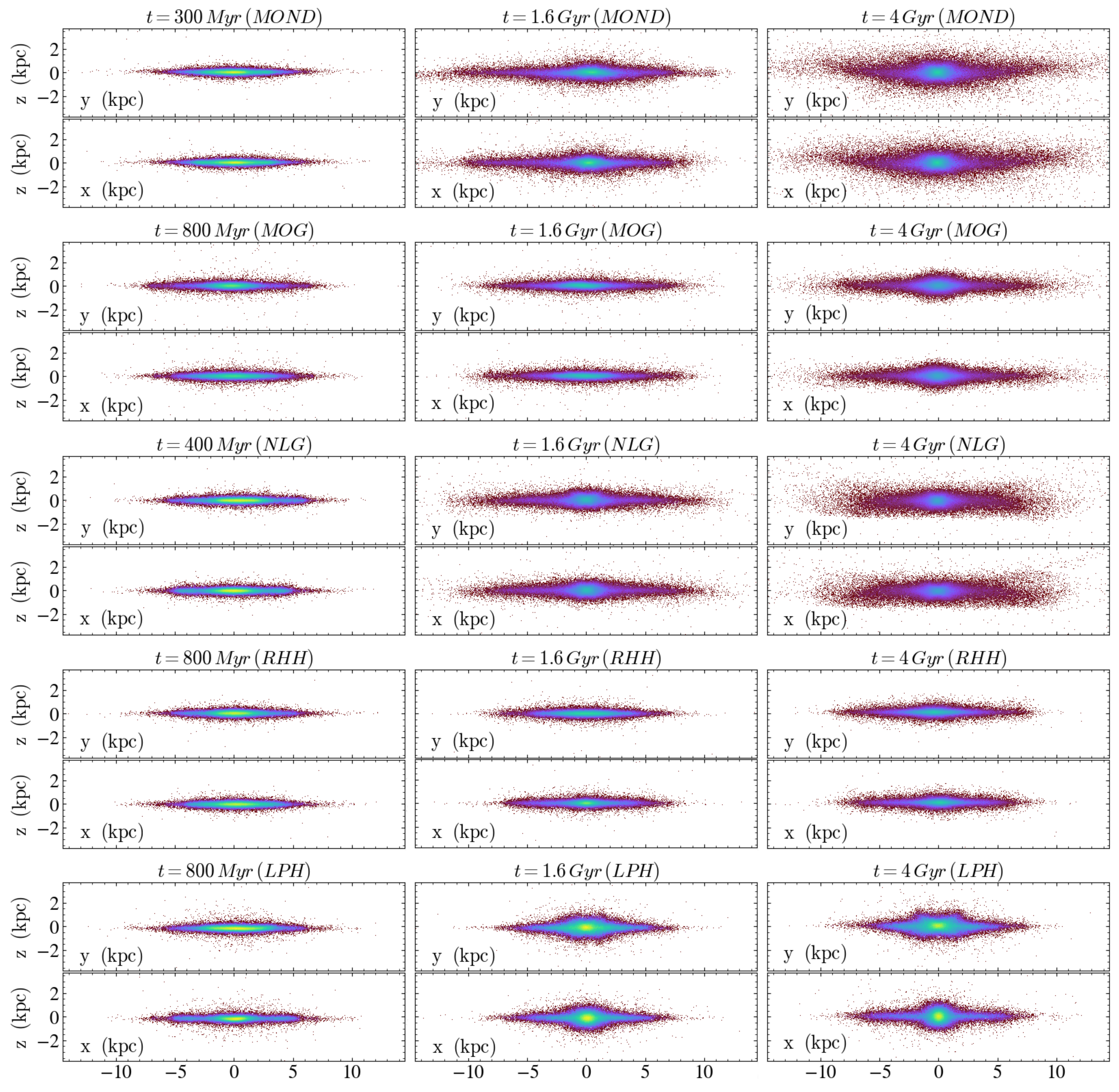}
	\caption{Edge on ($xz$ and $yz$) views of the $N = 10^6$ models at the same times used in Figure~\ref{pos_others}. Two successive rows are used for each model, showing the $xz$ and then the $yz$ view. The first six rows belong to our extended gravity models (in order: MOND, MOG, and NLG). The bottom four rows belong to the DM models (in order: RHH and LPH).}
	\label{edge-on}
\end{figure*}

\begin{figure*}
	\centerline{\includegraphics[width = 15cm]{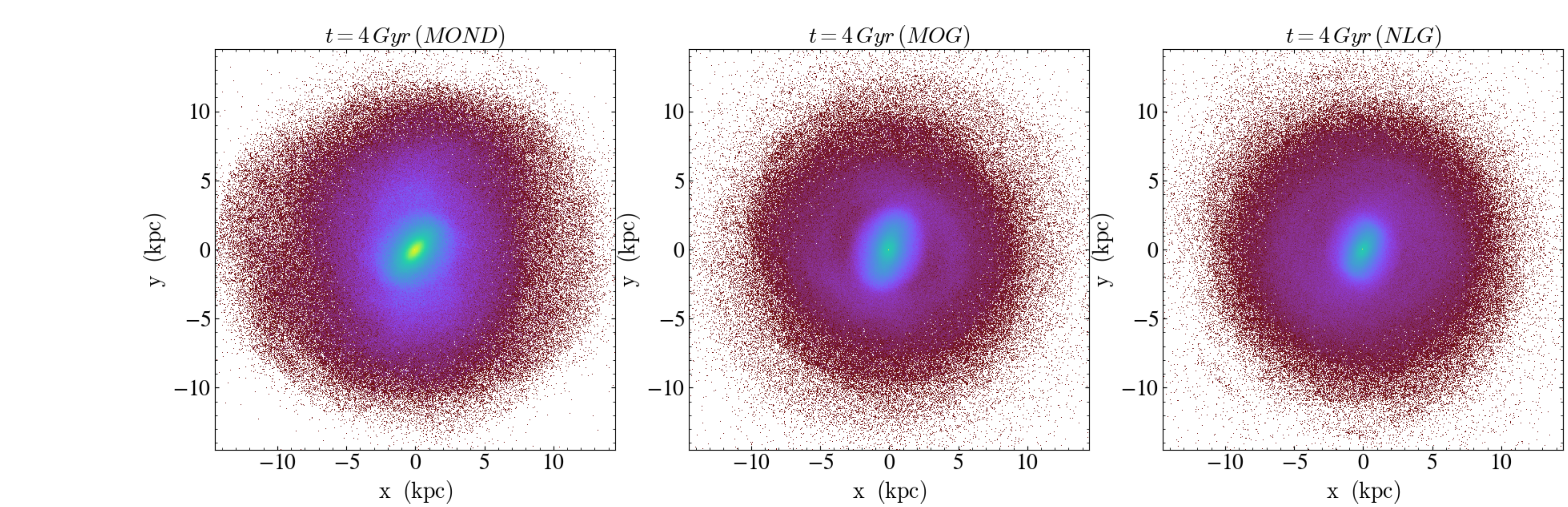}}
	\vspace{0.2cm}
	\centerline{\includegraphics[width = 10cm]{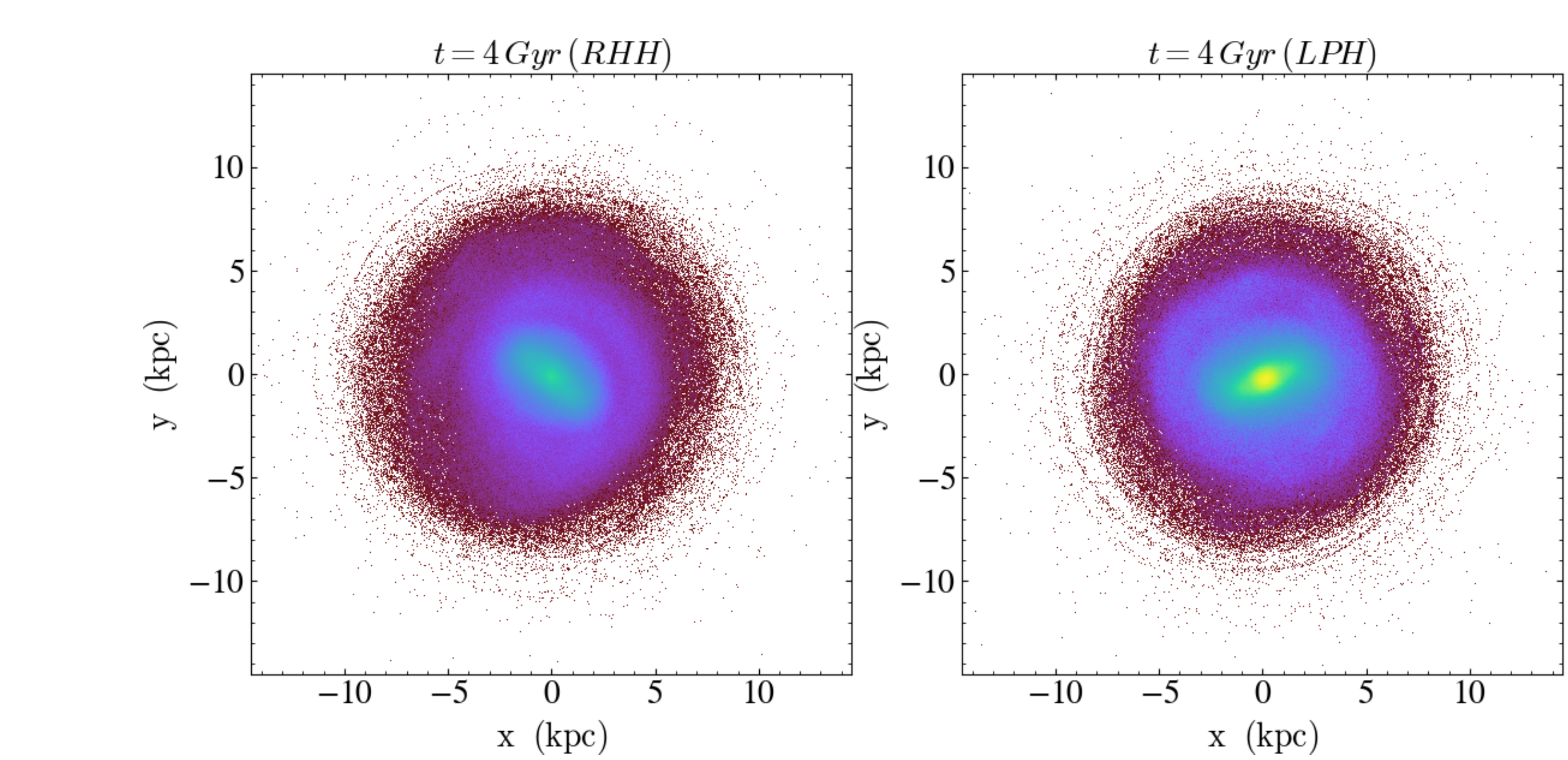}}
	\caption{Face-on $\left( xy \right)$ projections of the models with $N = 5 \times 10^6$ particles at the end of the simulations ($t = 4$~Gyr). Similarly to the low-resolution simulations, the extended gravity discs end up larger than the DM models (Figure~\ref{pos_others}).}
	\label{face}
\end{figure*}

\begin{figure*}
	\centerline{\includegraphics[width = 17.5cm]{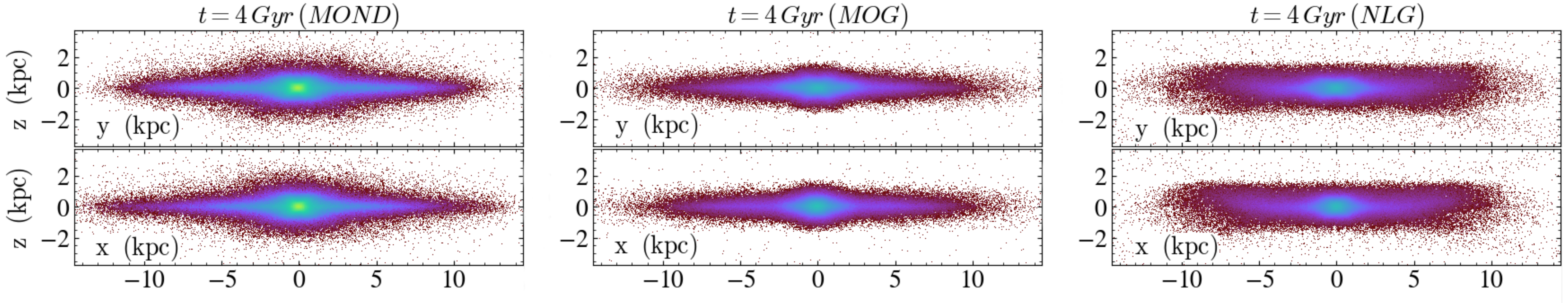}}\vspace{0.2cm}
	\centerline{\includegraphics[width = 12cm]{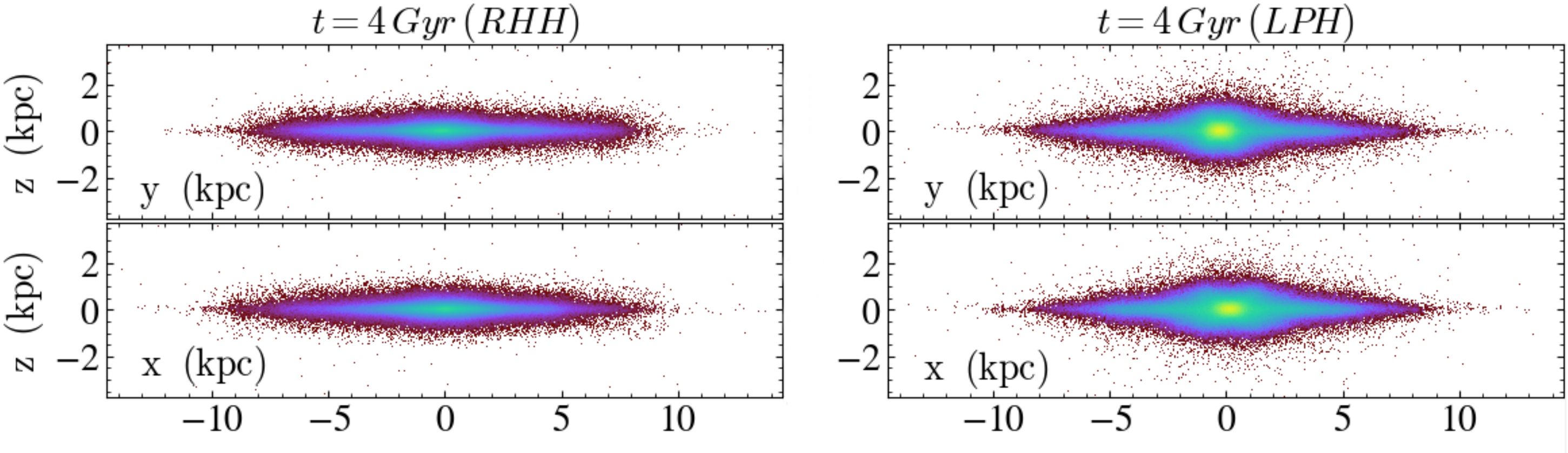}}
	\caption{The final projected positions of particles in our models with $N = 5 \times 10^6$. All the extended gravity models look thicker than the DM models, especially at large radii. The peanut shape evident in the LPH model is weaker in the extended gravity discs.}
	\label{edge}
\end{figure*}

The face-on projected positions of particles at different times are shown as snapshots in Figure~\ref{pos_others} for the models with $N = 10^6$ particles. The first snapshot in each model shows the time at which the bar magnitude reaches its maximum (Section \ref{Bar_instability}) and we see a two-fold symmetric spiral arm. It is interesting that in all models, the spiral arms are not permanent patterns \citep{Lin_1964} $-$ they rapidly fade to a stable pressure-dominated bar. The second snapshot for each model is roughly when the buckling instability occurs, causing the disc thickness to significantly grow. The last snapshot illustrates the end of each simulation.

The corresponding edge-on views are shown in Figure~\ref{edge-on}, with two rows used for each model to show $xz$ and $yz$ projections. The first six rows belong to extended gravity models, while the last four rows illustrate the DM models. Clearly, there are meaningful differences between the radial and vertical properties of the discs in extended gravity and DM models. We discuss these differences in the subsequent sections.

The results remain similar if $N = 5\times 10^6$, so we show only the final face-on and edge-on projections in this case (Figures~\ref{face} and \ref{edge}, respectively). Differences due to resolution are discussed further in Section \ref{Numerical_convergence}.

\subsection{Radial expansion}
\label{Radial_expansion}

An interesting feature of all extended gravity models is that the final discs are more radially extended compared to the DM case. It seems that a halo (if present) suppresses global radial expansion of the disc. This is probably linked to the enhanced role of disc self-gravity in extended gravity theories, which promotes the redistribution of angular momentum within the disc.

\begin{figure} 
	\centerline{\includegraphics[width = 8cm]{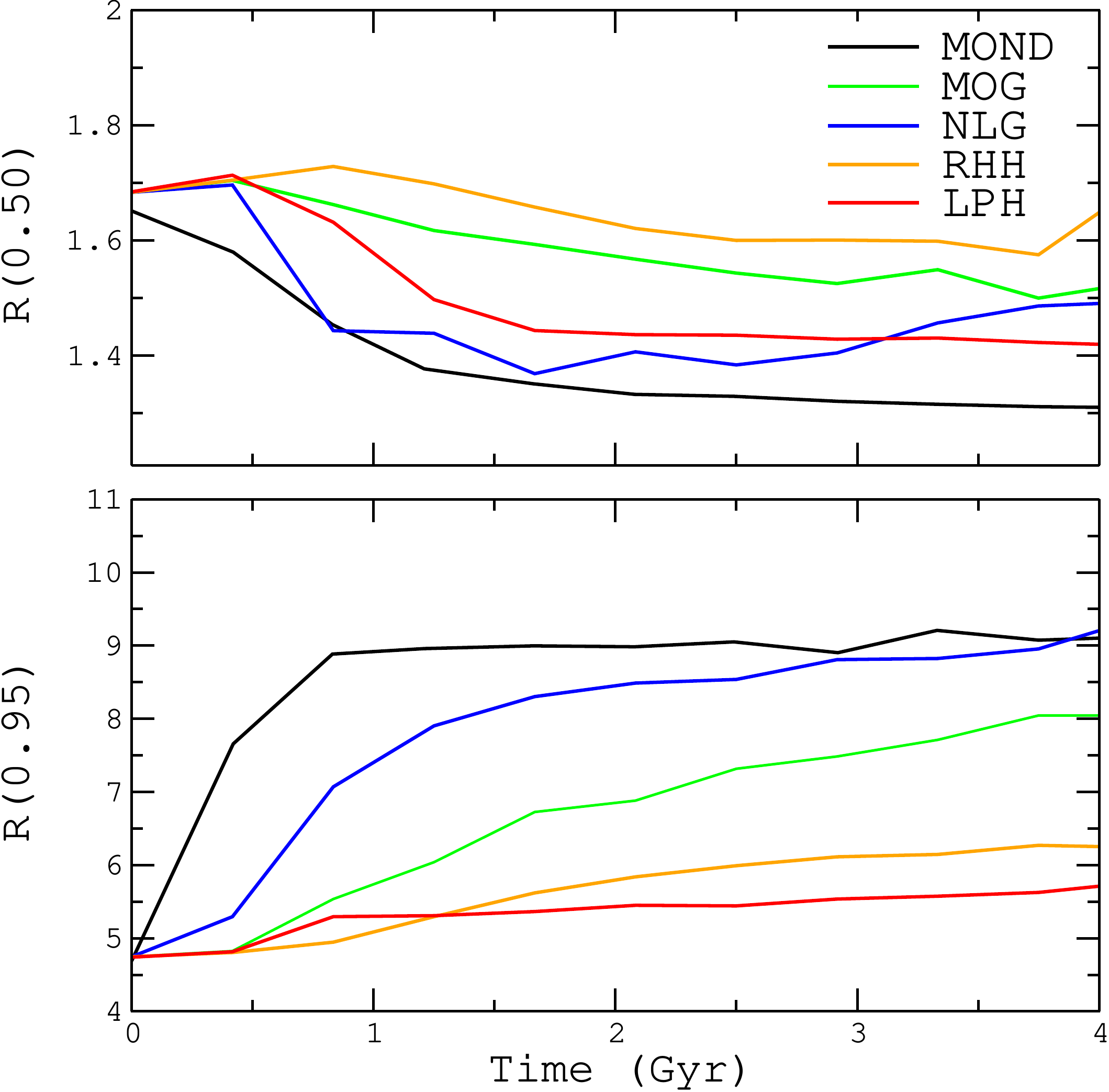}}
	\caption{Spherical Lagrange radii $R \left( X \right)$ in kpc as a function of time. The top and bottom panels belong to $X = 0.5$ and $X = 0.95$, respectively.}
	\label{lagR2}
\end{figure}

To quantify the radial expansion, we use Figure~\ref{lagR2} to show the Lagrange radius $R \left( X \right)$ at different times, where $X$ is the fraction of the baryonic mass inside spherical radius $R$, e.g. $R \left( 0.5 \right)$ denotes the half mass radius ($r_{\text{half}}$) of the disc. It is helpful to show the time evolution of $R \left( X \right)$ for two particular values of $X$, namely $X=0.5$ and $X=0.95$. It is clear that there are rapid variations in $R \left( X \right)$ near the beginning. This is expected as the discs are globally unstable in the early stages, most likely due to the initial conditions not being exactly in equilibrium. We see that in all models, $R \left(0.5 \right)$ decreases in the time interval when the bar instability happens (Section \ref{Bar_instability}), indicating contraction of the central region. There is no significant difference between extended gravity and DM models regarding the final magnitude of $r_{\text{half}}$. However, $R \left(0.95\right)$ grows with time, as expected from angular momentum conservation. This growth is tangibly higher in the extended gravity models. It is evident that $R \left( 0.95 \right)$ at the end of the MOND and NLG models is $>40$\% larger than for the LPH model. These changes arise mostly in the first half of the simulations $-$ in the second half, there is no substantial change in $R \left( X \right)$.

\begin{figure} 
	\includegraphics[width = 8cm]{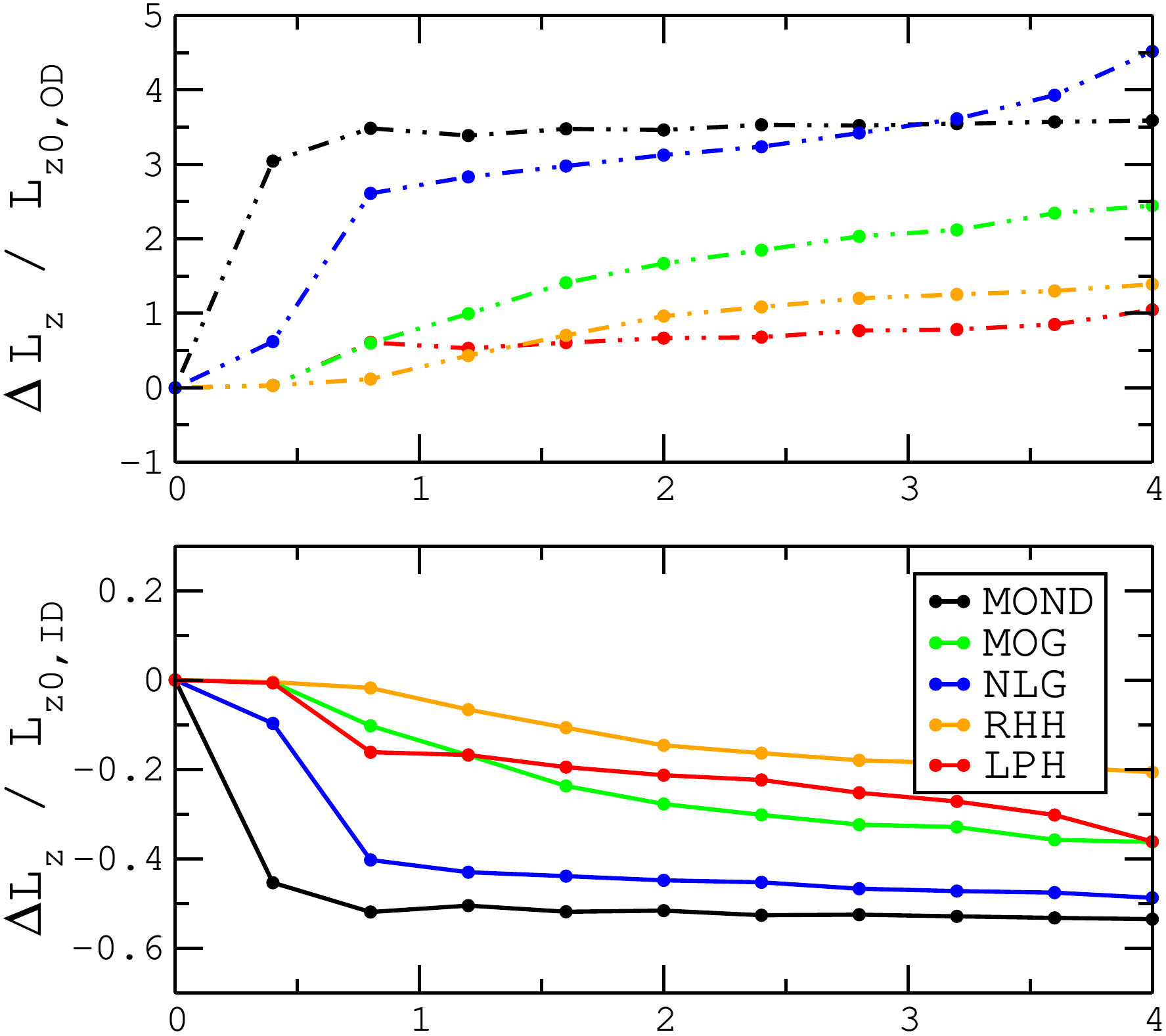}
	\includegraphics[width = 8cm]{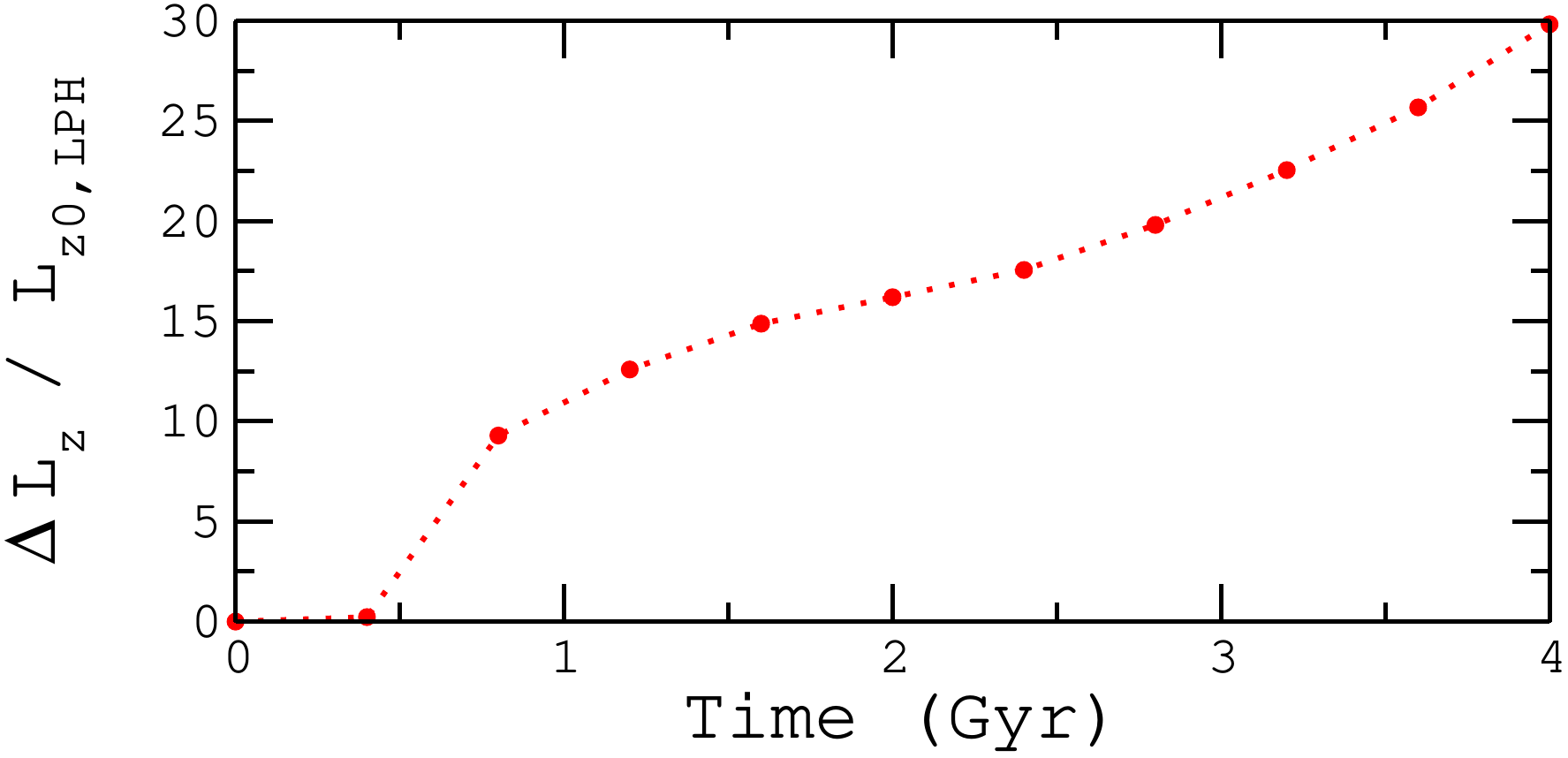}
	\caption{{The top and middle panels indicate the angular momentum change $\Delta L_z/L_{z0}$ in the outer (radii larger than 5 kpc, top panel) and inner regions of the disc (middle panel). $L_z$ is the angular momentum along the $z$ axis, and $L_{z0}$ is the initial angular momentum of the specified region (`ID', `OD', and `LPH' stand for the inner disc, outer disc, and live Plummer halo, respectively). The bottom panel shows the angular momentum change of the DM halo in the LPH model.}}
	\label{angular}
\end{figure}

To understand the radial expansion in extended gravity models, we explore the angular momentum transfer in different parts of the disc. Let us define the inner disc as that part of the disc inside radius $r_{\max}$. In all our models, we measure the angular momentum exchange between the inner and outer disc, with the boundary at $r_{\max} = 5$~kpc. It is clear from the top and middle panels of Figure~\ref{angular} that angular momentum exchange between the inner and outer discs is much more effective in extended gravity models compared to the DM case. In the LPH model, the halo absorbs angular momentum from the disc, so its angular momentum increases significantly. This disc to halo transfer does not exist in extended gravity models, where we instead see more effective angular momentum transfer throughout the disc. This is almost certainly related to the greater amount of phantom DM close to the disc compared with the amount of physical DM in the LPH model (Section \ref{Phantom_dark_matter}).

Although the greater amount of radial expansion in all our extended gravity models appears to be a clear signature of a departure from Newtonian dynamics, real galaxies in a cosmological context would expand due to additional processes not considered here, including accretion of gas from their environment. Such processes need to be considered before a comparison is possible with the observed size evolution of galaxies \citep[e.g.][]{Dokkum_2010, Mowla_2019, Yang_2021}. Once both secular evolution and gas accretion are considered, extended gravity theories in the cosmological context may predict even more radial expansion for galaxy discs than calculated here. Additionally, if the memory effect in NLG fades over cosmic time, the phantom DM fraction monotonically decreases \citep{Mashhoon_2017}. As a consequence of the weaker gravitational interaction, the size of galaxies would grow as $a^{1.4}$, where $a$ is the cosmic scale factor \citep[section 10.6 of][]{Mashhoon_2017}. This is rather similar to the observed size evolution of $a^{1.05 \pm 0.37}$ \citep{Yang_2021}. In Section \ref{Relation_to_cosmology}, we discuss how our results might differ once the cosmological context is considered.

\subsection{Bar instability}
\label{Bar_instability}

\subsubsection{Fourier amplitude}
\label{Fourier_amplitude}

As a suitable representative for the existence and intensity of the bar instability, we measure the bar amplitude $A_2(t)$. This is the third coefficient in the Fourier decomposition of the surface density in terms of azimuthal angle $\phi$. Therefore, $A_2/A_0 > 0$ indicates the existence of two-fold symmetric features (e.g. bar and spiral density waves) propagating in the system.

\begin{figure} 
	\includegraphics[width = 8.2cm]{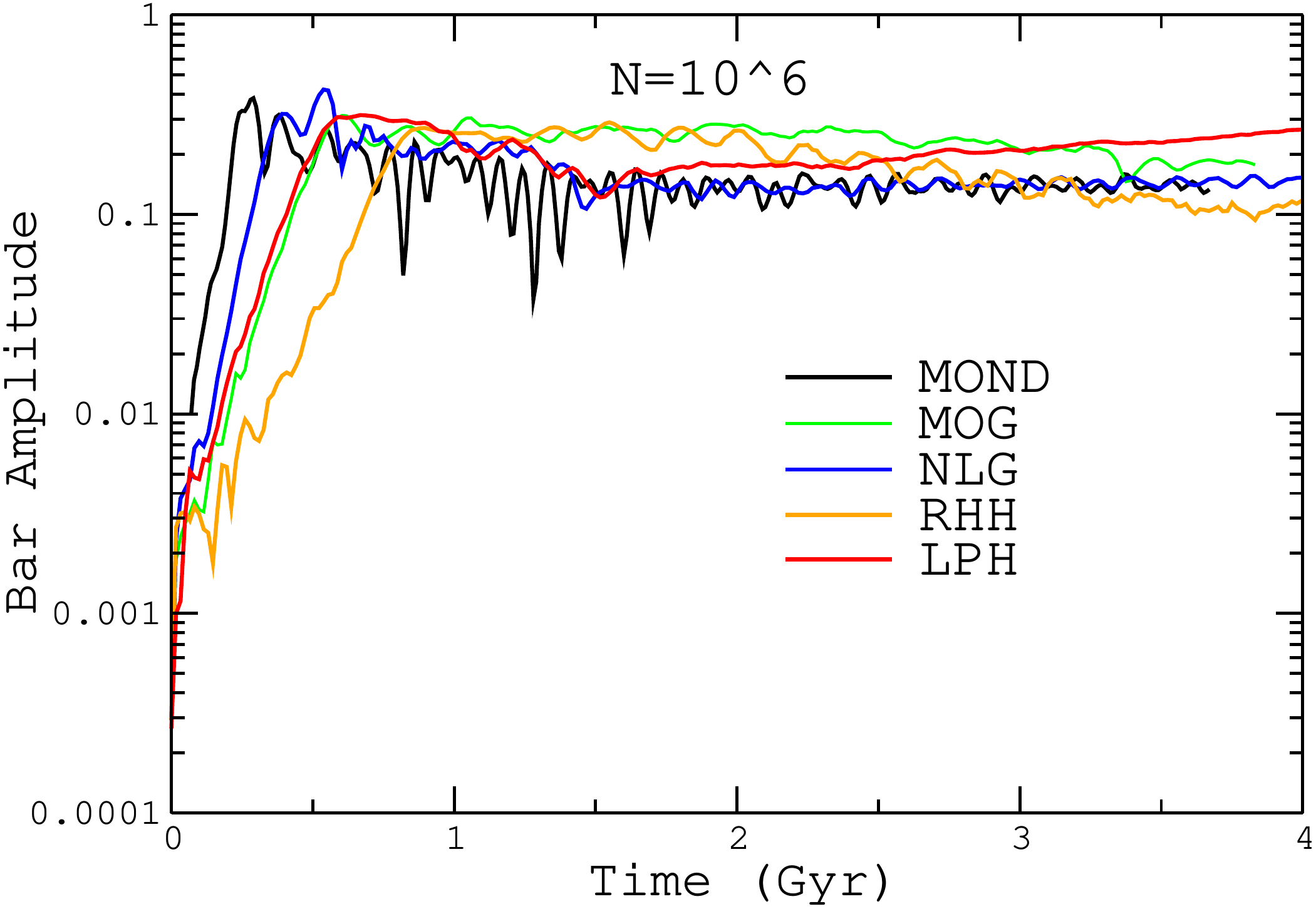}\vspace{0.3cm}
	\includegraphics[width = 8.2cm]{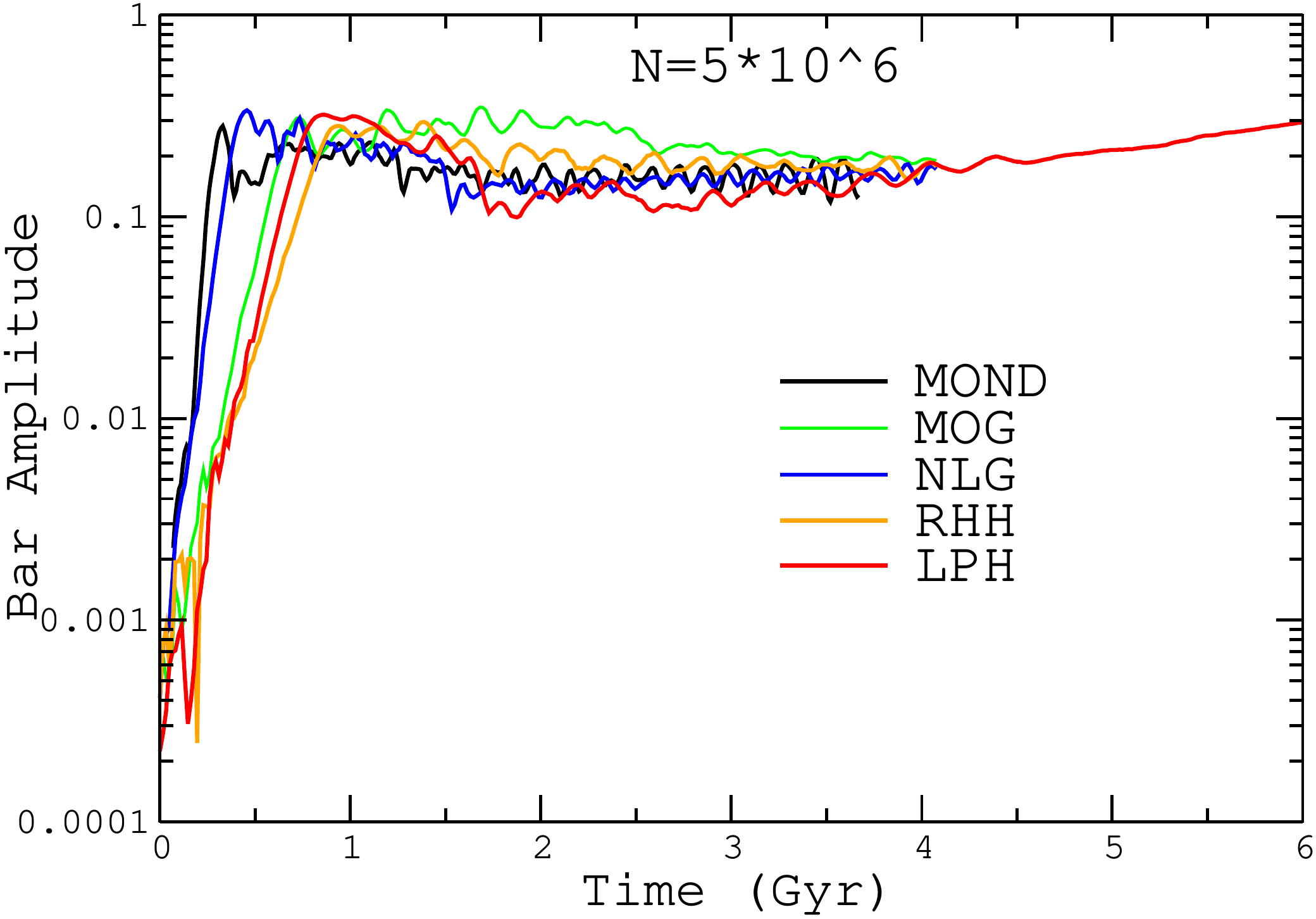}
	\caption{The bar amplitude $A_2/A_0$ {in} our five different models. For better visualization, the vertical axis is shown in logarithmic scale. The top (bottom) panel shows our low (high) resolution simulation.}
	\label{barhsb}
\end{figure}

We measure $A_2$ as a function of time for our models. The time evolution of this parameter is illustrated in Figure~\ref{barhsb}. Both panels show that none of the models can prevent the bar instability. However, the stabilizing effect of the rigid halo is clear (orange curves). This has been known since the seminal paper of \citet{Ostriker_1973}. It is also well-known that a live halo cannot suppress the bar instability, as is clear from the evolution of the bar amplitude for our LPH model (red curves).

The bar growth rate is substantially higher in MOND compared to other extended gravity models as well as the DM models. A rapid bar instability in MOND has also been reported by \citet{Tiret_2007}. In their model, the bar retains its maximum strength for a long duration (${\approx 4}$ Gyr) compared to the age of the Galaxy \citep[e.g.][]{Knox_1999}. After that, they report a sharp reduction in the bar magnitude. In our simulations, we see that the bar amplitude starts to decrease after a sharp maximum, and then stays almost constant. Observationally, this means that MOND predicts very strong bars for some spiral galaxies, but weakly barred galaxies are also expected \citep[see also][]{Banik_2020_M33}.

As is clear from Figure~\ref{barhsb}, bars are stronger in the DM case (the LPH model). The bar amplitude first experiences a minimum and then gradually starts to grow. This is not the case for any of the other models, implying that a live halo actually promotes a bar \citep{Athanassoula_2002}. At the end of our simulations, we see that the live DM model leads to the strongest bar, which is consistent with e.g. the simulations of M33 conducted by \citet{Sellwood_2019}.

It is interesting that the NLG model is somewhat similar to the MOND model and predicts a fast bar instability. On the other hand, the growth rate in MOG is very similar to the LPH model. Though the growth rate in MOND and NLG is much higher than in the standard LPH model, it would be seriously difficult to find evidence for this. The main reason is that we observe a single moment in the dynamical evolution of a galaxy rather than a time interval.

There are clear oscillations in the bar magnitude for all extended gravity models, while the DM case leads to smoother behaviour. These oscillations are not numerical artefacts related to the determination of the galactic centre \citep[see also][]{Roshan_2018, Roshan_2019}. Furthermore, they have been observed in lower resolution simulations of extended gravity which used a different approach for the time evolution of the system and an independent method for calculating the galactic centre \citep{Ghafourian_2017}. These oscillations are discussed further in Section \ref{Ultrafast_bars}.

\subsubsection{Power spectrum}
\label{Power_spectrum}

The surface density is a function of $\phi$, $r$, and $t$. Therefore the Fourier transform with respect to $\phi$ and $t$ gives Fourier coefficients $B(\omega,r)$, which are functions of the wave frequency $\omega$ and radius $r$. By looking at the power spectrum $|B(\omega,r)|^2$ of the density waves, it turns out that the extended gravity models host more density waves with different frequencies propagating throughout the disc. Differences with the DM models may result in observational discriminants, especially in LSB galaxies.

\begin{figure} 
	\centerline{\includegraphics[width = 4cm]{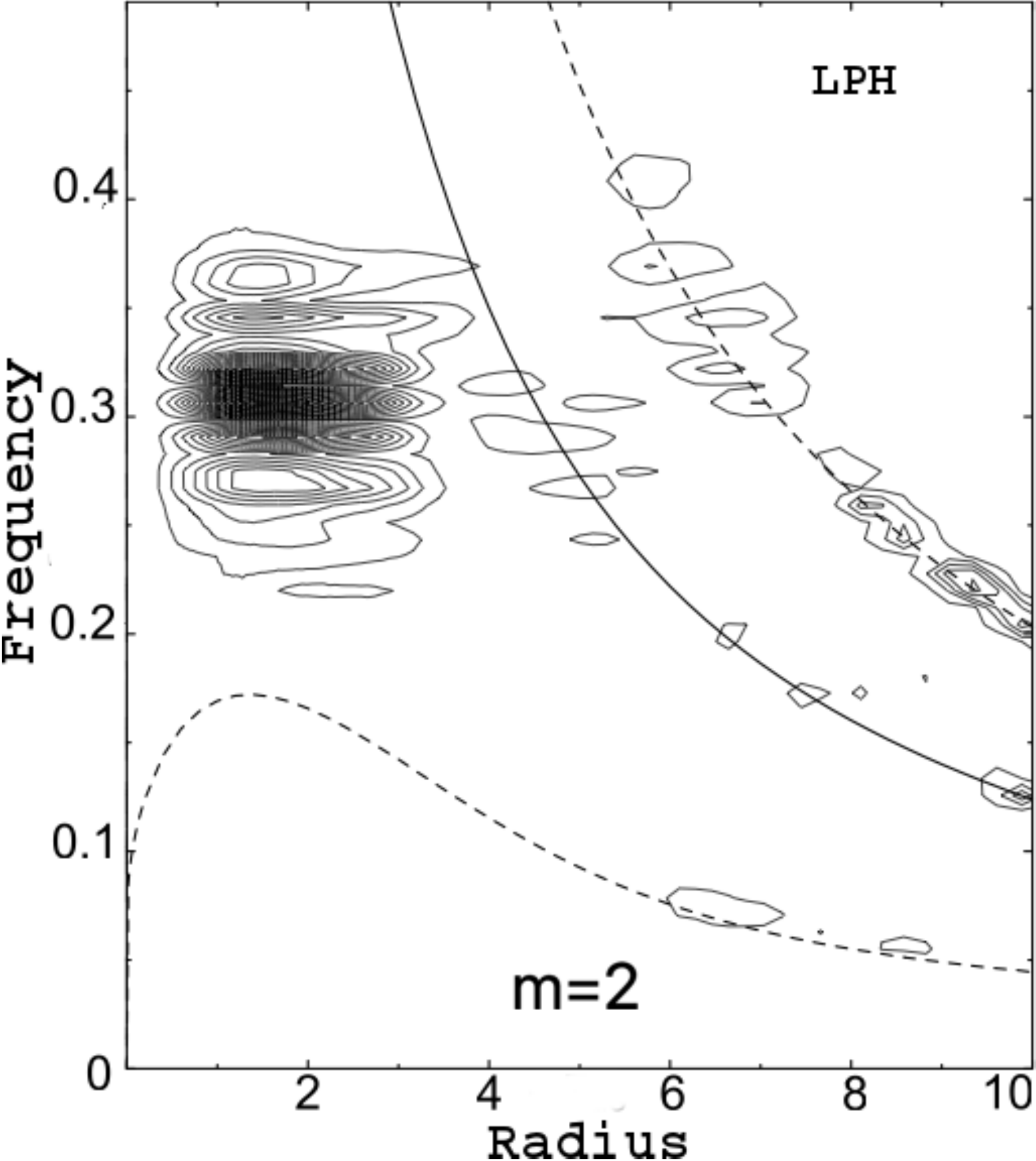}\hspace{0.3cm}
		\includegraphics[width = 3.89cm]{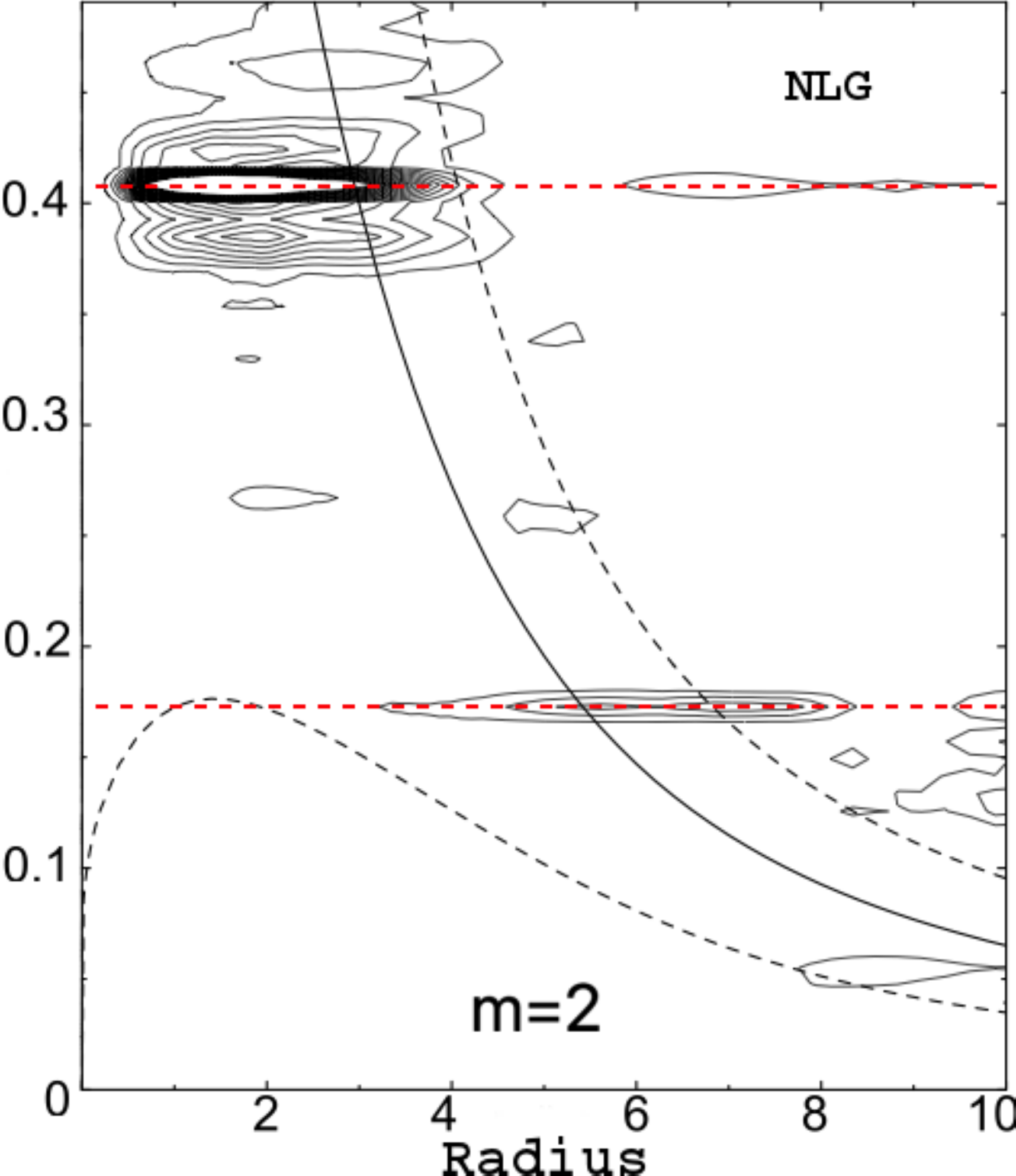}}\vspace{3pt}
	\centerline{\includegraphics[width = 4cm]{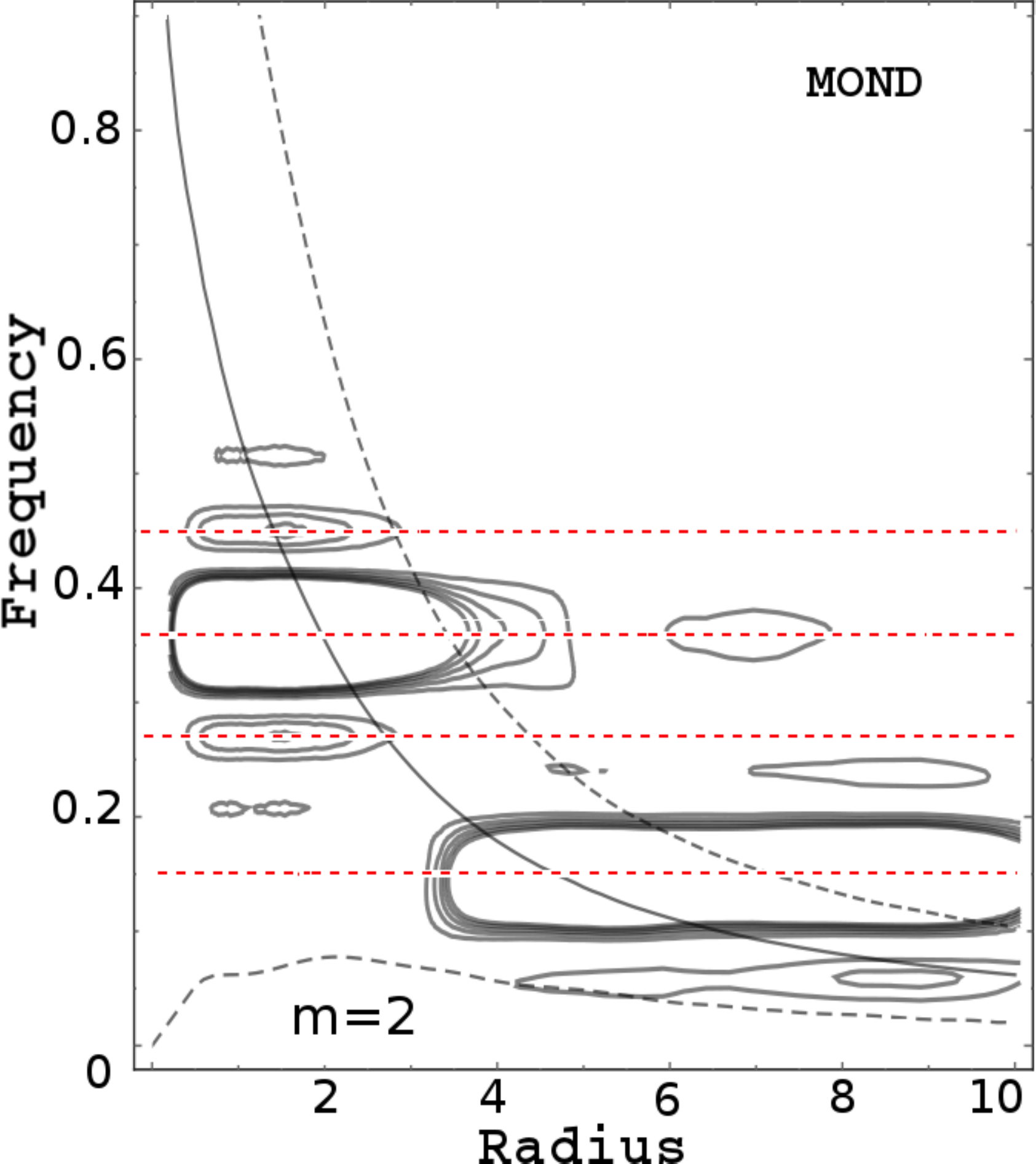}\hspace{0.3cm}
		\includegraphics[width = 3.89cm]{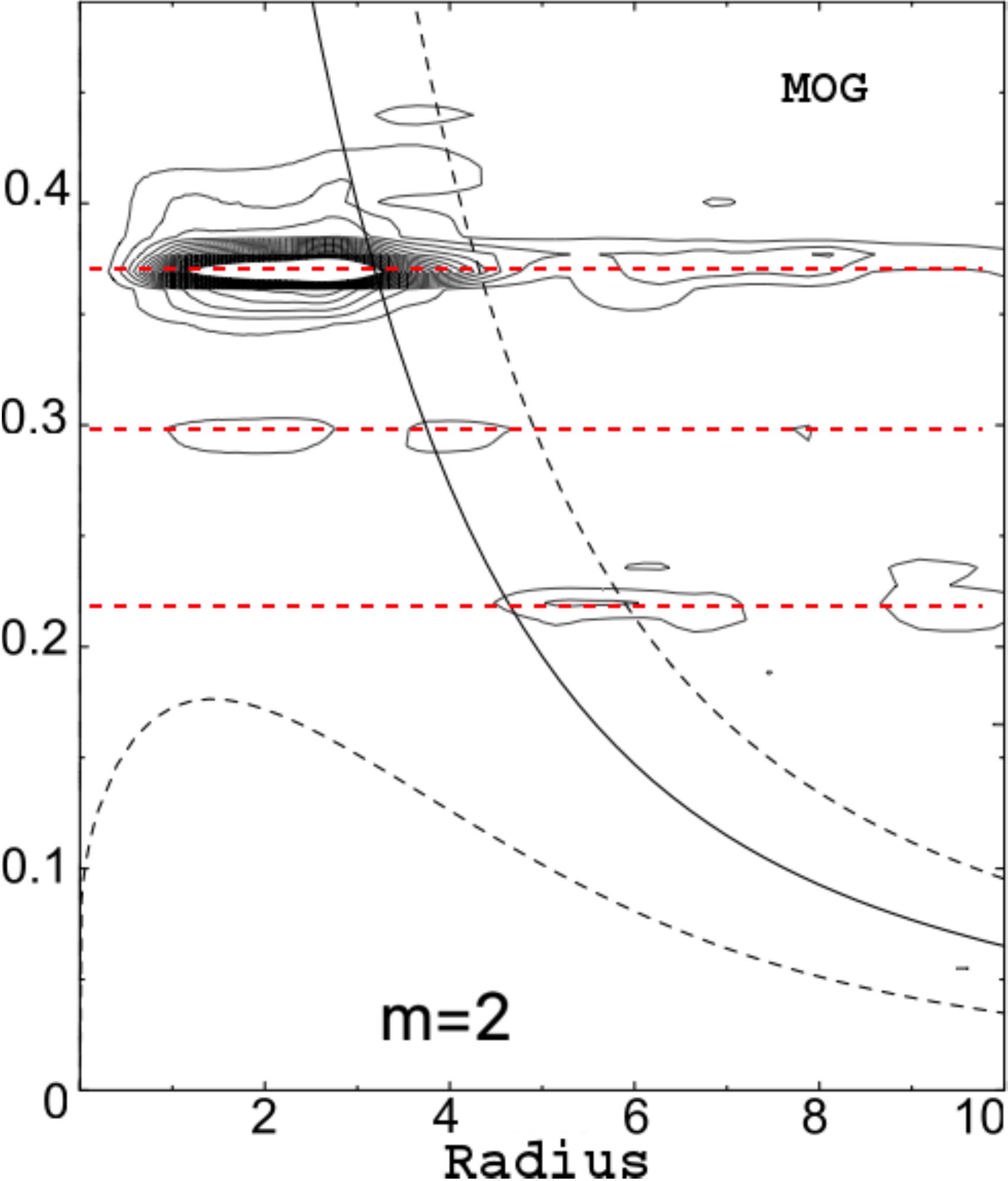}}
	\caption{The power spectrum for density waves in the LPH model (upper left), NLG (upper right), MOND (lower left), and MOG (lower right). The vertical axis is frequency in units of $192\, \text{km}\, \text{s}^{-1}\, \text{kpc}^{-1}$. The horizontal axis is the radius in kpc. Except the MOND case, the other panels were produced by the \textsc{galaxy} code.}
	\label{p_s}
\end{figure}

The dominant mode in the DM model is the bar mode ($\mathrm{m} = 2$). To quantify its strength, we plot contours of the power spectrum for $\mathrm{m} = 2$ (Figure~\ref{p_s}). In the case of LPH, NLG, and MOG models, the Fourier transform has been taken over $t = \left( 1.5 - 4 \right)$~Gyr to ensure that the density waves have been excited. Accordingly, for the MOND model, the interval $t = \left( 3 - 4 \right)$~Gyr has been used. Any horizontal line with contours concentrated around it indicates the existence of a density wave whose frequency is shown on the vertical axis. The upper left panel belongs to the standard LPH model. We see that there is one dominant $\mathrm{m} = 2$ mode with a time-varying frequency. We will discuss this case in more detail in subsequent sections. The other panels use red dashed lines to show the frequency of density waves for the NLG, MOG, and MOND models. We see that in NLG there are two waves and in MOG three waves with different intensities. The MOND model is even noisier. It seems that unlike in extended gravity, the DM halo suppresses the excitation of several modes on the surface of the disc, even though it cannot suppress the main bar instability mode. From this perspective, we see the expected stabilizing behaviour caused by the halo.

The above-mentioned features do not change with the number of particles (bottom panel of Figure~\ref{barhsb}). We see that by increasing the particle number $N$, the time evolution of the LPH model becomes a bit slower. Therefore, we extend the simulation duration to see the second increasing phase of the bar magnitude.

\subsection{Buckling instability}

\begin{figure} 
	\includegraphics[width = 8cm]{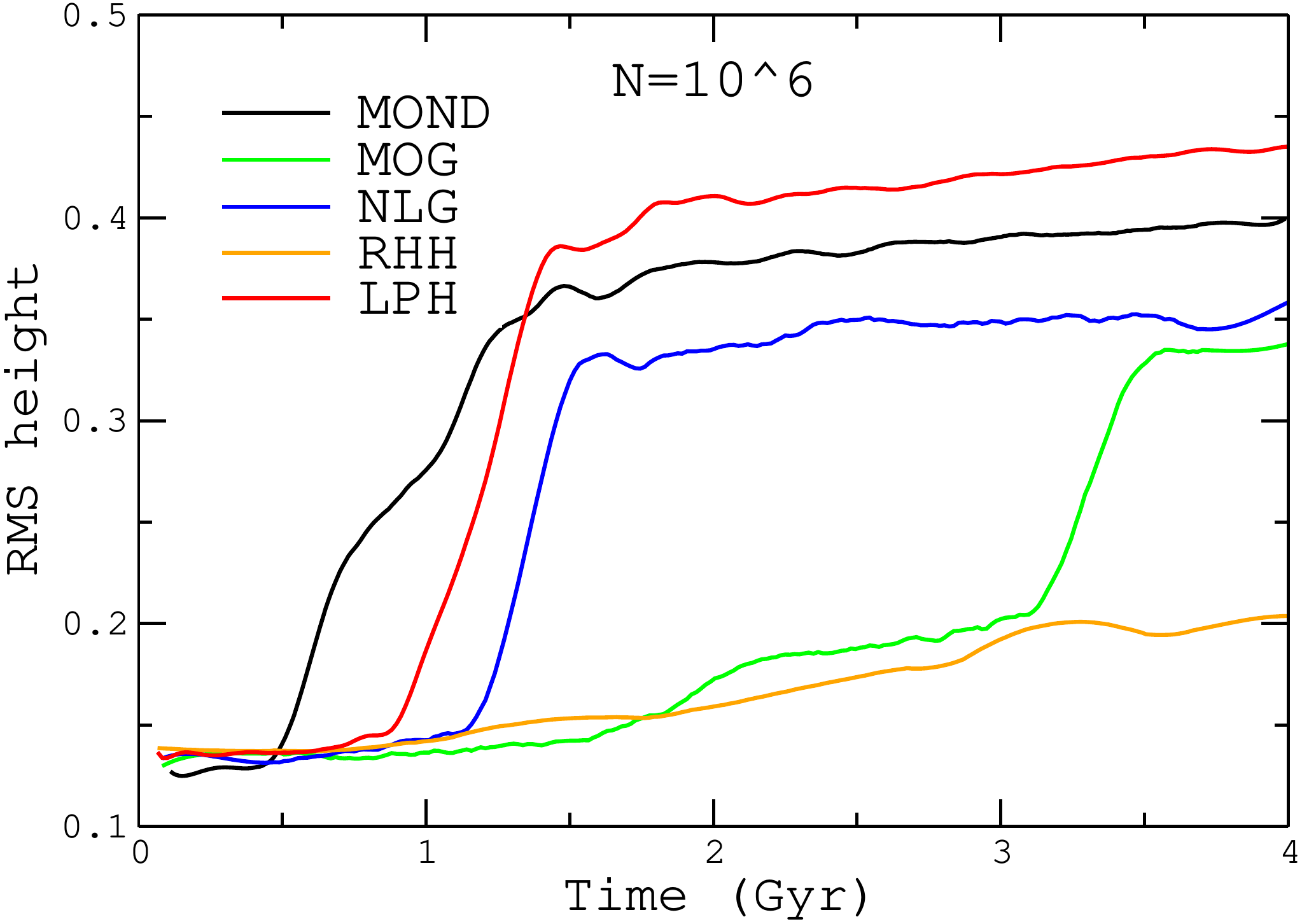}\vspace{0.3cm}
	\includegraphics[width = 8cm]{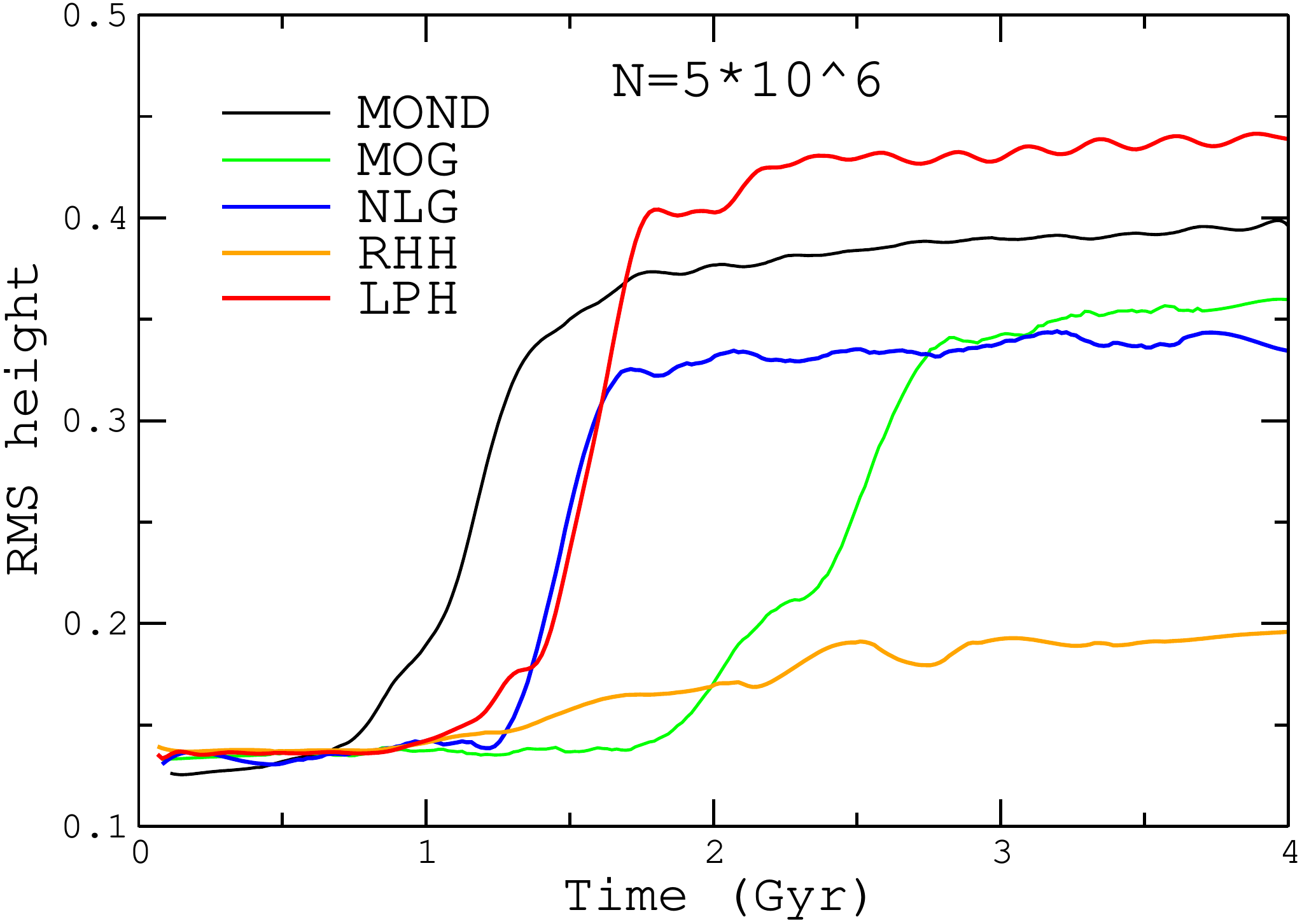}
	\caption{rms height in kpc {computed using particles with $\left| z \right| < 2$~kpc} for all models, measured at $R = 1.1$ kpc {where the} buckling instability happens more effectively.}
	\label{rmsth}
\end{figure}

It is also instructive to compare the \textit{buckling instability} in different models. The top panel in Figure~\ref{barhsb} shows that for our NLG and LPH models, the bar amplitude starts to decrease around $t \approx 0.5$~Gyr. It is well known that a rapid thickening of the disc can substantially weaken the bar. To see this behaviour, we have plotted the root-mean-square (rms) thickness at $R = 1.1$ kpc with respect to time (Figure~\ref{rmsth}). Some particles escape to large vertical distances and artificially increase the rms height. Therefore, we ignore the contribution of particles with $ \left| z \right| > 2$~kpc. The top and bottom panels belong to $N = 10^6$ and $N = 5\times 10^6$, respectively. We see the step-like behaviour for our NLG, LPH, and MOND models. This is related to the buckling instability through which the disc thickness increases rapidly.

\begin{figure} 
	\includegraphics[width = 0.238\textwidth]{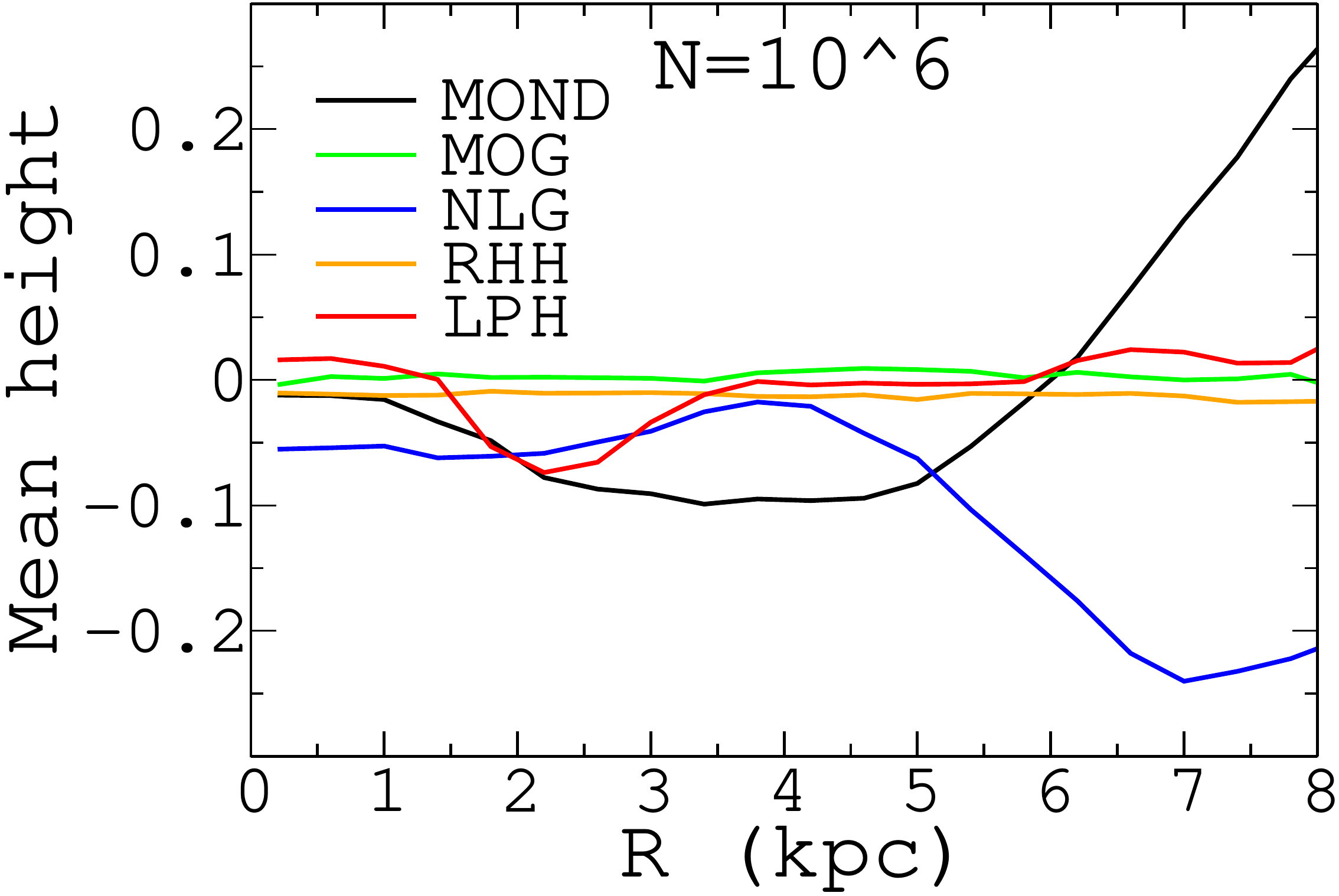}\hspace{0.07cm}
	\includegraphics[width = 0.225 \textwidth]{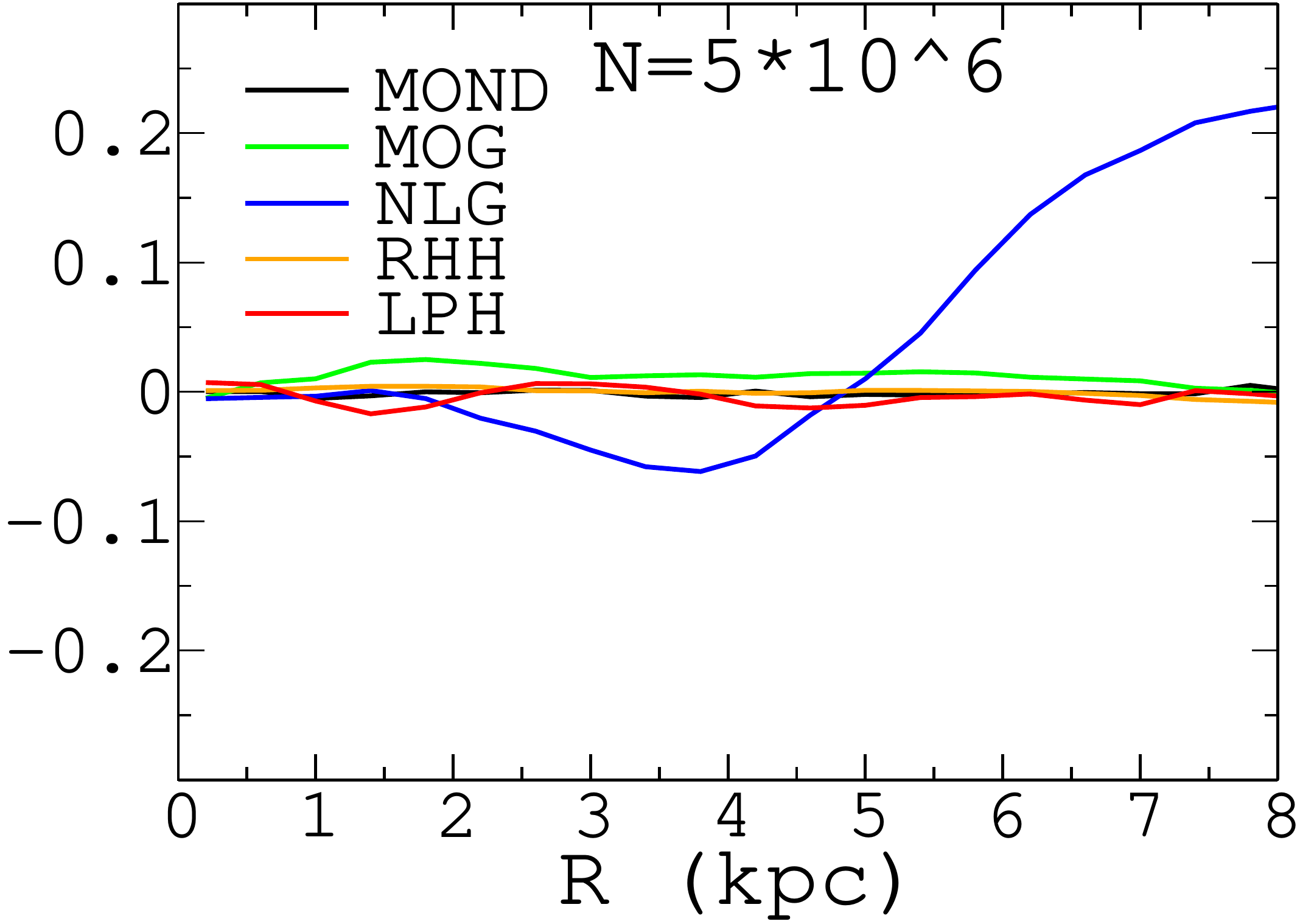}\\
	\includegraphics[width = 0.238 \textwidth]{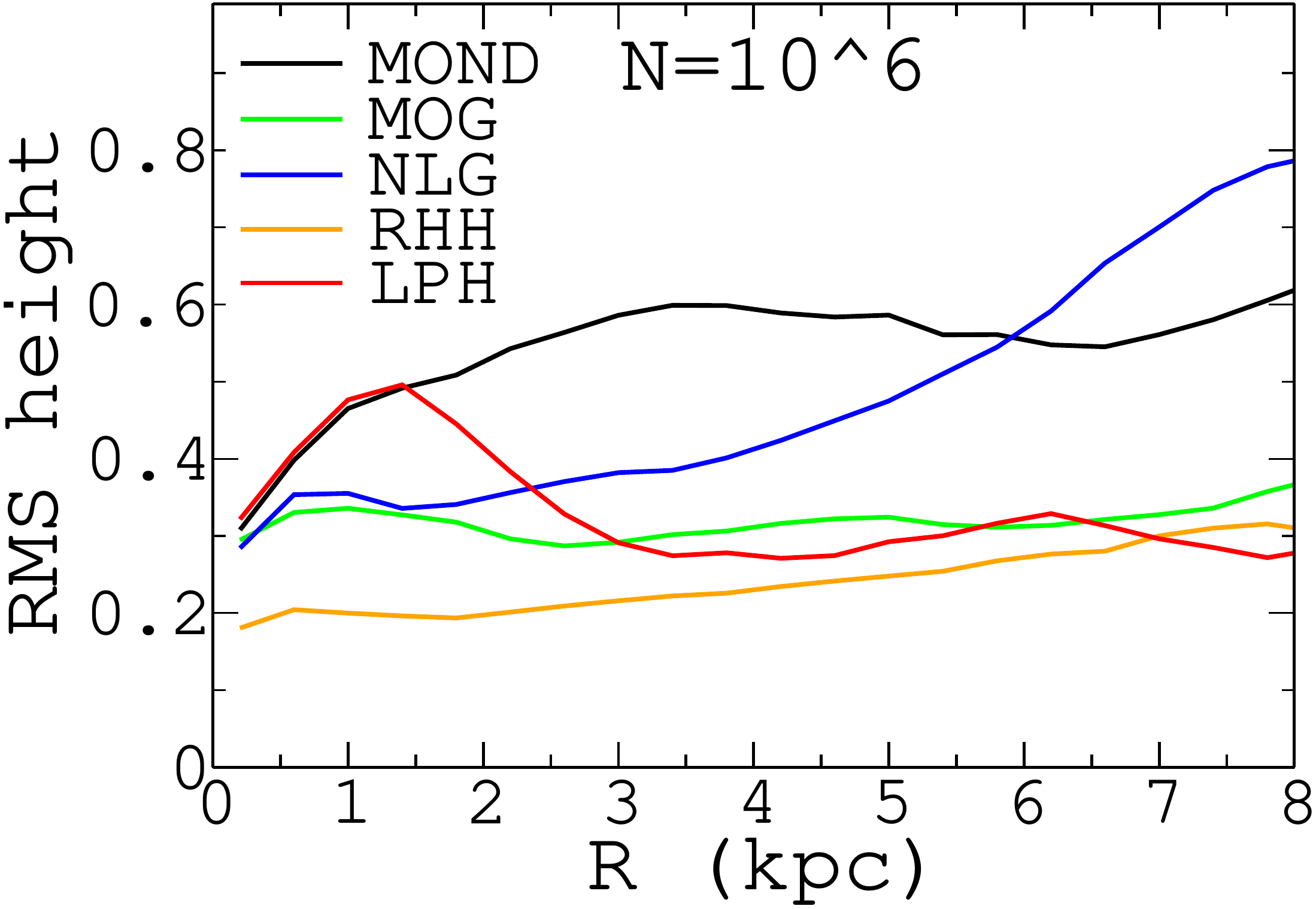}\hspace{0.07cm}
	\includegraphics[width = 0.225 \textwidth]{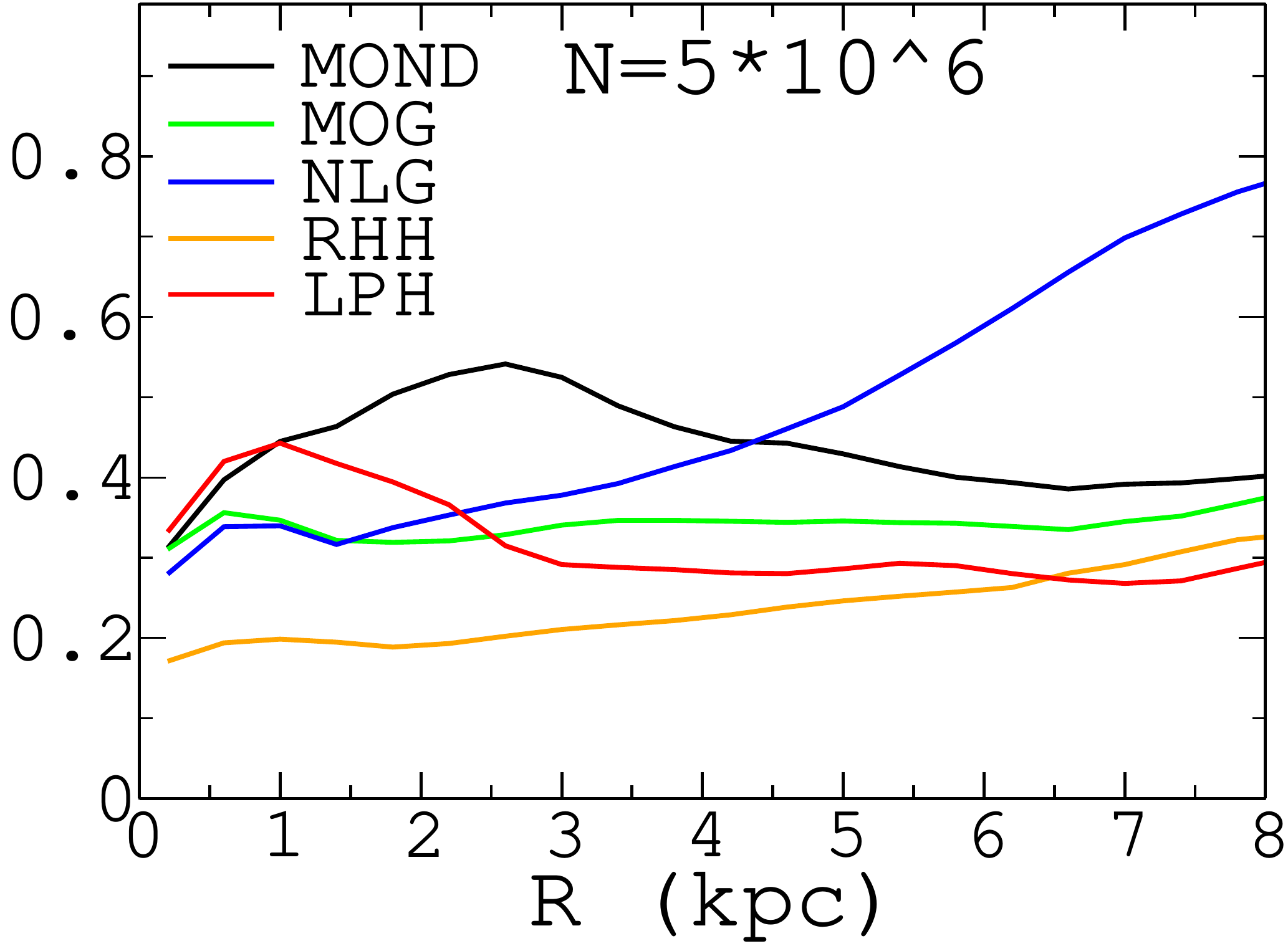}
	\caption{The mean and rms height as a function of radius at $t = 4$~Gyr. The left (right) column belongs to the $N = 10^6$ ($N = 5 \times 10^6$) models. Distances are given in kpc, and the contribution of particles with $ \left| z \right| > 2$~kpc is ignored.}
	\label{meanrms}
\end{figure}

It is interesting that although NLG and MOG lead to almost the same form for the point mass gravitational force, the evolution of galactic discs differs. For example, the buckling instability is postponed in MOG (Figure~\ref{rmsth}). The less violent buckling instability in MOG has also been reported in \citet{Roshan_2018}. It is clear that the rigid halo model is more stable against both bar and buckling instabilities.

In the MOND simulation, the buckling instability happens earlier. However, the growth rate of this instability is higher in the LPH DM model. Also, the rms height at $R = 1.1$ kpc in the LPH model is larger than for the extended gravity models for most of the simulation duration. Among these models, MOND mimics the DM behaviour more closely at small radii (see also Figure~\ref{meanrms}). Overall, we conclude that not only are stellar bars in spiral galaxies stronger in the LPH model, the rms height of discs at small radii is also larger in the CDM paradigm. Of course, a statistical investigation using cosmological simulations is required to reliably verify if extended gravity models predict a different morphology for the vertical structures of disc galaxies. 

To get a better understanding of the vertical behaviour of the discs under the effect of different gravity laws, it is helpful to plot the mean height (along $z$) and rms height as a function of radius at the end of our simulations. In Figure~\ref{meanrms}, we plot these quantities for each model with $N = 10^6$ and $N = 5 \times 10^6$. Recall that at $t = 0$, the mean height of all simulated discs is exactly 0 by construction (within numerical noise). The top panels in Figure~\ref{meanrms} show an increasing mean height for the NLG disc, directly proving that the disc is warped. The same behaviour also appears in our low-resolution MOND model (see also the edge-on projections in Figure~\ref{edge-on}). The existence of a warp in low-resolution MOND simulations has already been reported in \citet{Tiret_2007}, where $2 \times 10^5$ particles were used for the stellar disc. However, no warp is excited in our higher resolution MOND model. This is most apparent in the edge-on views of the final disc state in our high-resolution models (Figure~\ref{edge}). The other three models do not present a considerable change in mean height at either resolution setting.

The bottom panels of Figure~\ref{meanrms} show the rms height of each model, confirming that the vertical structure of the discs changes. Specifically, the inner region in the LPH DM model is thicker compared to all other models, though the thickness in MOND rather closely mimics LPH. According to this figure, a region of increased thickness is present in the LPH models around $R \approx \left( 1 - 1.5 \right)$~kpc, mirroring the peanut shape that appears in these models (readily apparent in edge-on projections, see Figure~\ref{edge}). Weaker and shorter peanuts appear in NLG and MOG around $R \approx \left( 0.5 - 1 \right)$~kpc. The MOND peanuts look longer $-$ in the high-resolution MOND model, the peanut appears at $R \approx 2.5$~kpc. In the low-resolution MOND model, although the rms height varies rather smoothly, the peanut is clearly visible in Figure~\ref{edge-on}. In both cases, there is no rapid change (or a sharp local maximum as in LPH) in the rms height. Therefore, we infer that the peanut is weak in MOND, which is consistent with the results of \citet{Tiret_2007}.

Another important feature is that although the inner regions in the LPH model are thicker, the outer disc ($R \ga 4$ kpc) is substantially thicker in extended gravity models, especially in MOND and NLG. Different behaviour in the vertical structure is expected $-$ starting with the same initial rotation curve means that at least on the disc surface, extended gravity and particle DM models lead to the same radial acceleration. However, in the vertical direction, the accelerations are not necessarily the same. For example, the NLG effects appear as an effective phantom $\rho_p$ surrounding the baryonic matter (Figure~\ref{efdens}). One can easily verify that $\rho_p$ is not spherically symmetric for an exponential baryonic distribution. This will lead to differences with a DM model that has a spherical halo component. Indeed, the presence of a phantom DM disc (Figure \ref{Sigma_p_MOND}) is an important prediction of MOND \citep{Bienayme_2009}. Since the vertical restoring force at large radii is mostly fixed by the rotation curve due to the low disc surface density, the thicker outer disc in MOND may indicate stronger secular disc heating due to enhanced self-gravity, which is evident in that $\Sigma_p > 0$ unlike the other models. However, it is not clear why the NLG model should flare so strongly. A complementary study is required to carefully investigate $\rho_p$ and its time evolution, and to relate this to the vertical structure of the disc. We leave this as a subject for future studies, though the initial $\rho_p$ and $\Sigma_p$ are discussed in Section \ref{Phantom_dark_matter} for an infinitely thin exponential disc.

\subsection{Pattern speed $\Omega_p$}

The bar pattern speed $\Omega_p$ is another important quantity which may help to discriminate between DM and extended gravity. Its evolution is directly correlated with the properties of any DM halo. Knowing $\Omega_p$ would help us understand how the stellar bar redistributes angular momentum throughout the disc and influences the secular evolution of the galaxy. Furthermore, the location of resonances depends on $\Omega_p$.

$\Omega_p$ can be measured rather precisely in simulations by finding the position angle of the bar axis at different times. Measuring $\Omega_p$ for real galaxies is not a simple task. There are several model-dependent methods \citep[e.g.][]{Perez_2004, Rautiainen_2008}. The so-called \citet{Tremaine_1984} method is the only one that is model-independent. For its limitations and practical difficulties, we refer the reader to \citet{Oehmichen_2019}.

Observations indicate that spiral galaxies host fast bars \citep{Aguerri_2015}. This is inconsistent with standard CDM-based isolated and cosmological simulations \citep{Debattista_2000, Algorry_2017}. It is well understood that dynamical friction between DM particles and baryonic discs substantially damps $\Omega_p$, causing a gross disagreement with observations. This issue is considered a challenge for the CDM paradigm in small scale systems ($\la 10$ kpc). We explore this further in Section \ref{R_statistics}.

These arguments do not necessarily challenge the existence of DM on galactic scales. In particular, superfluid DM haloes around galaxies would create very little dynamical friction \citep{Berezhiani_2016, Berezhiani_2019}. We discuss this model further in Section \ref{Broader_implications}. A similar argument applies to ultralight ($m \approx 10^{-22}$ eV/$c^2$) bosonic DM particles because they have a long de Broglie wavelength $\lambda \approx 1$ kpc. Consequently, their wave-like behaviour would appear on a galactic scale, preventing them from causing significant dynamical friction in spiral galaxies \citep{Hui_2016}. Moreover, it is claimed that this model can lead to a viable cosmic structure formation scenario \citep{Mocz_2019}. Those authors showed that it predicts serious deviations from $\Lambda$CDM at the large redshifts when the first stars formed. This is a smoking gun that should be tested by future telescopes like JWST.

\begin{figure}
	\includegraphics[width = 8cm]{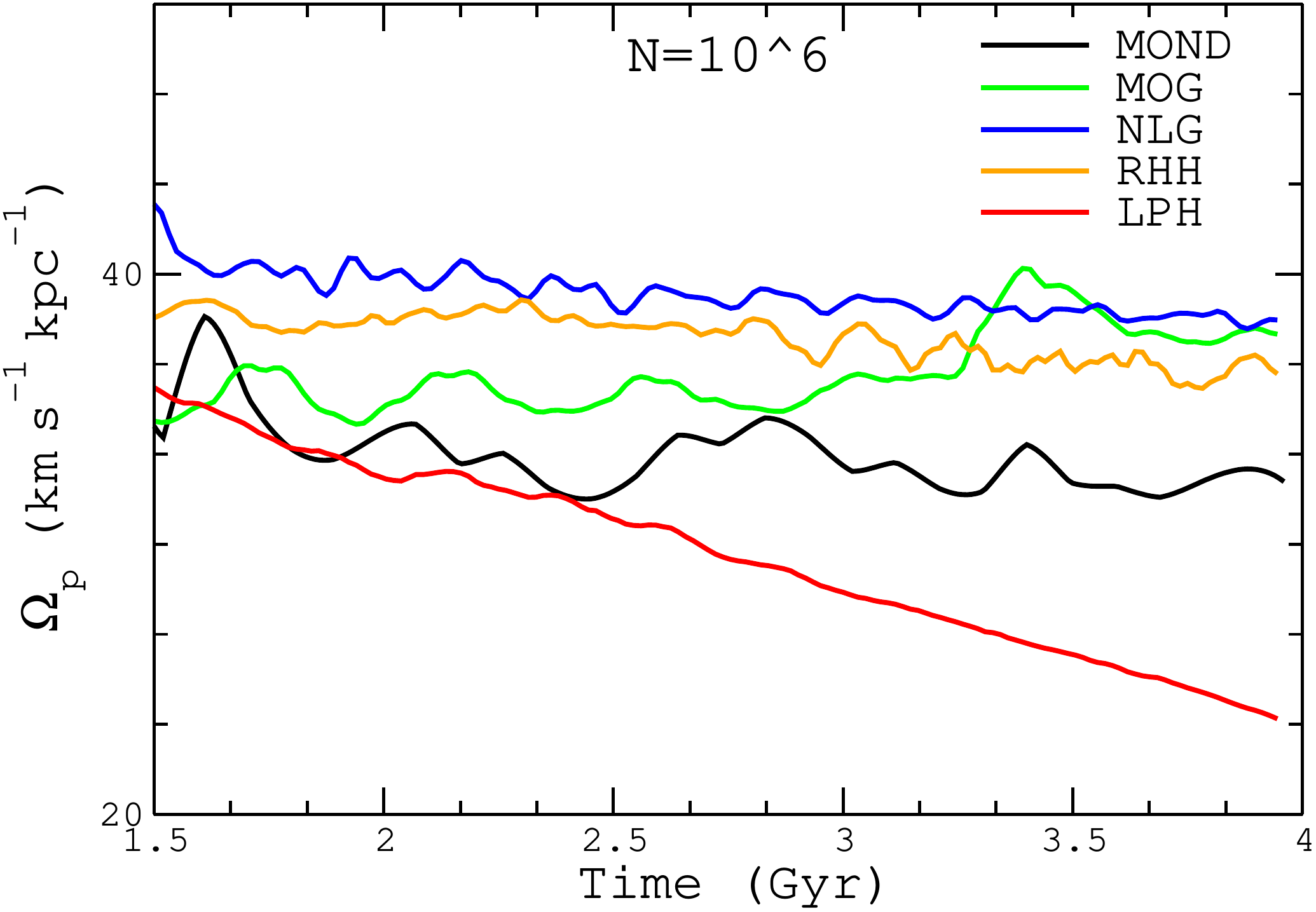}\vspace{0.3cm}
	\includegraphics[width = 8.2cm]{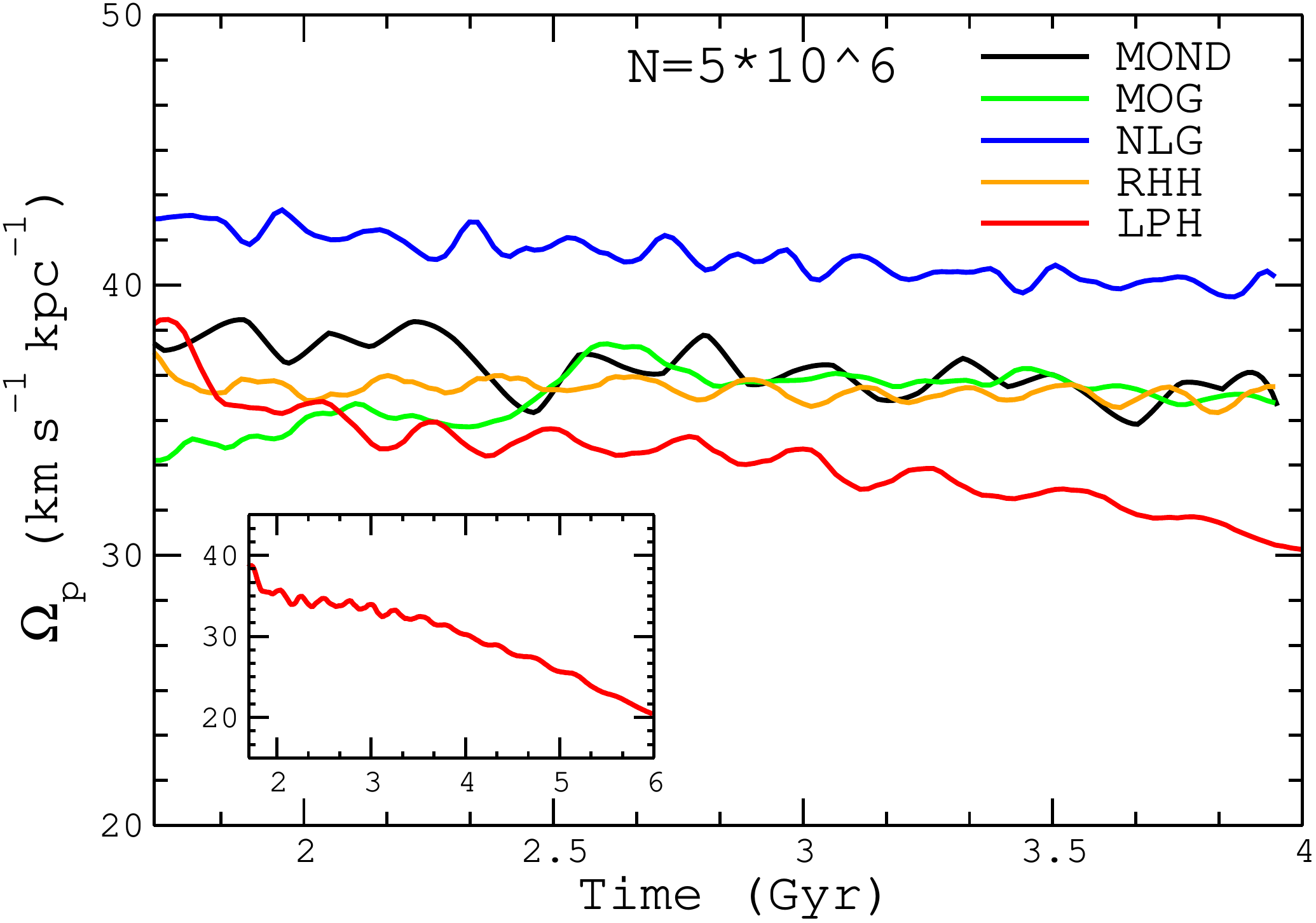}
	\caption{Time evolution of the bar pattern speed $\Omega_p$ for all our models. Different panels show results for a different number $N$ of particles. The inset in the lower panel shows $\Omega_p$ in the LPH model over a longer duration.}
	\label{pspeed}
\end{figure}

In extended gravity theories for the missing gravity problem, there is no DM slowing down the pattern speed. Therefore, one may expect fast stellar bars in extended gravity simulations. Bearing this in mind, we discuss the evolution of $\Omega_p$ in our simulations. The result is illustrated in Figure~\ref{pspeed}. In the top panel, we present $\Omega_p$ for $N = 10^6$, while the bottom panel belongs to $N = 5 \times 10^6$. There are other density waves excited in the discs, e.g. the $\mathrm{m} = 3$ mode excited at $t \approx 300$~Myr in the MOND model (Figure~\ref{pos_others}). To ensure that we have a well-settled bar rotating almost uniformly within the disc, we concentrate on $t > 1$~Gyr. We see that when the halo is rigid (the RHH model), the pattern speed remains constant with time. This is as expected since a perturber moving in a rigid halo potential cannot induce a perturbation to the halo density and pressure. Consequently, there is no wake behind the perturber to cause dynamical friction.

On the other hand, it is clear from both panels of Figure~\ref{pspeed} that the pattern speed decreases in the LPH model. This is because dynamical friction transfers angular momentum between the live halo and the stellar bar, as confirmed directly in the bottom panel of Figure~\ref{angular}. The pattern speed for the LPH model decreases almost linearly with time as $\Omega_p(t) \approx -at + b$, where $a$ and $b$ are positive. In the model with $N = 10^6$ we have $a = 4.775$/Gyr\textsuperscript{2}, $b = 42.632$/Gyr. For a point mass perturber moving inside a uniformly distributed medium, the dynamical friction can be expressed as Chandrasekhar's formula. It is not possible to find an exact analytic expression for the case of a stellar bar inside a differentially rotating disc and halo. However, as the pattern speed varies roughly linearly with time, we can estimate the magnitude of the friction. For a very crude estimation of the dynamical friction force, let us assume that the bar is rigid with length $R_b$ and mass $m_b$. Using Newton's second law, one may infer that the friction force is almost constant and given by $F_{d} \approx R_b m_b a$.

As expected, the lack of a halo causes $\Omega_p$ to remain nearly constant with time in all considered extended gravity models. Furthermore, we see that NLG has the highest $\Omega_p$. MOND gives a relatively low $\Omega_p$. A similar result for the time evolution of $\Omega_p$ in MOND has already been reported \citep{Tiret_2007}. {The MOND model also shows small oscillations in $\Omega_p$ by up to $\approx 10\%$ due to the coupling with other modes in the disc. This phenomenon was noted in the CDM context by \citet{Hilmi_2020}, and is also apparent in our LPH model.}

{Despite clear differences in the time evolution of $\Omega_p$, the long timescale involved makes this difficult to directly constrain. The present value of $\Omega_p$ can be determined, but by itself this is} not enough to compare galaxies with different properties and dynamical timescales. Instead, the ratio of the corotation radius $R_c$ over the bar semi-major axis $R_b$ provides an appropriate measure to compare the bar pattern speed in different galaxies. We next discuss this parameter for our simulations.

\subsection{\texorpdfstring{$\mathcal{R}$}{R} parameter} 
\label{R_statistics}

Bar pattern speed measurements help to find the corotation radius $R_c$. Combined with measurements of the bar length (semi-major axis) $R_b$, we can find the $\mathcal{R}$ parameter, defined as:
\begin{eqnarray}
	\mathcal{R} ~\equiv~ \frac{R_c}{R_b} \, .
	\label{R}
\end{eqnarray}
From an observational point of view, $\mathcal{R}$ has great importance since it reveals the above-mentioned contradiction between observations and $\Lambda$CDM simulations. The bar is `fast' when $\mathcal{R} \la 1.4$, while it is `slow' otherwise \citep{Galactic_Dynamics}.

To measure $\mathcal{R}$ at a given time $t$, we first calculate the corotation radius. To do so, we use the pattern speed $\Omega_p \left( t \right)$ to find the radius at which this matches the angular velocity $\Omega \left( R \right)$ from the rotation curve. Since both can be obtained very simply in our simulations, the corotation radius is measured with appropriate precision.

\begin{figure*}
	\centerline{\includegraphics[width = 8.8cm]{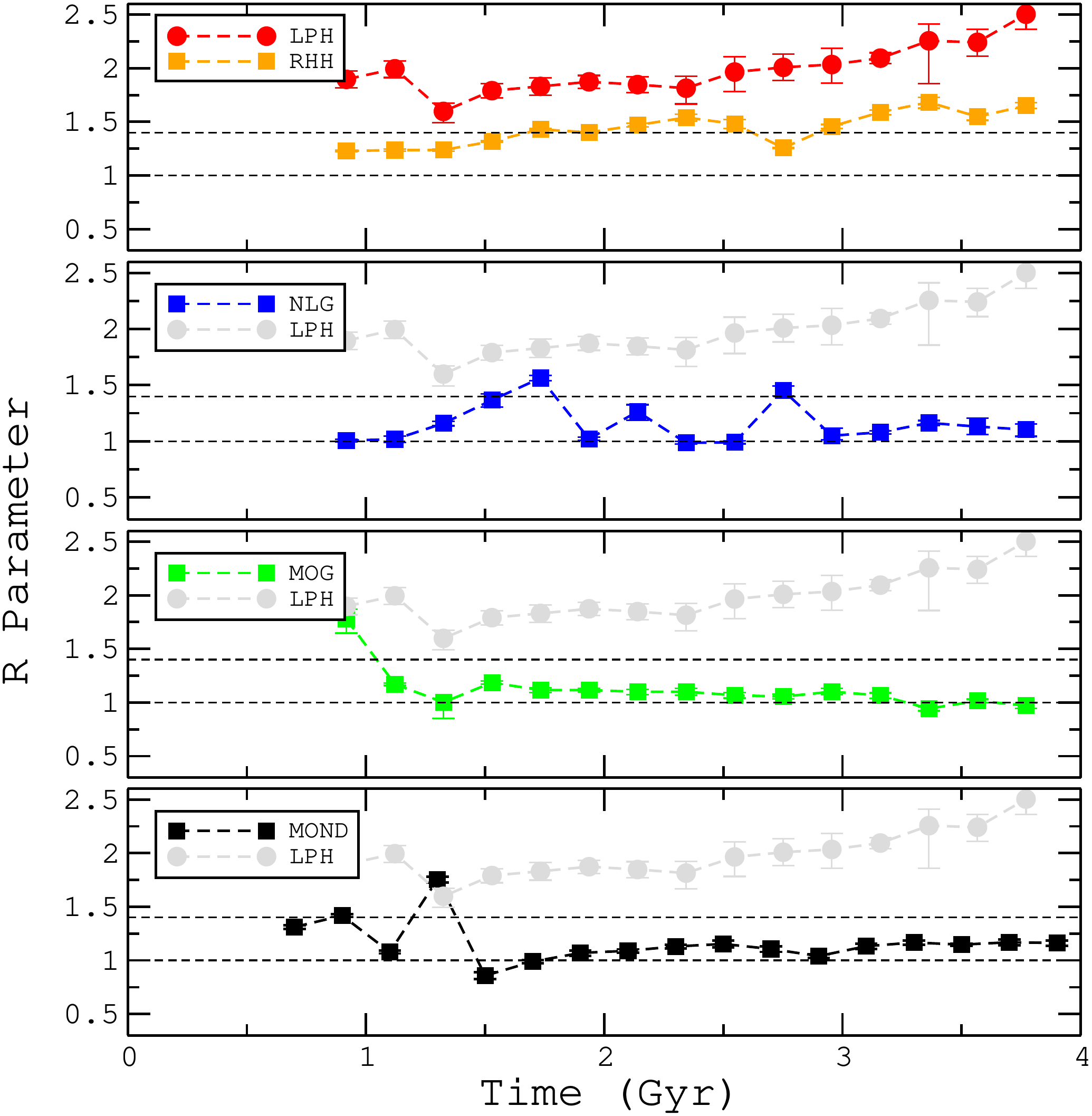}\hspace{0.3cm}\includegraphics[width = 8.8cm]{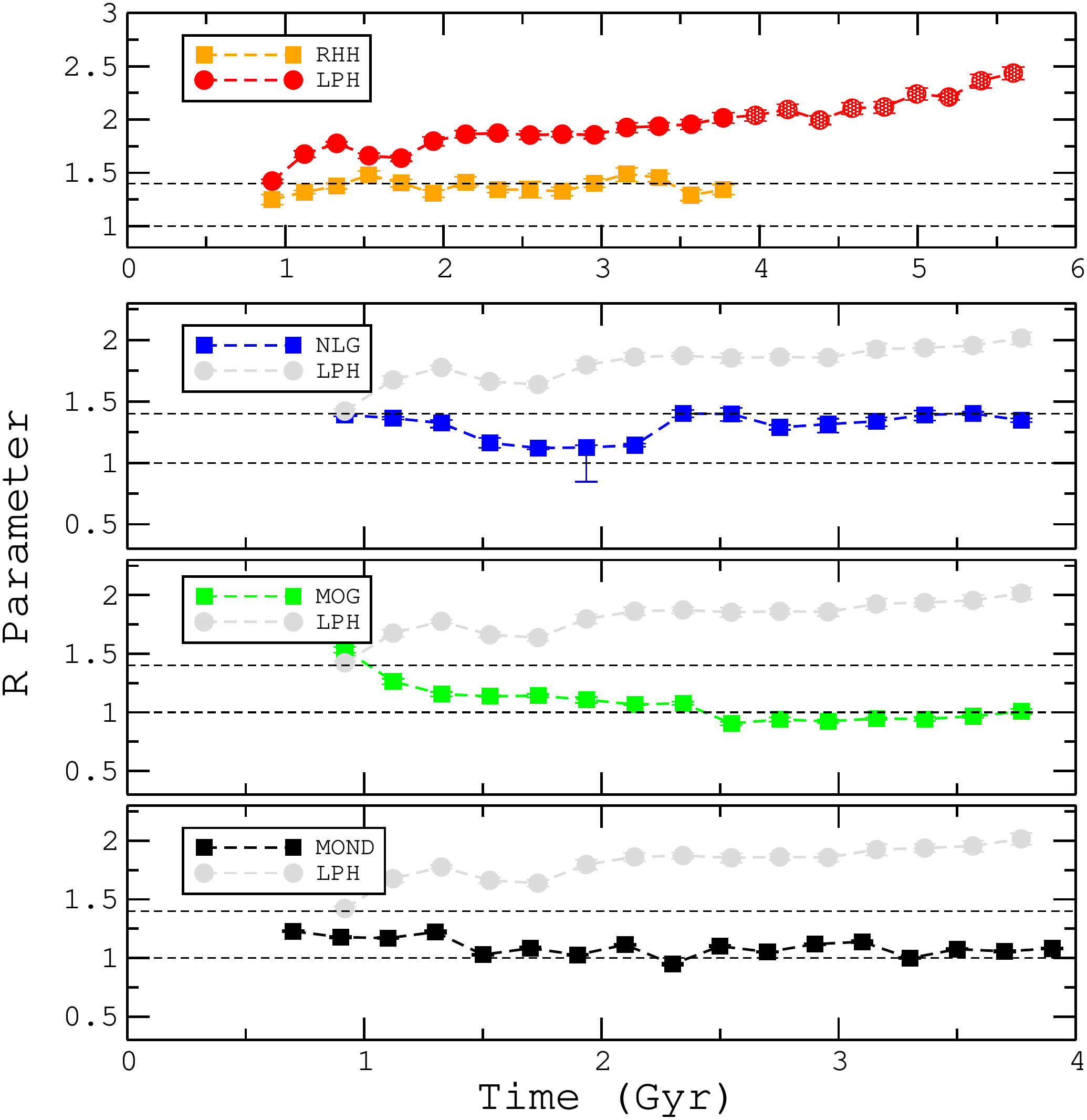}}
	\caption{Time evolution of the $\mathcal{R}$ parameter (Equation \ref{R}) for simulations with $N = 10^6$ (left panel) and $N = 5 \times 10^6$ (right panel). From top to bottom, the panels belong to RHH, NLG, MOG, and MOND. In all panels, we show the DM case (LPH) for better comparison. We divide the evolution period into time intervals of $\Delta t \approx 200$~Myr and choose the timestep closest to the middle of the interval as representative of the whole $\Delta t$ interval. The simulation duration is 4~Gyr except in the top right panel, which covers 6~Gyr. The rising $\mathcal{R}$ in the LPH model is due to a decreasing pattern speed (Figure~\ref{pspeed}).}
	\label{r1}
\end{figure*}

Unfortunately, measuring the bar length is not a trivial task. Various methods have been introduced in the literature to determine $R_b$ \citep[][and references therein]{Erwin_2005, Aguerri_2009, Aguerri_2015}. The method applied in this work makes use of the Fourier decomposition of the galaxy's surface density profile \citep{Elmegreen_1985, Ohta_1990, Aguerri_2000}. The bar radius is computed using the ratio of intensity in the bar ($I_b$) and inter-bar ($I_{ib}$) regions, where:
\begin{eqnarray}
	I_b &=& I_0 ~+~ I_2 ~+~ I_4 ~+~ I_6 \, , \\
	I_{ib} &=& I_0 ~-~ I_2 ~+~ I_4 ~-~ I_6 \, .
\end{eqnarray}
Here, $I_\mathrm{m}$ stands for the $\mathrm{m}$\textsuperscript{th} component of the azimuthal Fourier decomposition of the intensity, which depends on $R$. According to the definition of \citet{Aguerri_2000}, the bar length would be the outer radius beyond which
\begin{eqnarray}
	\frac{I_b}{I_{ib}} ~<~ 0.5 \left[ \left(\frac{I_b}{I_{ib}}\right)_{\text{max}} + \left(\frac{I_b}{I_{ib}}\right)_{\text{min}} \right] \, .
\end{eqnarray}
In this definition, the surface density profile is also considered. The error of using this method in numerical simulations has been reported as $\la 4\%$ except for very thin bars, where it reaches $\approx 8\%$ \citep{Athanassoula_2002}.

The results for $N = 10^6$ and $N = 5 \times 10^6$ are shown in the left and right panels of Figure~\ref{r1}, respectively. The fast bar regime is shown with horizontal dashed lines. In both panels, we see that $\mathcal{R}$ in the live DM model (LPH) is always above this regime. As already mentioned, this is a well-known fact formerly reported in several papers \citep[e.g.][]{Debattista_2000}. This is because $\mathcal{R}$ increases with time under the direct influence of dynamical friction. As expected, due to the absence of dynamical friction in the RHH model, bars are faster in this case. For MOND and MOG, we see almost the same behaviour as for RHH $-$ the bars lie in the desired fast bar regime. This is also true for NLG bars, though they are relatively slower.

Although the majority of spiral galaxies host fast bars \citep[see figure~8 in][]{Oehmichen_2019}, some appear to have the predicted ultrafast bars \citep[$\mathcal{R} < 1$,][]{Guo_2019}. In Section \ref{Ultrafast_bars}, we discuss in more detail how we are able to get apparently ultrafast bars in our extended gravity models despite theoretical arguments that they should be unstable \citep[e.g.][]{Contopoulos_1980}. We caution that for a meaningful comparison with observations, we still need more realistic simulations including gas components and a bulge. One should also use different techniques for measuring the bar length to derive an average value. We have used only one method that is rather precise, but it estimates a slightly larger value for $R_b$ compared to other methods \citep{Aguerri_2015}. However, we emphasize again that our main purpose in this paper is to compare different theories with each other, and not with real observations. Consequently, from the above discussion, we only conclude that bars are faster in extended gravity models.

To quantify the distribution of $\mathcal{R}$ more precisely, we assume $\mathcal{\widetilde{R}} \equiv \log_{10} \mathcal{R}$ is distributed as a Gaussian with mean $\overline{\mathcal{\widetilde{R}}}$ and intrinsic dispersion $\sigma_{\mathcal{\widetilde{R}}}$, thereby imposing the physical prior that $\mathcal{R} > 0$ despite a completely uninformative prior on $\mathcal{\widetilde{R}}$. We infer the population parameters $\left(\overline{\mathcal{\widetilde{R}}}, \sigma_{\mathcal{\widetilde{R}}} \right)$ from observations and using different theoretical models. Values of $\mathcal{R}$ calculated from barred galaxies in the EAGLE cosmological simulation at redshift $z = 0$ should be directly comparable to observations of nearby barred galaxies. {As an example, the MW has a fast bar with $\mathcal{R} = 1.22 \pm 0.11$ \citep[section 10.1 of][]{Portail_2017}.} \footnote{{The corotation radius of $6.1 \pm 0.5$ kpc in \citet{Portail_2017} is consistent with the $6.6 \pm 0.2$ kpc reported by \citet{Chiba_2021}. Note also that, as evident from our Figure~\ref{pspeed}, the bar pattern speed can oscillate over time by up to $\approx 10$\% in extended gravity theories due to couplings with other modes in the disc \citep[see also][]{Hilmi_2020}.}}

\begin{table}
	\caption{\emph{Top:} The observational sample of galaxies that we use to quantify the distribution of $\mathcal{R}$ (Equation \ref{R}). For \citet{Aguerri_2015}, we read off the results in figure~9 of \citet{Algorry_2017}. Much of our data comes from \citet{Guo_2019} $-$ we use the right panel of their figure~11. \emph{Bottom}: Number of galaxies in the EAGLE simulation used to quantify the expected distribution of $\mathcal{R}$ at the indicated redshift. We use all the EAGLE galaxies analysed by \citet{Algorry_2017}.}
	\centering
	\begin{tabular}{ccc}
		\hline
		Reference & Number & Number used \\
		& of galaxies & in our analysis \\
		\hline
		\citet{Corsini_2011} & 17 & 9 \\
		\citet{Cuomo_2019} & 16 & 2 \\
		\citet{Aguerri_2015} & 15 & 5 \\
		\citet{Guo_2019} & 17 & 13 \\
		\hline
		EAGLE ($z = 0$) & \multicolumn{2}{c}{48} \\
		EAGLE ($z = 0.27$) & \multicolumn{2}{c}{41} \\
		EAGLE ($z = 0.5$) & \multicolumn{2}{c}{32} \\
		\hline
	\end{tabular}
	\label{Galaxy_sample}
\end{table}

The likelihood $P$ of any $\left(\overline{\mathcal{\widetilde{R}}}, \sigma_{\mathcal{\widetilde{R}}} \right)$ combination is:
\begin{eqnarray}
	P \left( \overline{\mathcal{\widetilde{R}}}, \sigma_{\mathcal{\widetilde{R}}} \right) = \prod_i \frac{1}{\sqrt{{\sigma_{\mathcal{\widetilde{R}}}}^2 + {\sigma_i}^2}} \exp \left( {-\frac{\left(\overline{\mathcal{\widetilde{R}}} - \mathcal{\widetilde{R}}_i \right)^2 }{2 \left({\sigma_{\mathcal{\widetilde{R}}}}^2 + {\sigma_i}^2 \right)}} \right) ,
    \label{P_R_sigmaR}
\end{eqnarray}
where $i$ runs over different simulated or observed galaxies. In the simulations, we assume no measurement error $\sigma_i$ in the value of $\mathcal{\widetilde{R}}$, since any such uncertainty is expected to be very small compared to other uncertainties. We apply a similar analysis to the observational sample summarized in Table \ref{Galaxy_sample}. To begin with, we average the low and high error bars to come up with a single uncertainty $\delta R_b$ for each measured length. We then require $R_b$ to have a fractional uncertainty
\begin{eqnarray}
	\frac{\delta R_b}{R_b} ~<~ \epsilon \, ,
\end{eqnarray}
where the quality control parameter $\epsilon = \frac{1}{3}$. The analogous criterion is imposed on the corotation radius $R_c$ and its uncertainty $\delta R_c$. We find that $\epsilon = \frac{1}{3}$ achieves a good compromise between the quality and quantity of data, {with observational difficulties lying mainly in the determination of $R_c$}. We then estimate the fractional uncertainty in $\mathcal{R}$ as:
\begin{eqnarray}
	\alpha ~\equiv~ \frac{\delta \mathcal{R}}{\mathcal{R}} ~=~ \sqrt{\left(\frac{\delta R_b}{R_b} \right)^2 + \left(\frac{\delta R_c}{R_c} \right)^2} \, .
\end{eqnarray}
To further assure the quality of our dataset, we require that:
\begin{eqnarray}
	\alpha ~<~ \epsilon \, .
\end{eqnarray}
Despite this, $\alpha$ is sometimes not very small. Thus, we assume that a good estimate for $\sigma_i$ is:
\begin{eqnarray}
	\sigma_i ~=~ \frac{1}{2} \log_{10} \left( \frac{1 + \alpha}{1 - \alpha} \right) \, .
\end{eqnarray}

\begin{figure}
	\centerline{\includegraphics[width = 8.5cm]{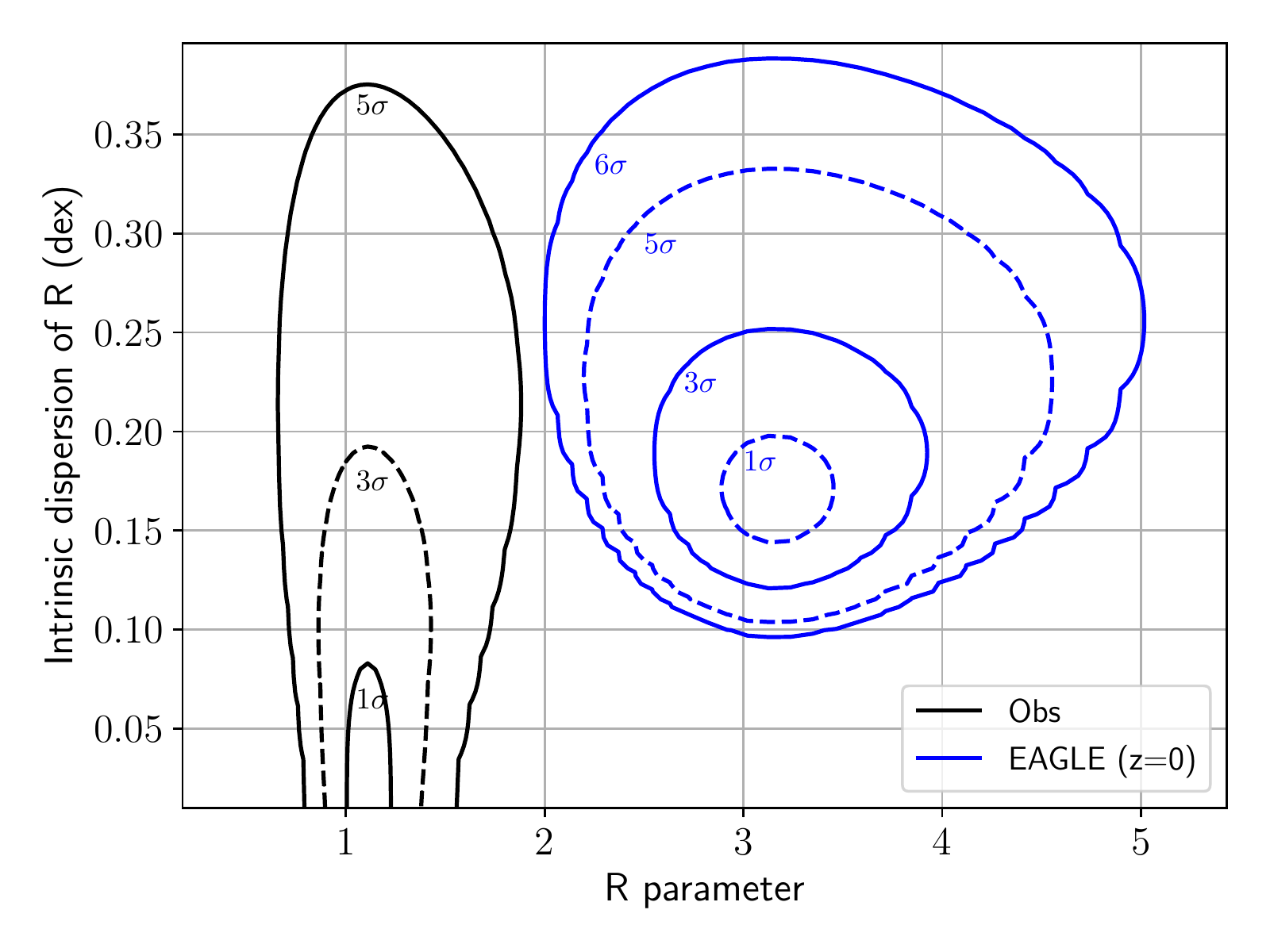}}
	\caption{The posterior inference on $\mathcal{R}$ and the intrinsic dispersion of $\log_{10} \mathcal{R}$, found by applying Equation \ref{P_R_sigmaR} to our compilation of observational results (Table \ref{Galaxy_sample}) and to the EAGLE simulation at ${z = 0}$ based on figure~9 of \citet{Algorry_2017}. Although the calculations are done in the space of $\log_{10} \mathcal{R}$, we change the $x$-axis to a linear scale when plotting so the results are more intuitive (i.e. we plot $10^{\overline{\mathcal{\widetilde{R}}}}$). The black (blue) contours correspond to ${1\sigma}$, ${3\sigma}$, and ${5\sigma}$ outliers from the observed (EAGLE) posterior. Due to the significant mismatch, the ${6\sigma}$ contour is also shown for the EAGLE simulation.}
	\label{EAGLE_data_comparison}
\end{figure}

Figure~\ref{EAGLE_data_comparison} shows our posteriors on $\left(\overline{\mathcal{\widetilde{R}}}, \sigma_{\mathcal{\widetilde{R}}} \right)$ based on a high-resolution grid in both parameters, with the resulting array then normalised to a sum of 1. There is a very significant mismatch between the EAGLE \citep{Algorry_2017} and observational posteriors, mainly because observations prefer $\mathcal{R} \approx 1$ while EAGLE galaxies prefer $\mathcal{R} \approx 3$ with more scatter. The $5\sigma$ allowed regions consistent with EAGLE and with observations represent distinct parts of parameter space, demonstrating that the two are incompatible at $>5\sigma$. Thus, we also show the $6\sigma$ confidence interval for the EAGLE galaxies. This still does not intersect the $5\sigma$ observational contour. Therefore, we expect that the level of disagreement is slightly above $\sqrt{5^2 + 6^2} = 7.8\sigma$.

To quantify the probability that EAGLE galaxy bars are compatible with observations, we pick some $\left(\overline{\mathcal{\widetilde{R}}}, \sigma_{\mathcal{\widetilde{R}}} \right)$ and draw the EAGLE contour through that point. We then find the probability that the EAGLE population parameters lie outside this contour, yielding a $P$-value. Since the observations do not uniquely specify $\left(\overline{\mathcal{\widetilde{R}}}, \sigma_{\mathcal{\widetilde{R}}} \right)$, we repeat this calculation for all different parameter combinations. Our final result is obtained by averaging the individual $P$-values, each weighted according to the observational probability of the corresponding $\left(\overline{\mathcal{\widetilde{R}}}, \sigma_{\mathcal{\widetilde{R}}} \right)$. In this way, we find that the EAGLE results shown in figure~9 of \citet{Algorry_2017} are incompatible with observations at $7.96\sigma$ confidence, in line with our previous estimate of slightly above $7.8\sigma$. Clearly, this $\Lambda$CDM model does not provide a good explanation for the observed distribution of $\mathcal{R}$  in barred spiral galaxies. The disagreement is so serious that we had to modify our algorithm so the $P$-value is not numerically rounded down to 0.

To compare EAGLE data with observations of nearby galaxies, one should consider simulated galaxies at $z = 0$, when the age of the universe is 13.8~Gyr \citep{Planck_2015}. Due to their 4~Gyr duration, our LPH simulations are expected to be most comparable to EAGLE galaxies at $z = 0.5$, when the age of the universe is $\approx 8.9$~Gyr. This is based on the average behaviour of bars in EAGLE \citep[figure~6 of][]{Algorry_2017} $-$ the bar instability happens around $z \approx 1.3$ ($t \approx 5$~Gyr), after which the bars enter a smooth and stable phase. Our 4~Gyr long simulations thus take us up to $t \approx 9$~Gyr. To make the comparison with observations more accurate, we also extend the high-resolution LPH simulation to 6~Gyr, confirming that the bar continues to slow down (Figure~\ref{r1}). 

\begin{figure}
	\centerline{\includegraphics[width = 8.5cm]{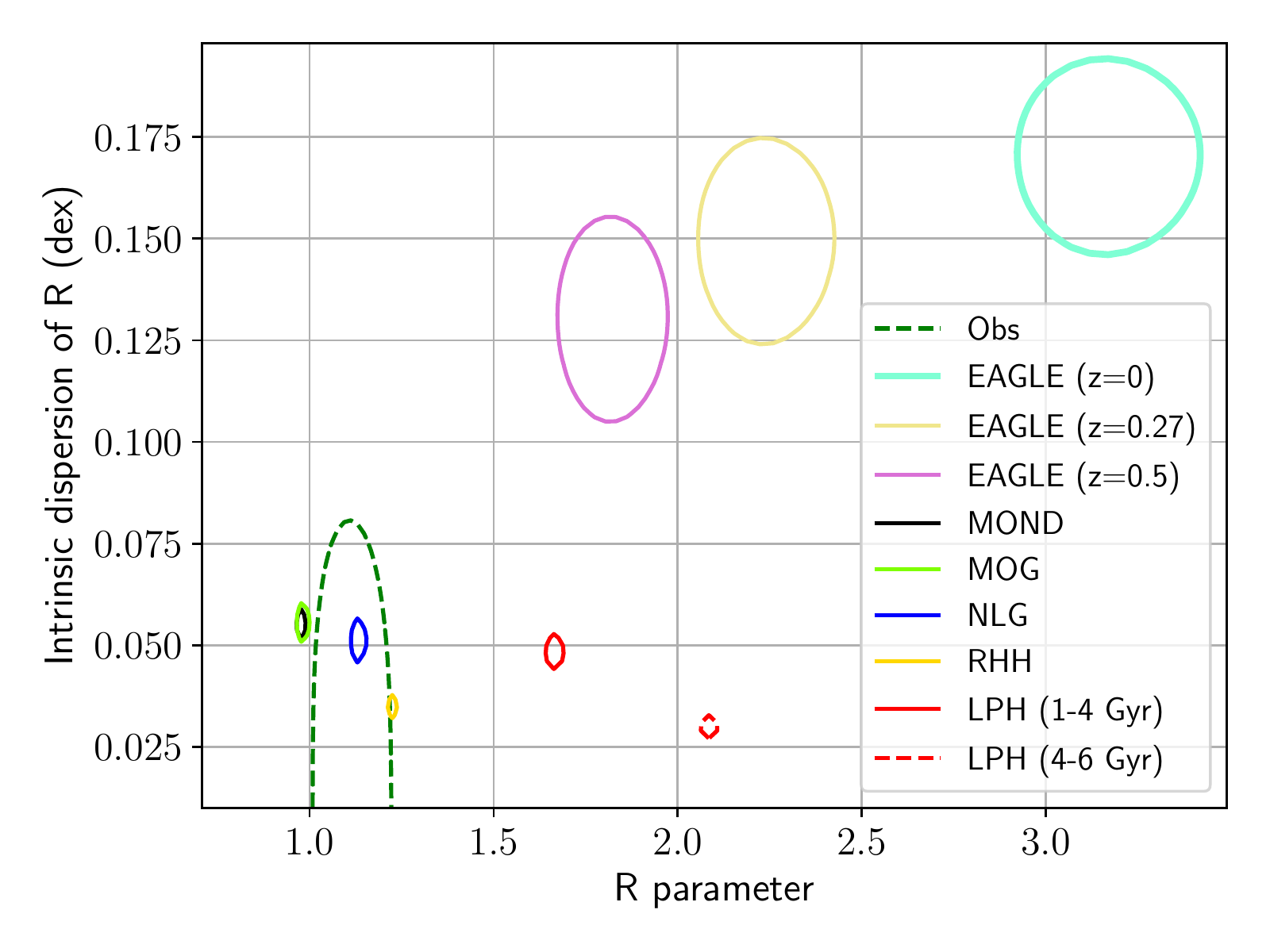}}
	\caption{Similar to Figure~\ref{EAGLE_data_comparison}, but now showing only the $1\sigma$ allowed region observationally and in different theories for our higher resolution simulations. Different snapshots are assumed to represent the diversity of galaxies at the same cosmic time. Our simulations are most comparable to EAGLE at $z = 0.5$ (see text). Notice that the EAGLE region moves to higher $\mathcal{R}$ as time passes, a consequence of dynamical friction. {This is also evident in our LPH model, results of which are shown separately for different periods in the simulation (solid and dashed red ellipses).} Our RHH and extended gravity models lack this process as there is no live halo, leading to significantly faster bars (lower $\mathcal{R}$) whose pattern speed changes little with time (Figure~\ref{r1}).}
	\label{Model_model_comparison}
\end{figure}

We treat $\mathcal{R}$ at different times in our simulations as representing the diversity of $\mathcal{R}$ in different galaxies at the same time. This is only an approximate approach, so a detailed comparison of our simulations with observations is not meaningful $-$ our main objective is to compare theories with each other. For this purpose, we use Figure~\ref{Model_model_comparison} to show the $1\sigma$ allowed regions of $\left(\overline{\mathcal{R}}, \sigma_{\mathcal{\widetilde{R}}} \right)$ for different theories and for EAGLE at $z = 0$, 0.27, and 0.5. As expected, our LPH model yields a similar $\overline{\mathcal{R}}$ to EAGLE at $z = 0.5$, though with less dispersion because we use only one galaxy sampled at different times. The effect of dynamical friction is apparent in that the EAGLE-preferred $\mathcal{R}$ increases with time, as also occurs in our LPH model (solid and dashed red contours in Figure \ref{Model_model_comparison}).

The LPH model is a clear outlier to both observations and the extended gravity models (Figure~\ref{Model_model_comparison}). This is due to its unique increasing behaviour of $\mathcal{R}$ at later times (top panel in Figure~\ref{r1}), which is related to the long-term decline in $\Omega_p$ (inset to Figure~\ref{pspeed}). Using more massive haloes (which makes the disc sub-maximal) would make $\mathcal{R}$ grow even faster since a higher DM density causes stronger dynamical friction. We check this by performing another simulation where $r_h$ is reduced from 12 kpc to only 8 kpc, but the truncation radius and halo mass are left unchanged. In this case, the bar is stronger and the dynamical friction is much more effective than in the maximal disc $-$ at the end of the simulation, $\mathcal{R} \approx 3.5$. This agrees with the general intuition that dynamical friction is enhanced when there is more DM.

Dynamical friction is absent in our RHH and extended gravity models, leading to much lower $\mathcal{R}$. This causes much better agreement with observations, which imply $\mathcal{R} \approx 1$ with little intrinsic scatter (Figure~\ref{EAGLE_data_comparison}). Therefore, our results on the $\mathcal{R}$ parameter strongly suggest that the anomalous rotation curves of galaxies are better understood as arising from a modification to gravity rather than from haloes of particle DM capable of exerting dynamical friction. It is important to mention that the lack of dynamical friction in MOND also implies galaxies evolve without merging much \citep{Renaud_2016}, which may explain the high observed frequency of thin bulgeless disc galaxies \citep{Kormendy_2010, Peebles_2020}.

\section{Discussion}
\label{Discussion}

\subsection{Numerical consistency tests}
\label{Numerical_convergence}

\begin{figure}
	\centerline{\includegraphics[width = 8.5cm]{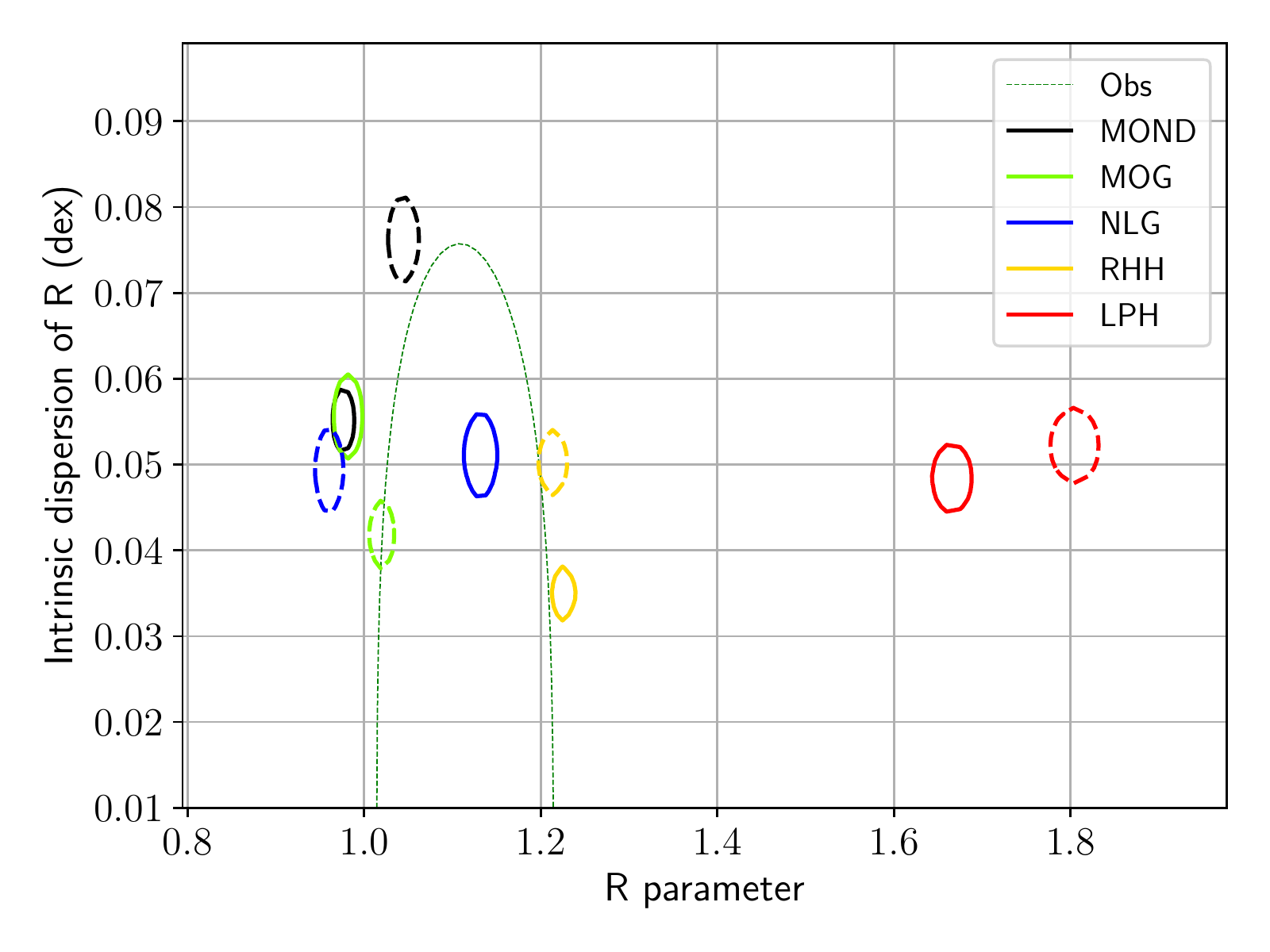}}
	\caption{Similar to Figure~\ref{Model_model_comparison}, but now showing the effect of numerical resolution. The $1\sigma$ allowed regions for simulations with $N = 10^6$ are shown as dashed contours, while the corresponding simulation with $N = 5 \times 10^6$ is shown using a solid contour of the same colour. For reference, the observationally allowed $1\sigma$ region is shown as a thin solid green line.}
	\label{Numerical_convergence_graph}
\end{figure}

To assure the integrity of our results, we checked that the energy and angular momentum are conserved for the LPH, RHH, NLG, and MOG models to an accuracy of better than $5\%$ throughout the full duration of the simulation. One should also check that the main results are unaffected by changing the particle number (i.e. $N$ should be large enough to suppress artefacts and shot noise). As already reported, changing the particle number from 1 to 5 million keeps the results consistent (Figure~\ref{r1}). Moreover, we decreased the time step $\delta t$ to increase the precision. To ensure that the results are independent of the adopted grid, we varied the number of grid points and the softening length (Table \ref{Galaxy_parameters}). These variations do not affect the overall behaviour of our models.

Our \textsc{por} simulations use a refinement condition based on the number of particles per cell (Section \ref{por}). Thus, using $5\times$ as many particles automatically increases the spatial resolution used by the potential solver in many parts of the simulated volume. Since the \textsc{por} results are not much changed by quintupling the number of particles, they appear to be numerically converged.

The main aim of this paper is to see if using an extended gravity theory to replace the role of CDM can reduce the typical value of $\mathcal{R}$ from $\approx 3$ to $\approx 1$, as required to explain observations. Figure~\ref{Numerical_convergence_graph} shows that our statistical analysis of the $\mathcal{R}$ parameter is not much affected by the choice of $N$ for any of our explored models. Thus, our main results are not dependent on the resolution $-$ though of course we generally focus on the models with $N = 5 \times 10^6$.

\subsection{Scaling results to other parameters}
\label{Scaling_results}

As discussed in Section \ref{Initial_conditions}, our simulations use a central disc surface density intermediate between the major Local Group galaxies. In MOND, the central surface density is the only dimensionless parameter of the matter distribution once the aspect ratio is fixed. This allows our results to be scaled to discs with the same central surface density but a different size and mass. This is also true for our LPH model because spiral galaxies fall on a rather tight RAR \citep{Lelli_2017}. As a result, the DM fraction within a fixed number of disc scale lengths must remain the same. This means that our LPH model addresses the evolution of maximal discs more generally than just discs with the parameters given in Table \ref{Galaxy_parameters}. The behaviour is more complicated in other extended gravity theories with a fundamental length scale, so results for these cannot simply be scaled to a galaxy with different mass and size.

We consider scaling the distances in our models by some factor $k$. Observationally, this is analogous to the effect of changing the heliocentric distance but keeping the same angular size. The scaling relations are given below, with primed (unprimed) variables indicating quantities in the scaled (original) version of our model:
\begin{eqnarray}
	\bm{r}' ~&\equiv&~ k\bm{r} \\
	\Sigma' \left( \bm{r}' \right) ~&=&~ \Sigma \left( \bm{r} \right) \\
	M' ~&=&~ k^2 M \\
	\bm{v}' \left( \bm{r}' \right) ~&=&~ \sqrt{k} \, \bm{v} \left( \bm{r} \right) \\
	t' ~=&=&~ \sqrt{k} \, t \\
	\rho' \left( \bm{r}' \right) ~&=&~ k^{-1} \rho \left( \bm{r} \right) \, .
\end{eqnarray}

To make the mass MW-like, we consider the case $k = 3$. The peak rotation velocity then rises from $\approx 150$~km/s (Figure~\ref{vhsb}) to almost 260~km/s. The 6~Gyr evolution of our LPH model now corresponds to a longer effective duration of 10.4~Gyr, which covers most of the Hubble time. Since our results are applicable to maximal discs more broadly, we argue that the steady increase in bar strength (Figure~\ref{barhsb}) and the $\mathcal{R}$ parameter (Figure~\ref{r1}) are likely generic features of maximal discs in $\Lambda$CDM, at least if they are not disturbed too frequently and lie on the empirical RAR.

\subsection{Ultrafast bars}
\label{Ultrafast_bars}

\begin{figure*}
	\centerline{\includegraphics[width = 7.8cm]{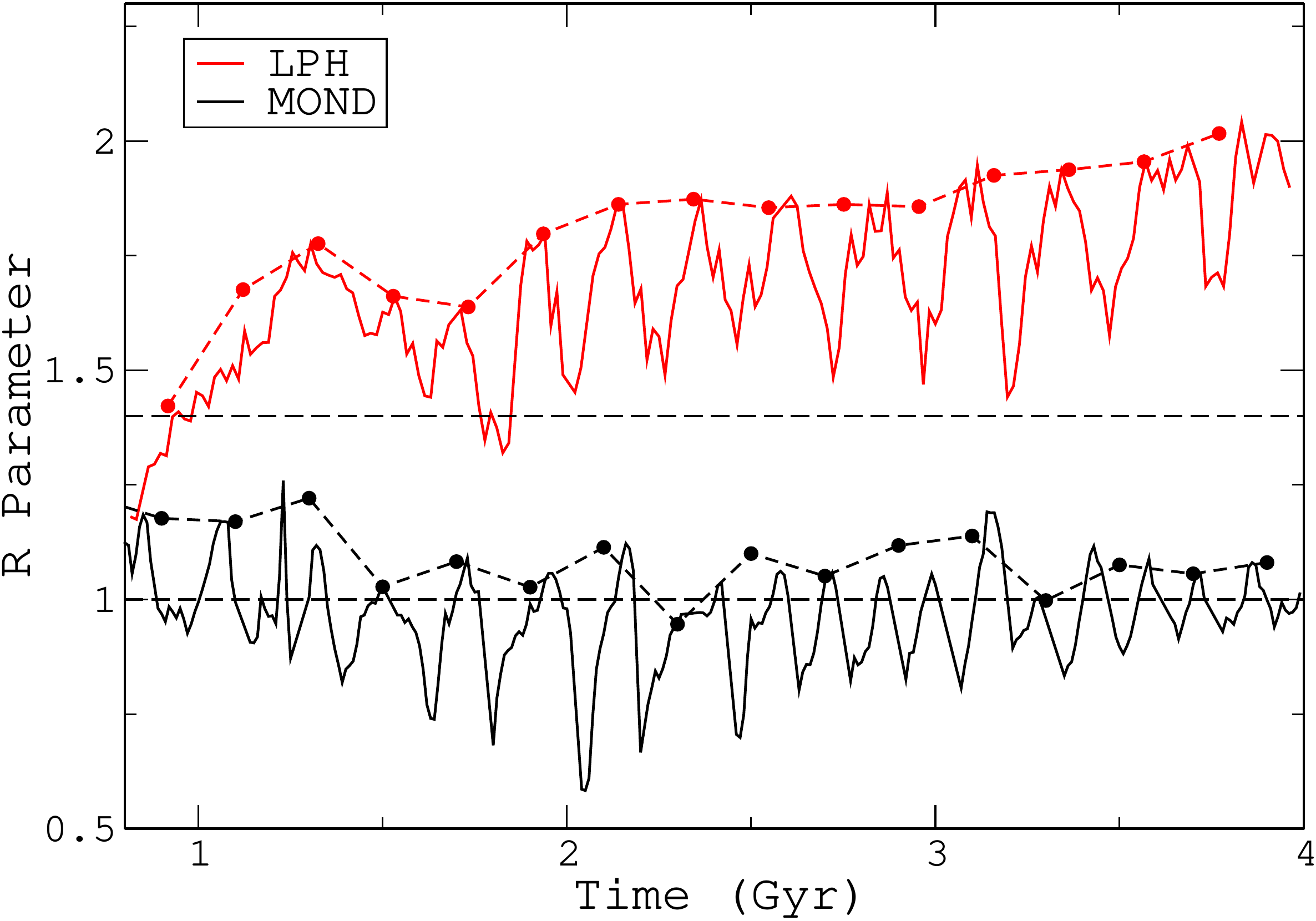}\hspace{0.9cm}\includegraphics[width = 7.6cm]{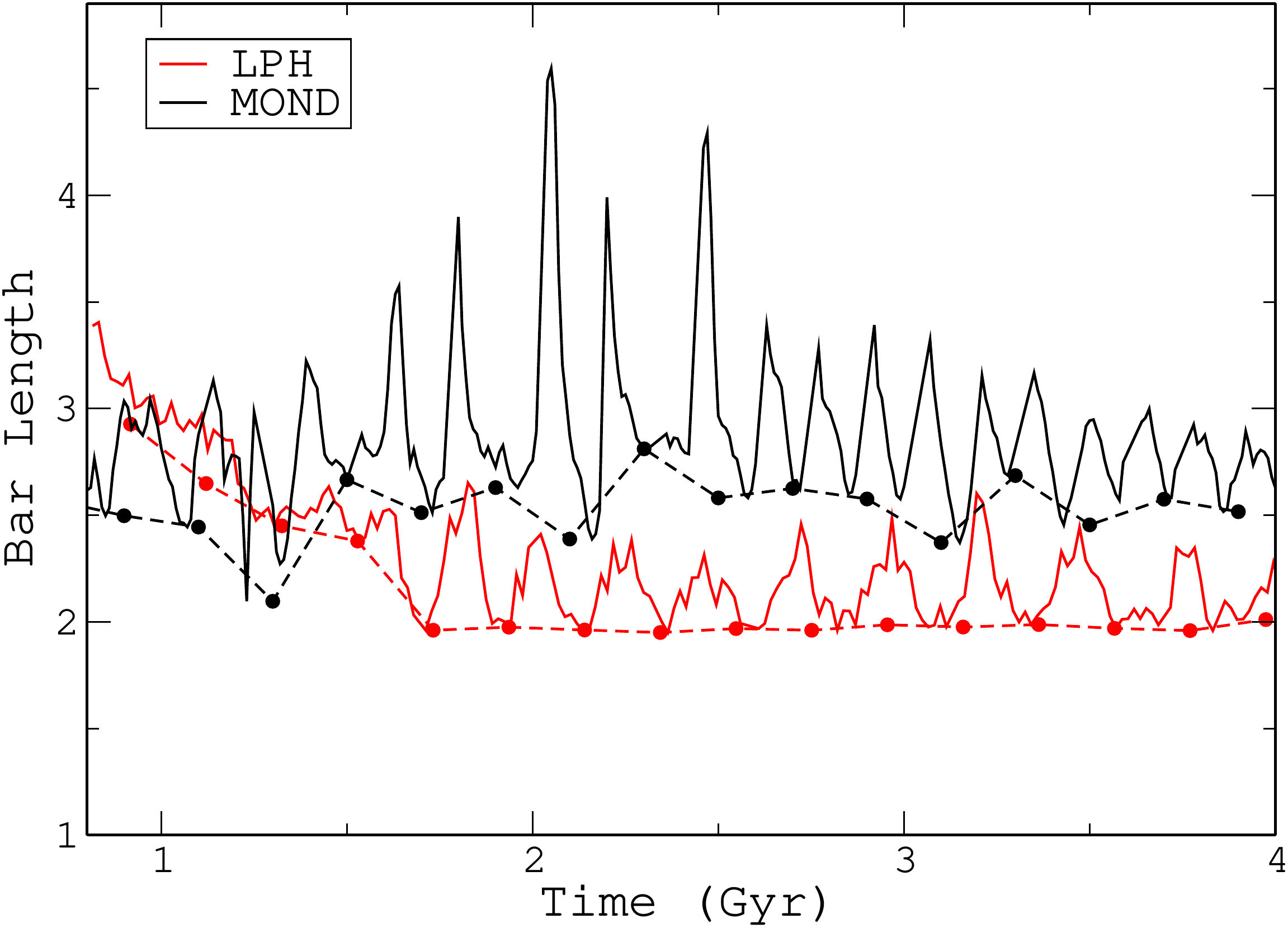}}
	\caption{Time evolution of the $\mathcal{R}$ parameter (left panel) and the bar length (right panel) for simulations with $N = 5 \times 10^6$. Solid curves show the complete time evolution, while dashed curves show the corresponding quantities at only those snapshots when the bar length is at a local minimum in time. The horizontal dashed lines in the left panel correspond to $\mathcal{R}=1$ and $1.4$, which demarcate the fast bar regime (ultrafast bars lie below 1).}
	\label{r2}
\end{figure*}

The bars in our extended gravity simulations spend some fraction of their time in the ultrafast regime (Figure~\ref{r1}). This is unexpected on theoretical grounds as a bar longer than its own corotation radius should be unstable \citep{Contopoulos_1980}. Since the bars in extended gravity are already in the fast regime, even a small error in calculating $\mathcal{R}$ could artificially push it into the ultrafast regime. In particular, the existence of several modes propagating on the disc induces apparent oscillations in the bar length \citep{Hilmi_2020}. In other words, the existence of spiral modes along with the bar in the central part of the disc artificially increases the bar length at times when they align, leading to a smaller $\mathcal{R}$ parameter $-$ possibly in the ultrafast regime.

This behaviour shows up in the MOND model. We illustrate this in Figure~\ref{r2}, where we have shown the complete time evolution of $\mathcal{R}$ and $R_b$ in our LPH and MOND models $-$ which we choose for illustration as other models behave similarly. It is clear that there are strong oscillations in $R_b$, and consequently in $\mathcal{R}$. To avoid this artefact, it is necessary to consider only minima in $R_b$ \citep{Hilmi_2020}. Therefore, we divide the evolution into time intervals of $\Delta t \approx 200$~Myr, choose the minimum value of $R_b$ in each interval, and compute the $\mathcal{R}$ parameter then. Different choices for $\Delta t$ do not change the main result, as long as $\Delta t$ is larger than the oscillation period.

Using Figure~\ref{r2}, one can imagine what happens if we instead choose the maxima in $R_b$. In this case, the $\mathcal{R}$ parameter would artificially drop. Although the bar still remains slow in the LPH case, in the MOND model the bar enters deep into the ultrafast regime.

\begin{figure}
	\centerline{\includegraphics[width = 8.4cm]{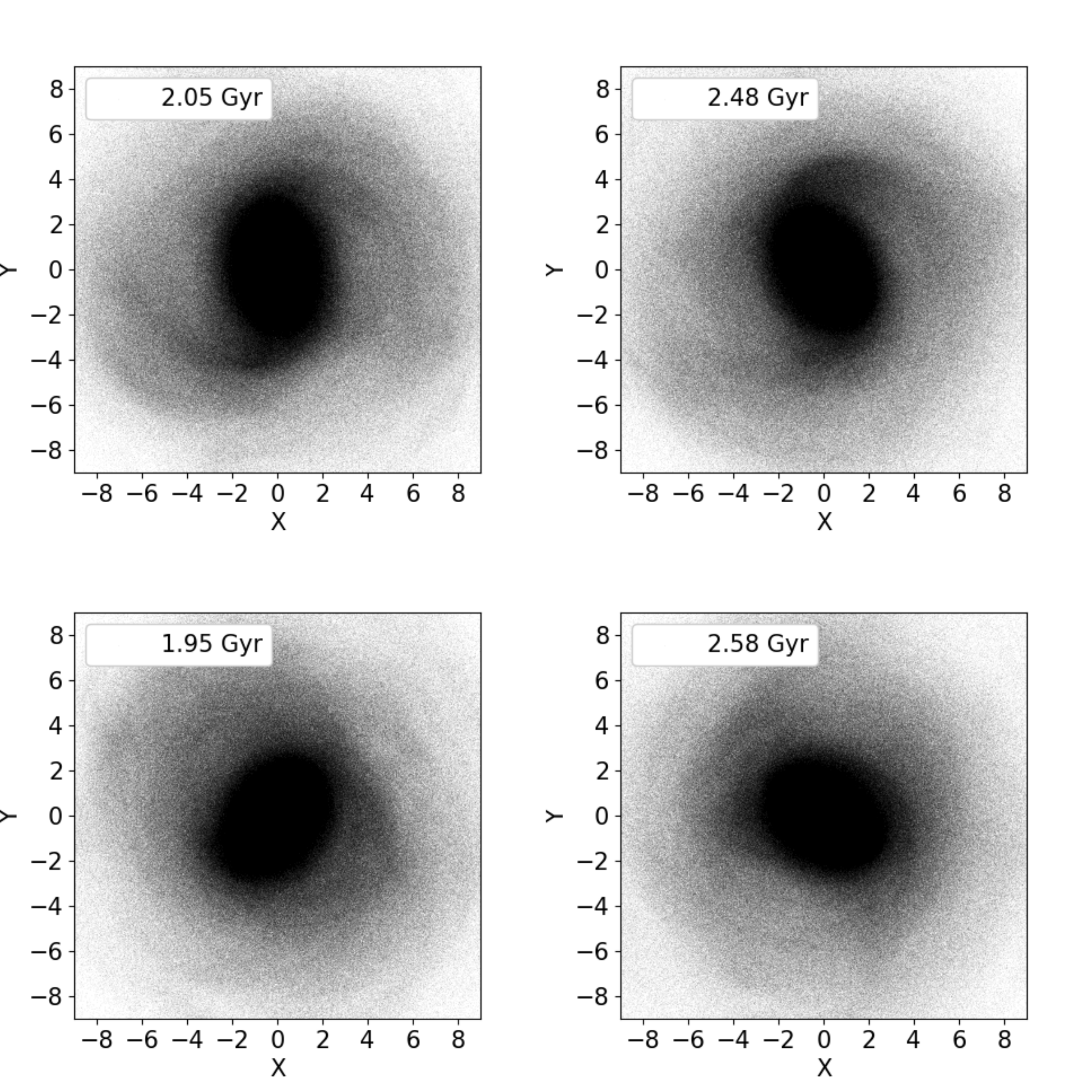}}
	\caption{Face-on projections of four snapshots in the MOND model. The presence of spiral arms (upper row) is simultaneous with the appearance of maxima in the right panel of Figure~\ref{r2}. The two snapshots in the lower row correspond to the minima in that figure. Distances are given in kpc.}
	\label{maxmin}
\end{figure}

For a better illustration, we also plot face-on projections of the system at four different times for the MOND model (Figure~\ref{maxmin}). The upper row shows that when $R_b$ has a peak in its time evolution, there are relatively strong spiral arms in the system. As a result, the calculated  $R_b$ is artificially large. However, as is clear in the lower row, such features are absent in the snapshots corresponding to minima in $R_b$.

As discussed in Section \ref{Power_spectrum}, the power spectrum is another powerful way to detect the different modes in a system. We have plotted the power spectrum of the MONDian disc in the lower left panel of Figure~\ref{p_s}. The intensities indicate that this model is even noisier than LPH and the other extended gravity models (other panels). The MOND disc has two prominent modes and two modes with less intensity. The lowest frequency mode appears in the disc outskirts, representing the spiral pattern illustrated in Figure~\ref{maxmin}. The period of this mode is $\approx 0.2$~Gyr, compatible with the $\approx 0.17$~Gyr oscillation period in Figure~\ref{r2}.

\begin{figure}
	\centerline{\includegraphics[width = 8.5cm]{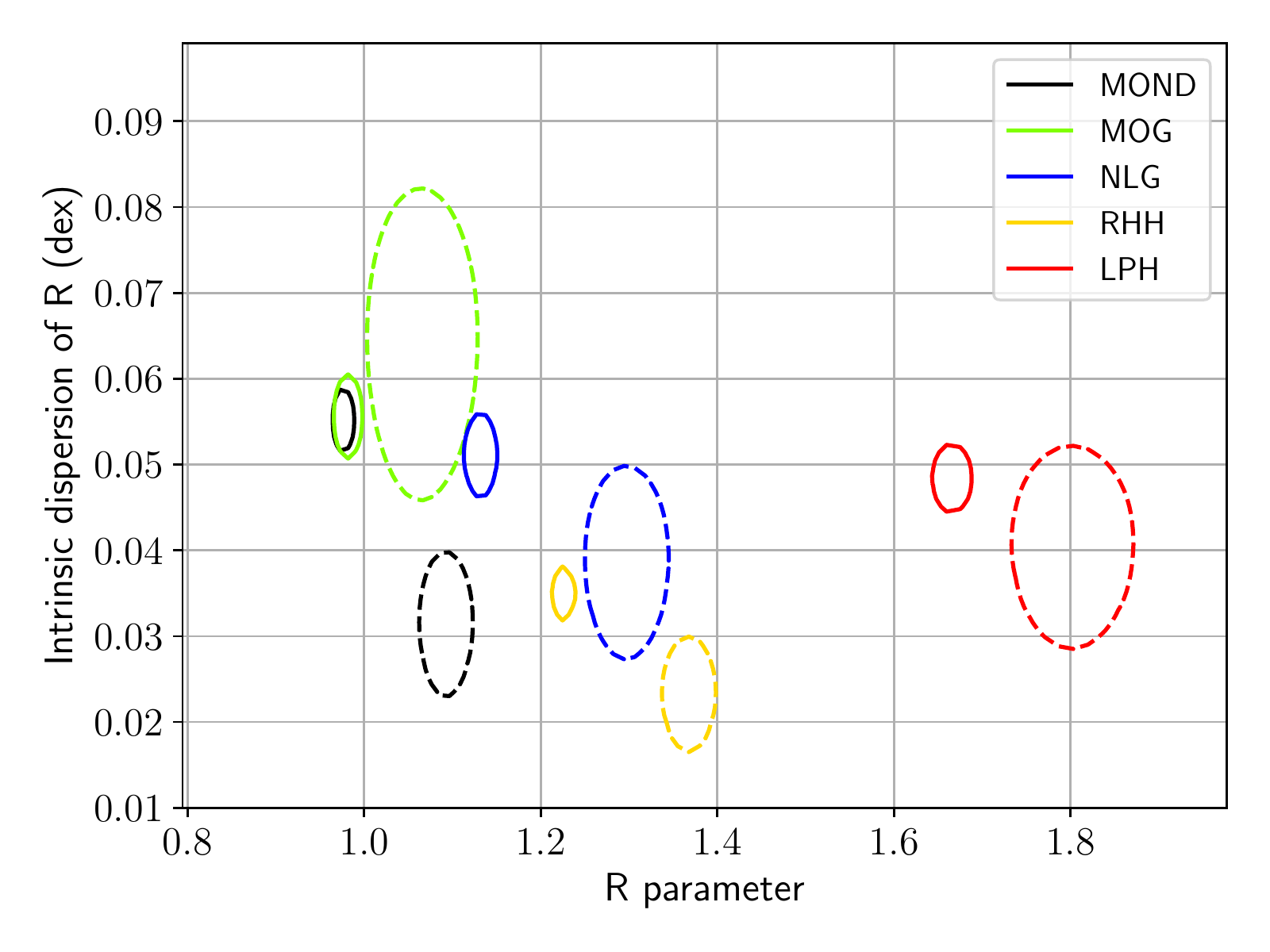}}
	\caption{Similar to Figure~\ref{Model_model_comparison}, but now showing the effect of considering only times at which the bar length is at a local minimum in its short-term oscillations. The solid $1\sigma$ confidence regions show results for all snapshots where $\mathcal{R}$ could be calculated. The dashed contours show the effect of considering only minima in $R_b$, with the same colour used for each model. Since the amount of data is $\approx 10\times$ smaller in this case, the error ellipse is much larger. Nonetheless, it is clear that the shift in $\mathcal{R}$ is enough to move the MOND and MOG models out of the ultrafast regime ($\mathcal{R} < 1$).}
	\label{R_true_inference}
\end{figure}

We can also consider how our statistical analysis (Section \ref{R_statistics}) would be affected if instead of using all snapshots where $\mathcal{R}$ can be calculated, we use only those corresponding to minima in $R_b$. This is done in Figure~\ref{R_true_inference}. Although uncertainties become larger due to the $\approx 10\times$ smaller amount of data, it is clear that this procedure removes the preference for $\mathcal{R} < 1$ in at least some timesteps previously apparent in Figure~\ref{Model_model_comparison}. Unfortunately, this procedure is very difficult to mimic in an observational sample, so it is unsuitable if the goal is to compare simulations with observations.

Our results show that oscillations in $R_b$ provide a plausible explanation for why some galaxies appear to have ultrafast bars $-$ though only if the bar is already in the fast regime. If the bar is deep in the slow regime, then we would need to very significantly over-estimate $R_b$, which is not very likely \citep{Hilmi_2020}. Moreover, the oscillations are much weaker in our LPH model (Figure~\ref{r2}). Thus, the issue should not affect our previous conclusion that barred spirals in EAGLE disagree very significantly with observations (Figure~\ref{EAGLE_data_comparison}).

\subsection{Higher Toomre parameter in the LPH model}
\label{Revised_Q_LPH}

We have argued that slow bars are expected in maximal $\Lambda$CDM discs because this is what happens in the EAGLE cosmological simulation and in our LPH model. The development of a bar could be inhibited by heating up the disc, which would involve a Toomre parameter $Q > 1$. Our nominal LPH models use $Q = 1.5$, but observationally there is some evidence that galaxy discs are dynamically overheated in the $\Lambda$CDM context such that higher values are appropriate \citep{Fuchs_2003, Saburova_2011, Das_2020}. A dynamically overheated disc with $Q = 2$ was able to suppress the bar instability in an idealised CDM-based Newtonian simulation of M33 \citep[section 4.3 of][]{Sellwood_2019}.

\begin{figure}
	\centerline{\includegraphics[width = 8.4cm]{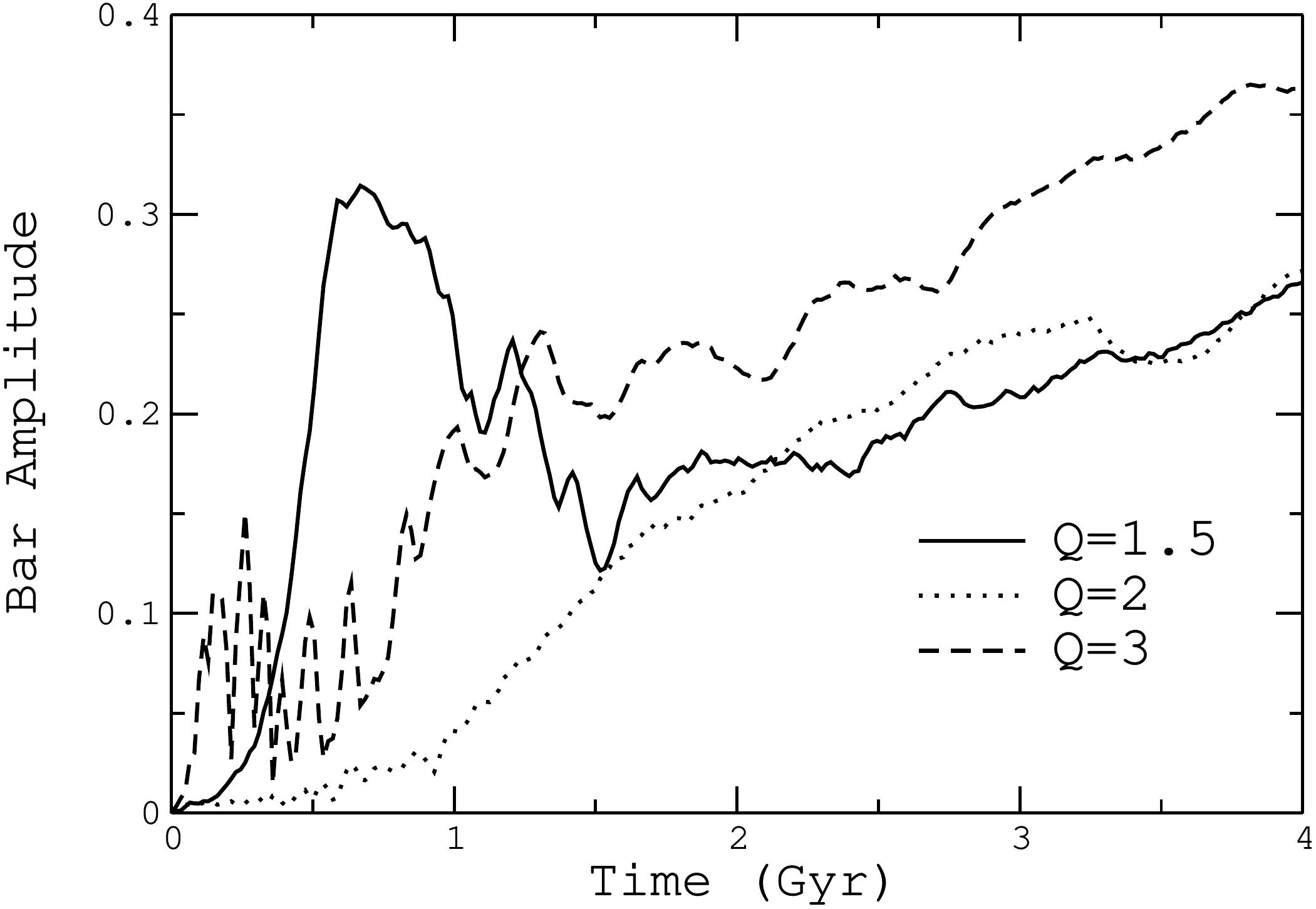}}
	\caption{Strength of the $\mathrm{m} = 2$ Fourier mode as a function of time in our LPH model with $N = 10^6$, shown here for three different values of $Q$ as indicated in the legend.}
	\label{Bar_strength_revised_Q_LPH}
\end{figure}

We therefore redo our LPH simulation with $N = 10^6$ for $Q = 2$ and $Q = 3$. {We implement a higher $Q$ by using a higher initial $\sigma_r$ (Equation \ref{Toomre_condition}). As before, $\sigma_{\phi}$ and $\sigma_z$ are set by solving the Jeans equations. Unlike in \citet{Sellwood_2019}, we keep the initial vertical scale height unchanged for models with higher $Q$.}

Figure \ref{Bar_strength_revised_Q_LPH} shows the evolution of the bar strength. All our LPH models have a rather strong bar by the end of the simulation. Interestingly, the model with the strongest bar after 4 Gyr is actually the $Q = 3$ model, which is the highest $Q$ model that we consider. This could be related to the bar being thicker in this case, which makes it less prone to the buckling instability \citep{Klypin_2009}. Those workers also found that models with a thicker disc (and presumably higher $Q$) have stronger bars. Our models with different $Q$ indeed exhibit differences in how the rms thickness evolves with time, but the values are very similar after 4 Gyr in all three cases considered {(Figure \ref{rms_z_HQ}).}

\begin{figure}
	\centerline{\includegraphics[width = 8.4cm]{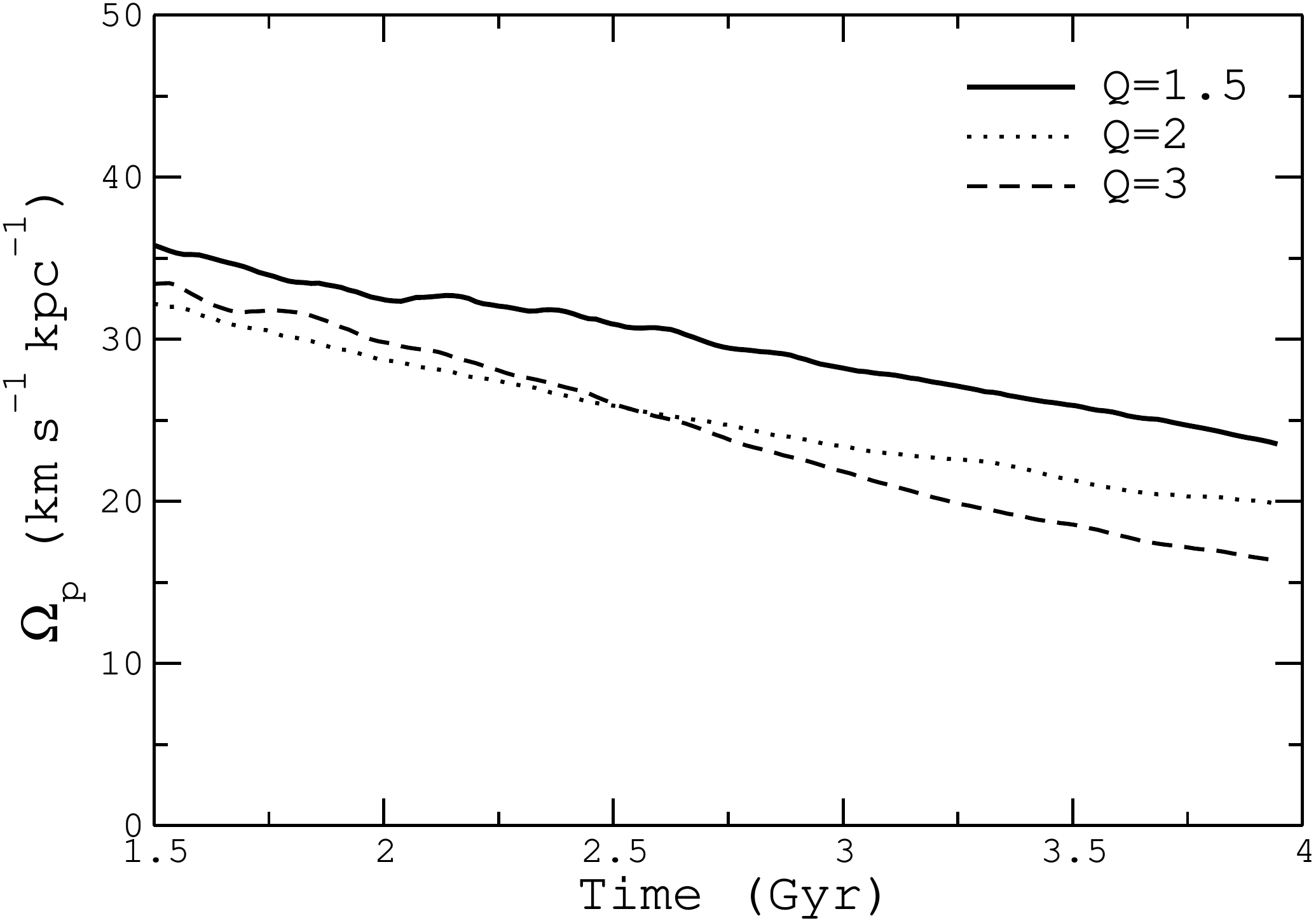}}
	\caption{{Similar to Figure \ref{Bar_strength_revised_Q_LPH}, but now showing the pattern speed $\Omega_p$ of the bar.}}
	\label{Pattern_speed_revised_Q_LPH}
\end{figure}

Having demonstrated that our higher $Q$ models also develop bars, we can analyse their pattern speed $\Omega_p$, which is shown in Figure \ref{Pattern_speed_revised_Q_LPH}. The bar slows down substantially in all cases, with a quite similar evolution regardless of the initial $Q$. This is consistent with {the earlier result of \citet{Widrow_2008}.}

\begin{figure}
	\centerline{\includegraphics[width = 8.4cm]{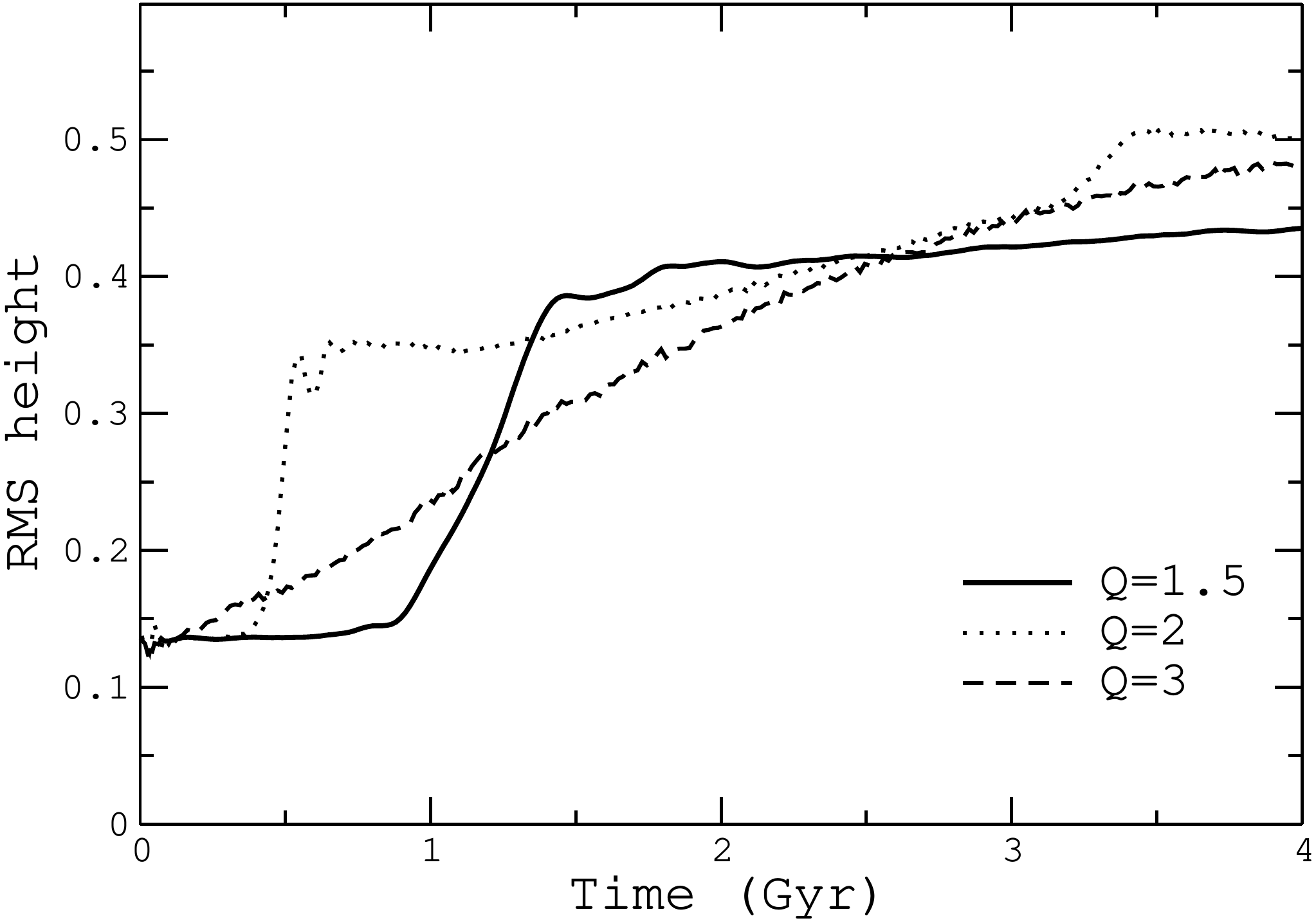}}
	\caption{Disc rms thickness in kpc computed for LPH models with different $Q$, measured at $R = 1.1$ kpc.}
	\label{rms_z_HQ}
\end{figure}

\begin{figure}
	\centerline{\includegraphics[width = 8.4cm]{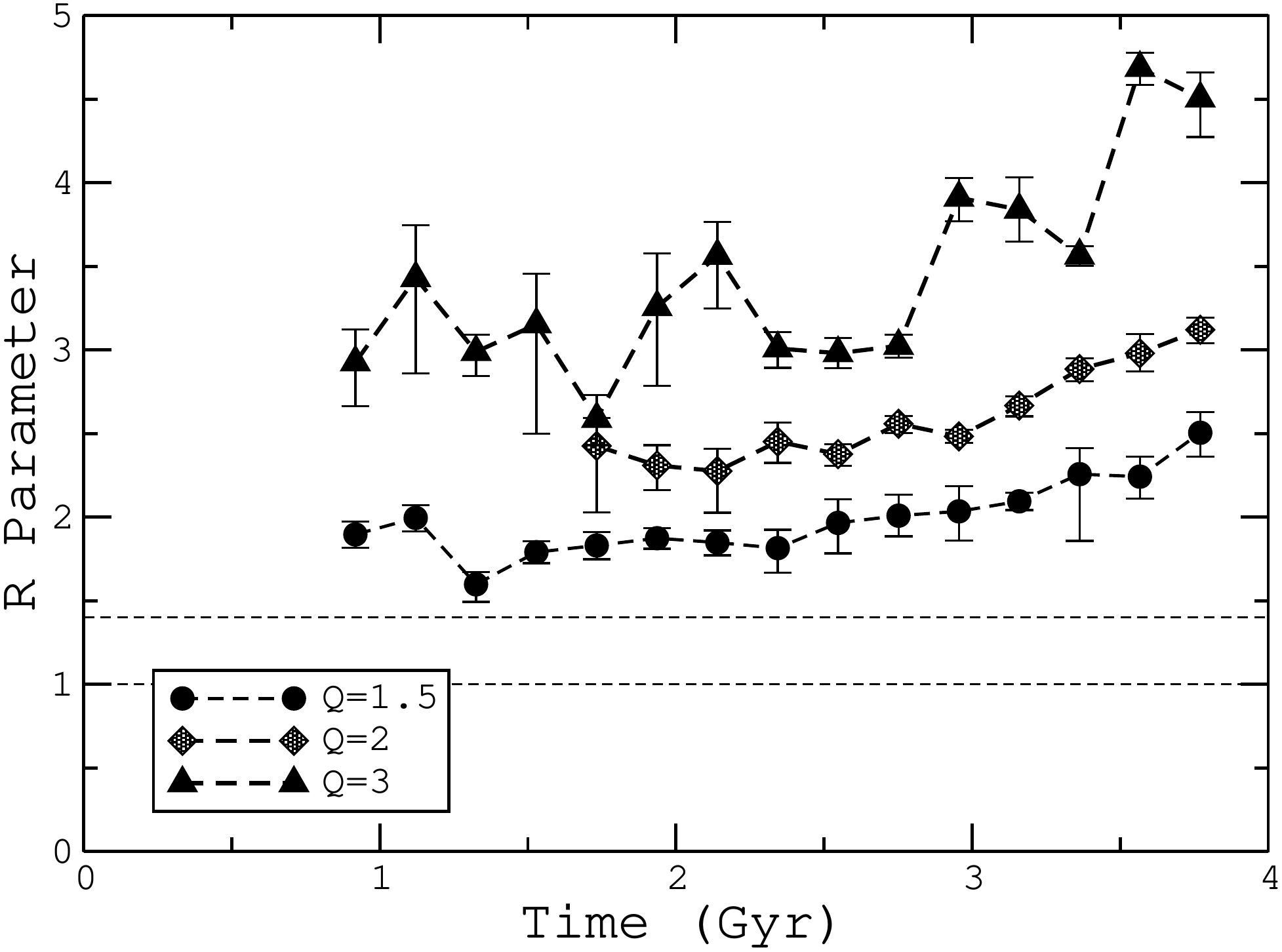}}
	\caption{Similar to Figure \ref{Bar_strength_revised_Q_LPH}, but now showing the $\mathcal{R}$ parameter (Equation \ref{R}). The horizontal lines demarcate the fast bar regime ($\mathcal{R} = 1 - 1.4$).}
	\label{R_revised_Q_LPH}
\end{figure}

We are now in a position to obtain the $\mathcal{R}$ parameter in our LPH models with higher $Q$. The results are shown in Figure \ref{R_revised_Q_LPH}. As expected from the decreasing $\Omega_p$, the $\mathcal{R}$ parameter rises well above the fast bar regime in all cases. Indeed, the $Q = 3$ model consistently has a higher $\mathcal{R}$ parameter than the other models. It is therefore clear that our LPH model yields a slow bar even if we start with a dynamically overheated disc. If anything, starting with higher $Q$ causes an even more significant tension with observations, which prefer $\mathcal{R} \approx 1$ with little intrinsic dispersion between galaxies (Figure \ref{EAGLE_data_comparison}). Moreover, self-regulated discs would be expected to have $Q \approx 1$ \citep{Silk_1997}.

\subsubsection{Comparison to previous isolated CDM models}
\label{Comparison_previous_models}

{The bar growth rate is lower in our LPH models with a higher $Q$, this being consistent with earlier results using 2D simulations \citep[figure 3 of][]{Athanassoula_1986}. Those authors were able to suppress the bar instability for $Q \ga 2-2.5$ (see their section 6.4). In contrast, our results indicate that higher $Q$ merely delays but does not prevent the bar instability, with a strong bar developing after 4 Gyr in all cases (Figure \ref{Bar_strength_revised_Q_LPH}). The difference could be due to motions out of the disc plane, and perhaps other details of the numerical implementation. Note also that many of their models do not have a DM halo.}

{The subsequent study of \citet{Athanassoula_2003} found that the bar slows down more gradually in models with higher $Q$, whereas we find $Q$ has little effect (Figure \ref{Pattern_speed_revised_Q_LPH}). However, their figure 4 shows the bar slowdown rate is nearly the same over the range $Q = 1.6-2.2$ (higher $Q$ models were not considered). This is broadly consistent with our LPH models, which only cover $Q \geq 1.5$. By considering models with even lower $Q$, \citet{Athanassoula_2003} showed that bars in galaxies with higher $Q$ slow down to a greater extent. This disagrees with the results of \citet{Klypin_2009}, possibly because the latter work used much shorter timesteps which are necessary to properly capture resonant bar-halo interactions that are crucial to slowing down the bar.}

{The halo density profile also plays a key role in the amount of friction. Unlike in this work, \citet{Athanassoula_2003} did not use a Plummer halo profile (see her equation 22). In a power-law profile where $\rho \propto r^{-\alpha}$ with $\alpha<0.5$, after a rapid dynamical friction phase at the beginning of the simulation, the rotating perturber experiences hardly any friction \citep{Read_2006}. As another example for the significance of the mass profile, we refer the reader to the model `LHH' (a responsive Hernquist halo model) in \citet{Ghafourian_2020}. The evolution of the pattern speed in this model is completely different from other models in the sense that the friction disappears for a relatively long time, even though all the models start from almost the same initial equilibrium state. The reasons for this difference is comprehensively explored in \citet{Ghafourian_2020}.}

{The distribution function of the halo particles also has a serious impact on the bar instability \citep{Sellwood_2016}. This means that the initial distribution function would indirectly influence the magnitude of the dynamical friction. In addition, the dynamical friction force on a moving perturber generally depends on the size of the host system. The truncation radius of our halos is generally larger than used in \citet{Athanassoula_2003}, since this work typically used 15 disc scale lengths whereas our LPH models use 24. We note that artificially truncated CDM haloes are not allowed in the $\Lambda$CDM model, so numerical experiments with CDM haloes that have such small radii are not physical.}

{Since our models have a baryonic surface density similar to the MW, they can be compared to the results of \citet{Widrow_2008}, who conducted 25 $N$-body simulations of MW-like galaxies. Their figure 19 shows that nearly all their models do have a strong bar that substantially slows down to a similar extent regardless of $Q$, which they varied over the range $1-2$. This is similar to our results in Figure \ref{Pattern_speed_revised_Q_LPH}.}

{Our results are also consistent with the work of \citet{Klypin_2009}, who considered models with a range of $Q$ with corresponding changes to the disc scale height. Those authors were careful to ensure the initial conditions were as realistic as possible for $\Lambda$CDM cosmology, and to use a high time resolution (see their sections 3.2 and 4.2, respectively). The main result of their work was that galaxies in dynamically hotter discs end up with longer bars that have a lower pattern speed and higher $\mathcal{R}$ parameter (see their table 2).}

{This contrasts with the M33 model of \citet{Sellwood_2019}, in which raising $Q$ to 2 suppressed the bar instability (see their figure 5). This could be due to their use of a sub-maximal disc (see their figure 2), as required in $\Lambda$CDM due to the low baryonic surface density and thus low acceleration. The combined effect of a dominant DM halo and an overheated disc might be able to suppress the bar, even if the latter alone cannot. The bar instability is also affected by the initial thickness of the disc, which was doubled in section 4.3 of \citet{Sellwood_2019} to suppress the instability. However, we do not increase the initial vertical scale height. The subsequent evolution of the disc rms thickness is also similar in all our LPH models (Figure \ref{rms_z_HQ}), so it may be that our discs are thinner than in \citet{Sellwood_2019}. Another difference is that we have not implemented any gas component, though the discussion in their section 3.2 suggests that this is not too crucial for a galaxy like M33.}

{The fact that our initial conditions are designed for consistency with the RAR (Section \ref{Initial_conditions}) might also underlie why our LPH model yields slow bars even though some previous galaxy simulations in the CDM context obtained fast bars for the maximal discs that we consider \citep{Athanassoula_2013, Athanassoula_2014}. Several problems have been identified with their conclusions \citep{Sellwood_2014b}, but the most important issue might be related to the halo properties and whether these are truly what one expects in the $\Lambda$CDM paradigm. Since we use a very similar algorithm to that used by \citet{Debattista_2000}, it is quite likely that this is the main reason for our simulations yielding slow bars for maximal discs. Ultimately, cosmological hydrodynamical simulations of $\Lambda$CDM must be used to check whether this model is consistent with the observed distribution of $\mathcal{R}$. We are currently investigating this using simulations other than EAGLE (Roshan et al. 2021, in prep).} 

{Due to the complex combination of underlying parameters that play a role in dynamical friction, the comparison between different simulations should be done with care. It seems that future work is still required to have a better understanding of how initial random motions affect the pattern speed of a stellar bar. An important constraint is that the statistical distribution of the initial conditions should follow from the cosmological model.}

\subsection{Relation to cosmological simulations}
\label{Relation_to_cosmology}

To fully understand how galaxies would behave in any theory, it is necessary to account for other processes not included in our simulations, which ultimately requires the cosmological context. Gas accretion from surrounding regions can rejuvenate the bar, whose strength could also be raised substantially by interactions with other galaxies \citep{Peschken_2019}. These processes are beyond the scope of our isolated $N$-body simulations, but we nonetheless consider their possible impact.

The amount of DM required in standard gravity is essentially fixed by the observational requirement for galaxies to lie on the empirical RAR \citep{Lelli_2017}. This is probably why although it may be possible to get fast bars in a $\Lambda$CDM context \citep{Fragkoudi_2021}, doing so causes tension with other constraints such as the stellar mass fraction inferred from abundance matching. Sitting on the RAR implies following the BTFR, which relates the baryonic mass $M_b$ to the asymptotic velocity $v_f$ according to a power law of the form $M_b \propto {v_f}^k$, with $k$ observationally very near to 4 \citep{McGaugh_Schombert_2015, Lelli_2019}. This is the expected value in MOND (Equation \ref{BTFR}). The value of $k$ in Auriga falls below the observed value \citep[figure 11 of][]{Grand_2017}. Consequently, at lower masses (and lower $v_f$), the baryonic mass will be larger than for galaxies on the observed BTFR. In a conventional gravity context, this would imply a lower amount of DM, which would reduce dynamical friction on the bar. This is highly relevant to the bar speed problem highlighted here since many of the galaxies used to obtain $\mathcal{R}$ empirically (listed in Table \ref{Galaxy_sample}) come from the study of \citet{Guo_2019}. This reaches down to $v_f \approx 100$ km/s, as is evident using circular velocities estimated from both Jeans anisotropic modelling and spatially resolved H$\alpha$ emission line measurements that show ``the average of the outer flat regions'' (see their figure 15). However, the Auriga galaxies all have $v_f \ga 160$ km/s \citep[figure 11 of][]{Grand_2017}. Comparing their table 1 to the absolute $r$-band magnitudes shown in figure 1 of \citet{Guo_2019} paints a consistent picture $-$ the observational sample in \citet{Guo_2019} reaches less massive galaxies than Auriga. Since less massive galaxies generally require a higher DM fraction in $\Lambda$CDM, the real challenge for this paradigm is to get fast bars at such low $v_f$ while still sitting on the RAR. Indeed, $\Lambda$CDM should reproduce ``fast bars across the Hubble sequence'' \citep{Aguerri_2015}, not just at the high mass end.

A reasonable fraction of strongly barred galaxies is required to reproduce observations \citep{Laurikainen_2002, Garcia_Gomez_2017}. Our simulated bars are rather weak (Figure \ref{barhsb}), with the more reliable higher-resolution runs indicating that the weakest bars occur in LPH for $t<4$~Gyr. However, strong bars are quite common in cosmological $\Lambda$CDM simulations \citep{Blazquez_2020}. This is no doubt related to processes like those mentioned above which are not included in our LPH model. These processes would also operate in extended gravity theories. Consequently, it is not presently clear whether the low bar strength in e.g. our isolated MOND model is a problem for MOND in general, or an issue that will be resolved with a cosmological simulation. Further work is required to address this issue, perhaps building on the MOND cosmology discussed in \citet{Haslbauer_2020}.

The evolution of the $\mathcal{R}$ parameter in our LPH model broadly agrees with that in the EAGLE cosmological simulation of $\Lambda$CDM (Figure \ref{Model_model_comparison}). Intuitively, this makes sense because the bar is like a normal mode in the disc, so external perturbations can change its amplitude but not its frequency. This suggests that our idealized LPH model captures the essence of the bar speed problem faced by $\Lambda$CDM, which we estimated to be at the $8\sigma$ level based on our analysis of results published elsewhere (Section \ref{R_statistics}). We can therefore postulate that the $\mathcal{R}$ parameter would remain similar in our extended gravity models if moving to a more advanced cosmological simulation. If this is correct, then these models would provide a good explanation for the observed distribution of $\mathcal{R}$ (Figure \ref{Numerical_convergence_graph}).

It is difficult to draw strong conclusions about the bar strengths given the significant differences between $\Lambda$CDM cosmological simulations and our idealized LPH model, which are both trying to represent $\Lambda$CDM. The similarity in bar strengths between $\Lambda$CDM and extended gravity theories (Figure \ref{barhsb}) suggests that this is not a very promising way to distinguish them. On the other hand, the significant differences between LPH and extended gravity theories suggests the distribution of the $\mathcal{R}$ parameter is a more promising test (Figure \ref{Model_model_comparison}). This is especially true given the similar population mean $\mathcal{R}$ and its rising trend between LPH and EAGLE, which suggests that our results are reliable with respect to the distribution of $\mathcal{R}$. In this regard, we conclude that the properties of galactic bars are likely better explained if galaxies lack CDM and obey non-Newtonian gravity, but this conclusion still needs to be checked by means of self-consistent cosmological simulations. These may reveal problems such as the bars being too weak, though it is important for a viable theory to sometimes produce a weak bar \citep[as demonstrated in a hydrodynamical MOND simulation of M33;][]{Banik_2020_M33}. Their model may be quite realistic as gas accretion may have been slowed down by M33 lying within the virial radius of M31, while a past pericentre (even if not very close) could have removed most of M33's satellites \citep{Patel_2018}. Moreover, the external gravitational field from M31 was included in section 4 of \citet{Banik_2020_M33}, which in MOND has a non-trivial effect beyond just moving M33 as a whole.

Turning to the case of $\Lambda$CDM, a somewhat promising aspect of our LPH models is that the bars are sometimes strong. This may be due to interactions, but our results in Figure \ref{barhsb} suggest that it could also arise in an isolated model if evolved for a long time due to bar-halo angular momentum exchange \citep[bottom panel of Figure~\ref{angular}, see also][]{Athanassoula_2002}. In addition to causing tension with isolated galaxies like M33 with a weak bar \citep{Sellwood_2019}, it is precisely this bar-halo interaction which causes the bar to slow down (Figure \ref{pspeed}), making $\mathcal{R} \gg 1$ and leading to strong tension with observations. If $\Lambda$CDM is the correct description of nature, some way should be found for the `clock' to be `reset' to avoid a similar fate in real galaxies. This would prevent the increasing bar strength at late times evident in Figure \ref{barhsb}, so strongly barred galaxies should be understood in some other way. We note that since such `reset' events are presumably caused by interactions and these are already included in cosmological simulations like EAGLE and Illustris, it is not at all clear why they would be much more frequent in a more realistic representation of $\Lambda$CDM. Rather, our results suggest that the problem it faces with regards to the $\mathcal{R}$ parameter is a fundamental consequence of having a live CDM halo, with the problem reproduced quite well in our idealized LPH simulation as it includes dynamical friction on the bar. As discussed in Section \ref{Comparison_previous_models}, this conclusion relies on choosing halo properties consistent with the observed rotation curves of galaxies, since otherwise one can always address the pattern speed problem by reducing the amount of DM and the resulting dynamical friction.

\subsection{Broader implications}
\label{Broader_implications}

As shown in Section \ref{R_statistics}, the properties of galactic bars are hard to reconcile with the latest cosmological simulations in a $\Lambda$CDM context. This is linked to the significant dynamical friction that a bar experiences when embedded in a live halo. Therefore, the slow bar problem is a generic failure of CDM-based models, as already reported in the literature \citep[e.g.][]{Algorry_2017, Peschken_2019}. Our results indicate that galaxies formed in the EAGLE simulations are strongly excluded observationally on the basis of their bar statistics. 

If similar statistics are recovered by other realisations of the $\Lambda$CDM paradigm, then any successes that it achieves on other scales should be viewed as a coincidence $-$ after all, an incorrect model with adjustable parameters can always be expected to match some observables. More generously, such successes can be viewed as a sign of partial correctness but with some fundamental missing ingredient(s), especially on galaxy scales. As an example, let us note that the CMB anisotropies, primordial nucleosynthesis, and cosmic expansion rate history can be explained in a MOND-based model with an additional collisionless matter component (e.g. sterile neutrinos) for much the same reasons as in $\Lambda$CDM \citep{Angus_2009, Haslbauer_2020}.

Another hybrid model is superfluid DM, where galaxies have DM haloes but dynamical friction is strongly suppressed for objects moving through the DM halo at subsonic velocities \citep{Berezhiani_2019}. Our results suggest that such models could also work with regards to bar pattern speeds. However, it is less clear how the Local Group satellite planes would be explained in this scenario $-$ the superfluid core for an MW-like galaxy is expected to have a radius of only $\approx 75$~kpc \citep[equation 18 of][]{Berezhiani_2016}, with more recent estimates also yielding similar values \citep{Hossenfelder_2020}. This is smaller than the radial extent of the MW satellite plane \citep{Pawlowski_2020, Isabel_2020}. If its members are tidal dwarfs (as required to explain their overall anisotropy), then any members $\ga 75$~kpc away would have extremely small internal velocity dispersions by virtue of lying outside the superfluid core. This would contradict the observed high dispersions \citep{McGaugh_2010}. A similar problem would arise around M31 since its satellites also have super-Newtonian velocity dispersions if they lack DM \citep{McGaugh_2013}, which is very likely for the 15 satellites which delineate a thin plane \citep{Ibata_2013, Sohn_2020}.

Note that the superfluid core size depends only weakly on galaxy mass once the theoretical parameters are fixed, so these results are rather robust. Moreover, a purely baryonic satellite whose eccentric orbit crosses the boundary of the superfluid core would have its internal gravity decrease (increase) by a very large factor when going out (in) through this boundary, likely leading to tidal disruption after a few orbits. This could be circumvented by altering the theoretical parameters to allow for larger superfluid cores, but it is not clear whether this is possible given constraints from other scales, e.g. galaxy clusters \citep{Hodson_2017}. In general, the MW and M31 satellite planes are more naturally understood in a MOND context as arising from a past MW-M31 flyby, which is required in MOND \citep{Zhao_2013} and likely reproduces the observed orientations \citep{Banik_Ryan_2018, Bilek_2018}.

\section{Summary and conclusions}
\label{conclusion}

In this paper, we used high-resolution $N$-body simulations to compare the dynamics of numerical galaxy models evolving in the context of four different gravity theories. Specifically, we constructed models in MOND, NLG, and MOG, and compared their evolution to the standard live DM model (LPH). Furthermore, we constructed a model with a rigid DM halo (RHH). The main purpose of this study is to find a way to discriminate between DM and extended gravity, especially by considering the bar. To explicitly quantify the angular speed of the stellar bar in our models, we measured $\mathcal{R}$ (Equation \ref{R}) at different times. The decreasing pattern speed in the LPH model appears as an increasing $\mathcal{R}$. Using the definition that $\mathcal{R}>1.4$ indicates a slow bar, the LPH model predicts slow bars, whilst all the extended gravity models studied in this paper lead to fast bars. Nearly all current measurements favour fast bars \citep[][]{Debattista_2000, Corsini_2011, Aguerri_2015, Cuomo_2019, Guo_2019}. Our main findings can be summarized as follows: 

\begin{enumerate}
	\item In the EAGLE implementation of $\Lambda$CDM, the average value of $\mathcal{R}$ in present-day barred spiral galaxies is $\approx 3$, whereas observations show $\approx 1$. By considering galaxies as having a log-normal distribution of $\mathcal{R}$ with some intrinsic dispersion, we show that the observationally inferred parameters differ from EAGLE at $7.96\sigma$ significance (Section \ref{R_statistics}). If confirmed in other suites of simulations \citep[e.g.][]{Peschken_2019}, this very serious discrepancy would rule out the $\Lambda$CDM paradigm as currently understood.

	\item The discrepancy could probably be alleviated by any of our explored extended gravity models without CDM (Figure~\ref{Model_model_comparison}). This is due to there being no effective dynamical friction in such models, causing the bar angular speed to remain constant with time. In the DM model, dynamical friction between halo particles and the bar (Figure~\ref{angular}) causes the pattern speed to undergo a clear decline with time (Figure~\ref{pspeed}), even if the disc is initialized with a higher Toomre parameter (Section \ref{Revised_Q_LPH}). But in extended gravity, there is no mechanism to remove angular momentum from the disc. This fact is directly responsible for the fast bars in all studied extended gravity models, including those which use a different disc surface density profile to the exponential law used here \citep{Ghafourian_2020}.

	\item The bar growth rate is higher in MOND and NLG compared to the DM model, so the bar instability happens much earlier. Consequently, one may conclude that discs are globally more unstable in these theories. This is well understood analytically in the case of MOND \citep{Milgrom_1989, Banik_2018_Toomre}, and is related to the phantom DM disc (Figure \ref{Sigma_p_MOND}).

	\item All extended gravity models predict weaker bars at the end of the simulations, though the bar spends a short time in the strong bar regime. Strong bars are frequently seen in real spiral galaxies \citep[e.g.][]{Laurikainen_2002, Garcia_Gomez_2017}. Therefore, this result may be a problem for extended gravity. However, encounters with other galaxies and more realistic simulations including gas accretion in a cosmological context are required to reach a reliable decision on the viability of extended gravity models (Section \ref{Relation_to_cosmology}). This work is currently under way for MOND (Wittenburg et al. 2021, in prep).

	\item The buckling instability happens earlier in the MOND model. Furthermore, all extended gravity models predict a smaller thickness for the inner parts of galactic discs compared to the DM case, though this is only marginally true for MOND (Figure \ref{meanrms}). It seems that resonances in the vertical direction happen more violently in the presence of a DM halo, even if they take longer to develop. This is related to disc-halo angular momentum exchange (Figure \ref{angular}) and the fact that disc self-gravity is very weak in $\Lambda$CDM compared to models with a completely self-gravitating disc. These differences are related to the weaker peanuts in extended gravity models, which may well lead to the weaker bars we obtain $-$ stronger peanuts seem to appear in the presence of stronger bars \citep{Martinez_Valpuesta_2008}. This means that the properties of the effective phantom DM halo in these models are significantly different from the DM model (Section \ref{Phantom_dark_matter}), leading to a different velocity dispersion at large radii. One consequence of stronger self-gravity is a thinner disc at fixed $\sigma_z$, or equivalently larger $\sigma_z$ at fixed thickness. Observational evidence for the latter was found by \citet{Das_2020}. We also mention that only NLG has strong flaring in the outskirts in the more reliable 5 {million} particle models.

	\item Galactic discs in extended gravity evolve to a larger radius than discs in the DM model initialized with the same baryonic content (Section \ref{Radial_expansion}). Combined with the above-mentioned findings, it seems that extended gravity models predict different morphologies for spiral galaxies. Future cosmological simulations in extended gravity should prove this claim. 

	\item Extended gravity models host more density waves propagating on the disc surface, especially in MOND (Figure~\ref{p_s}).
\end{enumerate}

Another important point is that bars sometimes appear to be ultrafast. Indeed, \citet{Guo_2019} found some cases with ultrafast bars, suggesting this may be consistent with observations. However, it is not expected theoretically \citep{Contopoulos_1980}. Following \citet{Hilmi_2020}, we found that apparently ultrafast bars arise due to bar-spiral arm alignment causing an overestimation of the bar length (Section \ref{Ultrafast_bars}). We addressed this issue by focusing on minima in the derived bar length, which undergoes short-term oscillations as the bar moves into and out of alignment with the spiral arms (Figure~\ref{maxmin}). This showed that our MOND model stays in the fast bar regime until the end of the simulation, and is not really ultrafast (Figure~\ref{r2}).

Bar length oscillations due to the existence of different density waves should also affect real observations. In other words, ultrafast bars reported in the literature may not really be ultrafast \citep{Hilmi_2020}. However, since the observational sample with reliable $\mathcal{R}$ measurements is already small, it would be very difficult to consider only those galaxies which are at a minimum in their bar length. Thus, we do not apply this procedure in our statistical comparison, which for consistency uses all timesteps where $\mathcal{R}$ can be reliably determined (Section \ref{R_statistics}). Figure~\ref{R_true_inference} shows that our results would not be much affected by restricting to only those timesteps where $R_b$ is at a local minimum in time, but the corresponding adjustment on the observational side is very unclear.

For a direct comparison with real galaxies, the simulations should include more baryonic physics and feedback. Important progress in this direction was recently achieved by the M33 model of \citet{Banik_2020_M33} and by \citet{Wittenburg_2020}, who simulated exponential disc galaxies forming out of a collapsing gas cloud in MOND. These isolated simulations are currently being extended to include the cosmological context based on a plausible MOND cosmology \citep{Angus_2009, Haslbauer_2020}.

As our final remark in this paper, let us reiterate that there are serious differences in the evolution of galactic discs with the same baryonic content and rotation curve depending on whether they are held together partly by DM particles or the detectable baryons alone in extended gravity. These deviations may help to discriminate between DM and extended gravity theories. As far as the $\mathcal{R}$ parameter is concerned, our results suggest a strong preference for the latter $-$ the EAGLE cosmological simulation in the $\Lambda$CDM context is in $8\sigma$ disagreement with observations \citep[see also][]{Algorry_2017, Peschken_2019}, while models where galaxies lack DM seem to fare much better due to the lack of dynamical friction on the bar (Figure~\ref{Model_model_comparison}). However, the absence of strong bars in our extended gravity models is not a satisfactory feature, an aspect which is better reproduced in our low-resolution LPH model \citep[though see][]{Sellwood_2019, Banik_2020_M33}. {This could be due to the idealized nature of our simulations, which lack processes like gas accretion and galaxy interactions that work to strengthen bars (Section \ref{Relation_to_cosmology}).} For the high-resolution LPH model, the strong bar appears at $t>4$ Gyr. To better assess which model is more consistent with observations, all the above-mentioned results in our extended gravity models should be carefully recovered in realistic galactic hydrodynamical simulations in a CDM-free context, and then compared with relevant observations. In addition, the EAGLE simulations are not the only realisations of the $\Lambda$CDM paradigm, so the bar statistics should be carefully investigated in other suites of simulations \citep[e.g. at least one very fast bar formed in the simulation of][]{Hilmi_2020}. The Illustris suite \citep{Vogelbserger_2014} would be well suited to this, with existing results suggesting that bars are slow here too \citep{Peschken_2019}. If these results continue to hold and it turns out that the $\Lambda$CDM paradigm must be replaced, then we must also bear other constraints in mind when deciding what the replacement should be (Section \ref{Broader_implications}). Such considerations must be done in an open-minded manner befitting the quest for the fundamental laws governing our Universe.

\section*{Acknowledgements}

IB is supported by an Alexander von Humboldt Foundation postdoctoral research fellowship. BF acknowledges funding from the Agence Nationale de la Recherche (ANR projects ANR-18-CE31-0006 and ANR-19-CE31-0017), and from the European Research Council (ERC) under the European Union's Horizon 2020 research and innovation programme (grant agreement number 834148). EA is supported by a stipend from the Stellar Populations and Dynamics Research Group at the University of Bonn. For NLG, MOG, RHH, and LPH simulations, this work made use of the Sci-HPC center of the Ferdowsi University of Mashhad. MR would like to thank Prof. Bahram Mashhoon for very insightful and constructive discussions on NLG. MR also appreciates Mohammad Hosseinirad for suggesting the \textsc{yt} project for visualizing the edge-on and face-on views of our models. The authors are grateful to the referee for suggesting additional ways of looking at our simulations, which greatly improved the manuscript.

\section*{DATA AVAILABILITY}
The data used in this article are available in the references given in the caption of Table \ref{Galaxy_sample}. Raw particle data for the high-resolution MOND simulation every 10~Myr are available upon request to IB or IT. Algorithms used to run the MOND simulations and extract data from them are discussed further in \citet{Banik_2020_M33}, which also contains links to download the algorithms. Installation and operating instructions are described in \citet{Nagesh_2021}.

\bibliographystyle{mnras}
\bibliography{short,galaxybonn} 
\bsp
\label{lastpage}
\end{document}